\documentclass[11 pt, a4paper]{article}
\pdfoutput=1
\usepackage{graphicx}
\usepackage{a4wide}
\usepackage{longtable}
\usepackage{amsmath}
\usepackage{amssymb}
\usepackage{mathabx}
\usepackage{pifont}% http://ctan.org/pkg/pifont
 \usepackage{hyperref}
\usepackage{color}
\usepackage{stackengine}

\title{}
\author{}

\numberwithin{equation}{section}

\newcommand{\mpl}{{M_{\rm {pl}}}}

\newcommand{\dd}{\, {\rm d}}
\newcommand{\gsim}{\;\mbox{\raisebox{-0.5ex}{$\stackrel{>}{\scriptstyle{\sim}}$}
}\;}
\newcommand{\lsim}{\;\mbox{\raisebox{-0.5ex}{$\stackrel{<}{\scriptstyle{\sim}}$}
}\;}

\newcommand{\pn}{\Phi_{\rm N}}

\newcommand{\oo}{\mathcal{O}}

\newcommand{\tg}{\tilde{g}}

\newcommand{\nm}{{\mu\nu}}

\newcommand{\eff}{_{\rm eff}}

\newcommand{\mmm}{{_{\rm m}}}

\newcommand{\GN}{G_{\rm N}}

\newcommand{\mc}{M_{\rm c}}
\newcommand{\lc}{\lambda_{\rm c}}
\newcommand{\pmi}{\phi_{\rm min}}
\newcommand{\ms}{M_{\rm s}}
\newcommand{\rst}{\rho_\star}
\newcommand{\brho}{\bar{\rho}}
\newcommand{\rs}{r_{\rm s}}
\newcommand{\cmark}{\ding{51}}%
\newcommand{\xmark}{\ding{55}}

\begin{document}

%\title{Tests of Chameleon Gravity}
%\author{Clare Burrage}
%\email[Email: ]{clare.burrage@nottingham.ac.uk}
%\affiliation{School of Physics and Astronomy, University of Nottingham, Nottingham, NG7 2RD, UK}
%\author{Jeremy Sakstein}
%\email[Email: ]{sakstein@physics.upenn.edu}
%\affiliation{Center for Particle Cosmology, Department of Physics and Astronomy, University of Pennsylvania 209 S. 33rd St., Philadelphia, PA 19104, USA}

%-------Note from JS: I took this part from the Penn/Chicago MG review. We can change it to suit our own needs later. I think it's better than using APS or anything else due to referencing conflicts and missing footnotes

%\setlength{\oddsidemargin}{0.1in}
%\setlength{\evensidemargin}{0.1in}
\setlength{\textwidth}{6.5in}

\renewcommand{\thefootnote}{\fnsymbol{footnote}}
%~
%\vspace{.01cm}
\begin{center}
{\Huge \bf{Tests of Chameleon Gravity}}
\end{center} 

\vspace{1.2truecm}
\thispagestyle{empty}
\centerline{{\Large Clare Burrage,${}^{\rm a,}$\footnote{\tt \href{mailto:clare.burrage@nottingham.ac.uk}{clare.burrage@nottingham.ac.uk}} and Jeremy Sakstein${}^{\rm b,}$}\footnote{\tt \href{mailto:sakstein@physics.upenn.edu}{sakstein@physics.upenn.edu}}}
\vspace{.5cm}

\centerline{\it${}^{\rm a}$School of Physics and Astronomy}
\centerline{\it University of Nottingham, Nottingham, NG7 2RD, UK}

\vspace{.5cm}

\centerline{\it${}^{\rm b}$Center for Particle Cosmology, Department of Physics and Astronomy}
\centerline{\it University of Pennsylvania, Philadelphia, PA 19104, USA}

%----------------

\begin{abstract}
Theories of modified gravity where light scalars with non-trivial self-interactions and non-minimal couplings  to matter---chameleon and symmetron theories---dynamically suppress deviations from general relativity in the solar system. On other scales, the environmental nature of the screening means that such scalars may be relevant. The highly-nonlinear nature of screening mechanisms means that they evade classical fifth-force searches, and there has been an intense effort towards designing new and novel tests to probe them, both in the laboratory and using astrophysical objects, and by reinterpreting existing datasets. The results of these searches are often presented using different parametrizations, which can make it difficult to compare constraints coming from different probes. The purpose of this review is to summarize the present state-of-the-art searches for screened scalars coupled to matter, and to translate the current bounds into a single parametrization to survey the state of the models. Presently, commonly studied chameleon models are well-constrained but less commonly studied models have large regions of parameter space that are still viable. Symmetron models are constrained well by astrophysical and laboratory tests, but there is a desert separating the two scales where the model is unconstrained. The coupling of chameleons to photons is tightly constrained but the symmetron coupling has yet to be explored. We also summarize the current bounds on $f(R)$ models that exhibit the chameleon mechanism (Hu \& Sawicki models). The simplest of these are well constrained by astrophysical probes, but there are currently few reported bounds for theories with higher powers of $R$. The review ends by discussing the future prospects for constraining screened modified gravity models further using upcoming and planned experiments.
\end{abstract}

\newpage
\tableofcontents
\newpage
\renewcommand*{\thefootnote}{\arabic{footnote}}
\setcounter{footnote}{0}

\parskip=1.5pt
\normalsize

% Introduction
\section{Introduction}

Since its publication in 1915, Einstein's theory of general relativity (GR) has withstood the barrage of observational tests that have been thrown at it over the last century. From Eddington's pioneering measurement of light bending by the Sun in 1919 to the first detection of gravitational waves by the LIGO/VIRGO consortium in 2015 \cite{Abbott:2016blz,TheLIGOScientific:2016src}, its predictions have been perfectly consistent with our observations. To test the predictions of any theory requires alternatives with differing predictions and, for this reason, alternative theories of gravity have a history that is almost as rich and varied as that of GR itself.

The zoo of modified gravity theories is both vast and diverse (see \cite{Clifton:2011jh,Joyce:2014kja,Koyama:2015vza,Bull:2015stt} for some compendia of popular models) but all have one thing in common: they break one of the underlying assumptions of general relativity. From a theoretical standpoint, GR is the unique {low-energy} theory of a Lorentz-invariant massless spin-2 particle \cite{Weinberg:1965rz}, and any modification must necessarily break one of these assumptions. Several interesting and viable Lorentz-violating theories exist that may have some insight for the quantum gravity problem \cite{Blas:2014aca}, and, similarly, healthy theories of massive spin-2 particles have recently been constructed \cite{deRham:2014zqa}. 

An alternative to these approaches is to introduce new fields that couple to gravity. One of the simplest possible options is to include a new scalar degree of freedom. These scalar-tensor theories of gravity are particularly prevalent, and are natural extensions of general relativity. Scalars coupled to gravity appear in many UV completions of GR such as string theory and other higher-dimensional models, but the cosmological constant problem and the nature of dark energy, two modern mysteries that GR alone cannot account for, are driving a vigorous research effort into infra-red scalar-tensor theories, with much of the effort focussing on light scalars (with cosmologically relevant masses)
%mass of order $10^{-33}$ eV (the Hubble scale) 
coupled to gravity.

Typically, the existence of such scalars are in tension with experimental bounds. If the scalar is massless, or has a Compton wavelength larger than the size of the solar system (which is certainly the case for Hubble-scale scalars), the theory's predictions typically fall within the remit of the parameterized post-Newtonian (PPN) formalism for testing gravity in the solar system (see \cite{Will:2004nx} and references therein). Scalars whose Compton wavelengths are smaller than $\sim $ AU predict deviations from the inverse-square law inside the solar system, which has been tested on interplanetary scales using lunar laser ranging (LLR) \cite{Williams:2004qba}, and down to distances of $\oo(\mu\textrm{m})$ using laboratory-based experiments such as the E\"{o}t-Wash torsion balance experiment \cite{Adelberger:2003zx}. In many cases, scalar-tensor theories spontaneously break the equivalence principle so that objects of identical mass but differing internal compositions fall at different rates in an external gravitational field. This too can be tested with LLR and terrestrial searches.

{Recently, the simultaneous observation of gravitational waves and a gamma ray burst from a binary neutron star merger (GW170817 and GRB 170817A) \cite{GBM:2017lvd,TheLIGOScientific:2017qsa} by the LIGO/Virgo collaboration and the Fermi and INTEGRAL satellites  has placed a new and stringent bound on modified gravity theories. The close arrival time of the gravitational wave and photon signal ($\delta t<1.7 $s) constrains the relative difference speed of photons ($c$) and gravitons ($c_T$) to be close to unity at the $10^{-15}$ level ($-3\times10^{-15} <|c_T^2-c^2|/c^2<7\times10^{-16} $)\cite{Sakstein:2017xjx,Ezquiaga:2017ekz,Creminelli:2017sry,Baker:2017hug,Crisostomi:2017lbg,Langlois:2017dyl,Dima:2017pwp,Bartolo:2017ibw} where the upper and lower bounds correspond to a $\sim 10$ s uncertainty in the time between the emission of the photons and the emission of the gravitational waves \cite{GBM:2017lvd}. Many scalar-tensor theories predict that the difference between the speeds of gravitons and photons is of order unity for models that act as dark energy \cite{Bellini:2014fua,Brax:2015dma} and so this bound represents a new hurdle for them to overcome.}

These stringent bounds imply that the simplest  theories with light scalars have couplings to matter that must be irrelevant on cosmological scales. Theories that try to avoid this problem using a large mass to pass solar system tests must have a Compton wavelength $\le\oo(\mu\textrm{m})$, in which case they too are cosmologically inconsequential. Ostensibly, it seems that scalar-tensor theories are trivial in a cosmological setting, but the link between solar system tests of gravity and cosmological scalar-tensor theories can be broken. Indeed, the last decade of scalar-tensor research can aptly be epitomized by two words: \emph{screening mechanisms}.

Screening mechanisms utilize non-linear dynamics to effectively decouple solar system and cosmological scales. At the heart of screening mechanisms lies the fact that there are 29 orders-of-magnitude separating the cosmological and terrestrial densities and 20 orders of magnitude separating their distance scales. As a result, the properties of the scalar can vary wildly in different environments. The {quintessential} example of a screening mechanism being used to ensure a dark energy scalar avoids solar system constraints is the chameleon mechanism \cite{Khoury:2003rn,Khoury:2003aq} {(earlier predecessors include \cite{Gessner:1992flm,Pietroni:2005pv,Olive:2007aj})}. In chameleon models, the mass of the scalar is an increasing function of the ambient density. This allows it to have a sub-micron Compton wavelength in the solar system but be light on cosmological scales. Later, a  closely related second dark energy screening mechanism was discovered: the symmetron mechanism \cite{Hinterbichler:2010es,Hinterbichler:2011ca}. Earlier work had studied a similar model but with a different motivation \cite{Pietroni:2005pv,Olive:2007aj}, and string-inspired models with
similar phenomenology have also been proposed \cite{Damour:1994zq,Brax:2011ja}. Unlike the chameleon, the symmetron has a light mass on all scales and instead screens by driving its coupling to matter to zero when the density exceeds a certain threshold. A third mechanism, the environment-dependent dilaton was subsequently discovered that screens in a similar manner \cite{Brax:2010gi}.

In this work we will only discuss screening mechanisms of this type, which rely on non-linear self-interaction terms in the potential.  A final class of screening, which relies on non-linearities in the kinetic sector screen through what is known as the Vainstein mechanism \cite{Babichev:2013usa,Joyce:2014kja}. These theories will not be discussed here as the phenomenology of these models, and therefore the most constraining observables, are very different to that of the chameleon and symmetron models. {Similarly, we will not discuss massive gravity \cite{deRham:2010kj,Hinterbichler:2011tt,deRham:2014zqa,deRham:2016nuf}, which screens using the Vainshtein mechanism, for the same reason. We note however that many models that do screen using the Vainshtein mechanism (as well as those that predict a mass in the graviton dispersion relation such as massive gravity) are severely constrained by the new bounds from the observation of gravitational waves and photons from GW170817/GRB 170817A discussed above if they are to simultaneously act as dark energy \cite{Sakstein:2017xjx,Baker:2017hug,Ezquiaga:2017ekz,Creminelli:2017sry,Crisostomi:2017lbg,Langlois:2017dyl}. (In the case of massive gravity, solar system tests are stronger than the LIGO/Fermi bound \cite{Baker:2017hug}.) The models we will discuss in this review (chameleon/symmetron/dilaton) predict that $c_T=c$ identically and so this bound is irrelevant for them.}

{Scalar fields with screening mechanisms cannot simultaneously screen and self-accelerate cosmologically \cite{Wang:2012kj} but they can act as a dark energy quintessence field \cite{Copeland:2006wr} i.e. they require a cosmological constant term to drive the cosmic acceleration and they are capable of producing deviations from GR on linear and non-linear cosmological scales as well as astrophysical scales (see \cite{Jain:2013wgs,Sakstein:2015oqa} and references therein). In addition to this, many candidate UV completions of GR such as string theory predict a multitude of scalars that couple to matter and screening mechanisms are a convenient method of hiding such additional degrees of freedom. For these reasons screening mechanisms are considered interesting and novel paragon for alternative theories of gravity and,} as such, there is an ongoing experimental search for screened scalars. Being designed to evade conventional tests of gravity, screening mechanisms have inspired novel and inventive approaches to search for them experimentally. These range from reinterpreting the results of experiments not designed to look for them, to designing instruments specifically adapted to testing their unique properties, to using astrophysical objects that have never before been used to test gravity, such as Cepheid stars and galaxy clusters. In many cases, new and imaginative scenarios have been concocted.

These searches typically use different parametrizations, making them difficult to compare with one another. The purpose of this review is to collect the state-of-the-art constraints coming from laboratory and astrophysical tests, and to combine them into a single parametrization. This not only makes it clear which models are ruled out by different experiments, but also aides in deciding the optimum search strategy for exploring the remaining models. In many cases, we will extend the experimental results  to models to which they have not previously been applied.

This review is organized as follows. In section \ref{sec:screening_mechanisms} we will introduce the different screening mechanisms we will consider in this review, outline their salient features, and present the parameters we will use to compare constraints. In section \ref{sec:screening} we will discuss how screening works in both astrophysical and laboratory settings. Section \ref{sec:experiments} contains a brief description of the experiments that have been used to constrain screening mechanisms, and translates the constraints into our parametrization. The crux of this review is presented in section \ref{sec:constraints}, where we combine all of the contemporary constraints from various experiments into a series of diagrams that show which regions of parameter space are ruled out, and how different experiments compare in the same parametrization. We do this for chameleon and symmetron modes. In section \ref{sec:conclusions} we conclude by discussing the implications of the constraints for screened modified gravity, and future prospects for constraining the remaining parameter space.

% Screened modified gravity
%\include{screened_MG}

% Screening Mechanisms --- JS: I merged this with the previous section to improve the flow.
\section{Screening Mechanisms}
\label{sec:screening_mechanisms}
%\subsection{Conformal Scalar-Tensor Theories}

The screening mechanisms that we consider in this review are all specific subsets of the general scalar-tensor theory
\begin{equation}
\label{eq:STtheoriesgeneral}
S=\int\dd^4 x\sqrt{-g}\left[\frac{R}{16\pi G}-\frac{1}{2}\nabla_\mu\phi\nabla^\mu\phi-V(\phi)\right]+S\mmm[\tg_{\mu\nu},\phi],
\end{equation}
which describes a canonically normalised scalar field $\phi$ subject to a potential $V(\phi)$ and conformally (Weyl) coupled to matter through the \emph{Jordan frame metric}
\begin{equation}
\label{eq:Weylgen}
\tg_\nm=A^2(\phi)g_\nm.
\end{equation}
It is this non-minimal coupling described by the \emph{coupling function} $A(\phi)$ that results in deviations from GR\footnote{Note that one could consider a more general theory where each particle species $i$ is conformally coupled to a different metric $\tg^{(i)}_\nm=A_i^2(\phi)g_\nm$ although we will not consider such theories here since they are not well-studied in the context of screened modified gravity. An even more general metric includes disformal terms $\tilde{g}_{\mu\nu}=A(\phi)g_{\mu\nu} +B(\phi)\partial_{\mu}\phi\partial_{\nu}\phi$ \cite{Bekenstein:1992pj}. Constraints on disformal couplings to matter can be found in \cite{Brax:2014vva,Sakstein:2014isa,Sakstein:2014aca,Ip:2015qsa,Sakstein:2015jca}.}. In particular, the \emph{Einstein frame metric}, $g_{\mu\nu}$, is a solution of Einstein's equations sourced by both matter and the scalar stress energy tensors, but matter moves on geodesics of the Jordan frame metric, $\tilde{g}_{\mu\nu}$. {In what follows we work only with the Einstein frame version of the theory.  Classically, all observable quantities will be independent of the choice of frame and our choice of the Einstein frame is purely for calculational convenience.  In the Jordan frame there would be no direct coupling between the scalar fields and matter, but instead the gravitational action will depend non-trivially on the scalar field.  In this frame matter particles travel on geodesics of the Jordan frame metric, but the evolution of the gravitational potentials is modified by their mixing with the scalar field.  }

As an example {of motion in the Einstein frame}, consider a non-relativistic particle in the Newtonian limit. This particle moves on geodesics of $\tg_\nm$ and so, defining the tensor $\mathcal{K}^\alpha_\nm\equiv\tilde{\Gamma}^\alpha_\nm-\Gamma^\alpha_\nm$, the Newtonian limit of the geodesic equation is \cite{Sakstein:2015oqa,Burrage:2016bwy}
\begin{equation}
\label{eq:geoNewt}
\ddot{x}^i+\Gamma^i_{00}=-\kappa^i_{00}=-\frac{\beta(\phi)}{\mpl}\nabla^i\phi,
\end{equation}
where a dot denotes a derivative with respect to proper time and we have calculated $\mathcal{K}^i_{00}$ using \eqref{eq:Weylgen} (see \cite{wald2010general,Zumalacarregui:2013pma}). The \emph{coupling} is
\begin{equation}\label{eq:betadef}
\beta(\phi)\equiv\mpl\frac{\dd\ln A}{\dd\phi}.
\end{equation}
The Christoffel symbol $\Gamma^i_{00}=\partial^i\Phi_{\rm N}$ contains the Newtonian force and so we can interpret
\begin{equation}
\label{eq:F5gen}
F_5=-\frac{\beta(\phi)}{\mpl}\nabla\phi
\end{equation}
as a new or \emph{fifth-}force. In this review we do not consider scalars with non-minimal kinetic terms which screen through the Vainshtein mechanism. 

Another important consequence of the coupling to matter is that the field is sourced by the trace of the energy-momentum tensor so that its equation of motion is
\begin{equation}
\label{eq:EOMgen}
\Box\phi=\frac{\dd V(\phi)}{\dd\phi}-\frac{\beta(\phi)T}{\mpl}.
\end{equation}
Note that $T=g_{\mu\nu}T^{\mu\nu}$ where $T^\nm=2/\sqrt{-g}\delta S\mmm/\delta g_\nm$ is the Einstein frame energy-momentum tensor. This is not covariantly conserved ($\nabla_\mu T^\nm\ne0$) because matter moves on geodesics of $\tg$; it is the Jordan frame metric $\tilde{T}^\nm=2/\sqrt{-\tg}\delta S\mmm/\delta \tg_\nm$ that satisfies $\tilde{\nabla}_\mu\tilde{T}^\nm=0$. The two are related via $T^\nm=A^6\tilde{T}^\nm$ \cite{wald2010general,Sakstein:2015oqa}. For non-relativistic matter, one has\footnote{There are three commonly used densities in the literature: the Jordan frame density $\tilde{\rho}=-\tilde{T^{0}_0}$, the Einstein frame density $\rho=-T^0_0=A^6(\phi)\tilde{\rho}$, and the `conserved Einstein frame density' $\rho_{\rm conserved}=A(\phi)\rho$. The Jordan frame density is the result of statistical physics calculations and it is in this frame that one may specify an equation of state. The Einstein frame density is what arises naturally in equation \eqref{eq:EOMgen} as a result of varying the action \eqref{eq:STtheoriesgeneral} and the conserved density is a quantity that is useful in cosmological contexts \cite{Khoury:2003rn,Hui:2009kc,Brax:2011aw,Brax:2011ta,Brax:2012gr,Sakstein:2015oqa}. In particular, the conserved density satisfies a conservation equation that makes the cosmological equations look similar to those of GR. Since this review is concerned with experimental tests of chameleon theories, we have opted to work with the Einstein frame density. At the Newtonian level (weak-field limit), these densities are equivalent \cite{Hui:2009kc,Sakstein:2015oqa} and so the choice is somewhat arbitrary, but we note that one must work with the Jordan frame pressure and density if one is interested in compact objects such as neutron stars \cite{Babichev:2009fi,Minamitsuji:2016hkk,Babichev:2016jom,Sakstein:2016oel,Brax:2017wcj}. We will not consider such objects here.} $T=-\rho\approx-\tilde{\rho}\approx\tilde{T}$, where we have ignored post-Newtonian terms \cite{Hui:2009kc,Sakstein:2014isa}. The equation of motion is then
\begin{equation}
\label{eq:EOM_rho}
\Box\phi=\frac{\dd V(\phi)}{\dd\phi}+\frac{\beta(\phi)\rho}{\mpl}=\frac{\dd V\eff}{\dd\phi},
\end{equation}
which defines a density-dependent effective potential\footnote{Several definitions of the effective potential exist in the literature. If one uses the conserved Einstein frame density then one has $V_{\rm eff}(\phi)=V'(\phi)+\rho A(\phi)$ \cite{Khoury:2003rn,Sakstein:2015oqa}. Furthermore, one often sees the effective potential written as $V_{\rm eff}(\phi)=V'(\phi)+(A(\phi)-1)\rho$ (using the conserved Einstein frame density). This is motivated by models that have $A(\phi)=1+\beta(\phi_0)\phi/\mpl+\cdots$ and the factor of $-1$ is used to subtract the matter density from the chameleon energy density in order to avoid double counting in cosmology. (The equation of motion \eqref{eq:EOM_rho} which governs the dynamics is unchanged by including such a factor.) Since we do not consider cosmology here we will not include this factor.   }
\begin{equation}
\label{eq:veff_gen}
V\eff(\phi)=V(\phi)+\rho\ln A(\phi).
\end{equation}
It is this that governs they dynamics of $\phi$ and not $V(\phi)$ solely.

In order to classify different screening mechanisms it is instructive to consider the field profile sourced by a spherical object of mass $\mathcal{M}$ and radius $\mathcal{R}$ embedded in a medium of background density $\rho_0$. If the effective potential has a minimum then the field in this medium will assume the value $\phi_0=\pmi(\rho_0)$ where this is achieved. 
%
%so that the equation of motion for the scalar \eqref{eq:EOMgen} is 
%%\begin{equation}
%\nabla^2\phi=\frac{\dd V}{\dd\phi}+\beta(\phi)\frac{\mathcal{M}}{\mpl}\delta^{(3)}(r).
%\end{equation}
Expanding the field about this background value $\phi=\phi_0+\delta\phi$, where $\delta\phi$ is the field sourced by the point mass, and $\phi_0$ the uniform background value (i.e. we have performed a background-object split), we have the equation of motion for a massive scalar
\begin{equation}
\nabla^2\delta\phi-m_{\rm eff}^2(\phi_0)\delta\phi=\frac{\beta(\phi_0)\rho(r)}{\mpl},
\end{equation}
where the effective mass 
\begin{equation}\label{eq:meff_def}
m_{\rm eff}^2(\phi)\equiv V\eff''(\phi)
\end{equation}
is the mass of small fluctuations about the minimum of the effective potential. The scalar potential outside the source is then
\begin{equation}\label{eq:varphi}
\delta\phi=\frac{\beta(\phi_0)}{4\pi\mpl}\frac{f(\mathcal{M},\mathcal{R})}{r}e^{-m\eff r},
\end{equation}
where the undetermined function $f(\mathcal{M},\mathcal{R})$ is a model dependent function of the source mass parameters.
When the source is a point mass one simply has $f(\mathcal{M},\mathcal{R})=\mathcal{M}$ but in general the effective mass may vary inside the object and the object may have a non-trivial density profile. 
From equation \eqref{eq:varphi}, it is clear that there are three ways one can suppress the effects of the scalar. Either
\begin{enumerate}
\item The mass $m\eff r\gg1$ so that the force is short ranged,
\item The coupling to matter $\beta(\phi_0)\ll1$, or 
\item Not all of the mass sources the scalar field.
\end{enumerate}
Of course, one could simply choose the parameters such that either of the first two conditions is satisfied but this leads to a trivial situation where the modifications of gravity are negligible on all scales, and requires fine-tuning the parameters. We are interested in theories where solar system tests are satisfied trivially but strong modifications may appear on other scales, producing new and interesting phenomena that may be relevant to small-scale physics or dark energy and cosmology. Said another way, we would like to construct theories that exhibit some environmental dependence of the screening, for example  so that conditions 1. or 2. are only satisfied locally. The density-dependence of the effective potential aids us here because the ambient density of different objects can vary over many orders of magnitude. For example, there are 29 orders of magnitude separating the mean cosmic density from the density on Earth. The essence of screening mechanisms is that the effective potential is chosen such that the minimum is density-dependent so that the field  can assume different values in different environments so that the scalar potential can be dynamically suppressed.
%inside different objects so that whether or not the conditions are satisfied depends on the situation in question. 

It is possible to construct models with the requisite density-dependent minimum such that one or more of the conditions above are satisfied. Models that utilize a combination of the first and third condition are typically known as \emph{chameleon} models\footnote{Chameleon models were the first example of screening mechanism that screens using this effect. The scalar \emph{blending in with its environment} inspired the name.} \cite{Khoury:2003rn} and models that utilize the second are known as either symmetron \cite{Hinterbichler:2010es} or dilaton models \cite{Brax:2010gi}. 
%The latter are poorly constrained observationally and so we will not consider them in this article at the present time.

%\subsection{Screening Mechanisms}

\subsection{Chameleon Screening}

As remarked above, the chameleon is constructed to give an effective mass that increases with the density. 
A wide variety of potentials and coupling functions can achieve this; here we follow the existing literature and focus on power law potentials and exponential couplings,
%This is achieved using the scalar potential and coupling function
\begin{equation}
\label{eqSM:chameleons}
V(\phi)=\frac{\Lambda^{n+4}}{\phi^n},\quad A(\phi)=e^{\frac{\phi}{\mc}},
\end{equation}
so that the effective potential is then
\begin{equation}\label{eqSM:Veffcham}
V\eff(\phi)=\frac{\Lambda^{n+4}}{\phi^n}+\rho\frac{\phi}{\mc},
\end{equation}
where $M_c = M_P/\beta$.  Theories with $M_c\sim M_P$, $\beta\sim 1$ have gravitational strength couplings to matter. The effective potential  has a density-dependent minimum given by
\begin{equation}\label{eqSM:phimincam}
\pmi(\rho)=\left(\frac{n\mc\Lambda^{4+n}}{\rho}\right)^{\frac{1}{n+1}}.
\end{equation}
The effective mass about this minimum is
\begin{equation}\label{eq:eqSM:meffcham}
m_{\rm eff}^2=V''_{\rm eff}(\phi)=n(n+1)\Lambda^{n+4}\left(\frac{\rho}{n\mc\Lambda^{n+4}}\right)^{\frac{n+2}{n+1}}.
\end{equation}
For $n>-1$ this certainly satisfies our requirement that the mass is an increasing function of the density, with the exception of $n=0$, which does not admit a minimum. Negative even integers i.e. $n=-4,-6,-8,\ldots$ also have this property with the exceptions $n=-1,\,-2$, which do not allow the mass to vary with the density. There is no minimum when $n=-3,-5,-7,\ldots$ and so there are no viable chameleon mechanism in these cases. 

\begin{figure}[ht]
\stackunder{
{\includegraphics[width=0.45\textwidth]{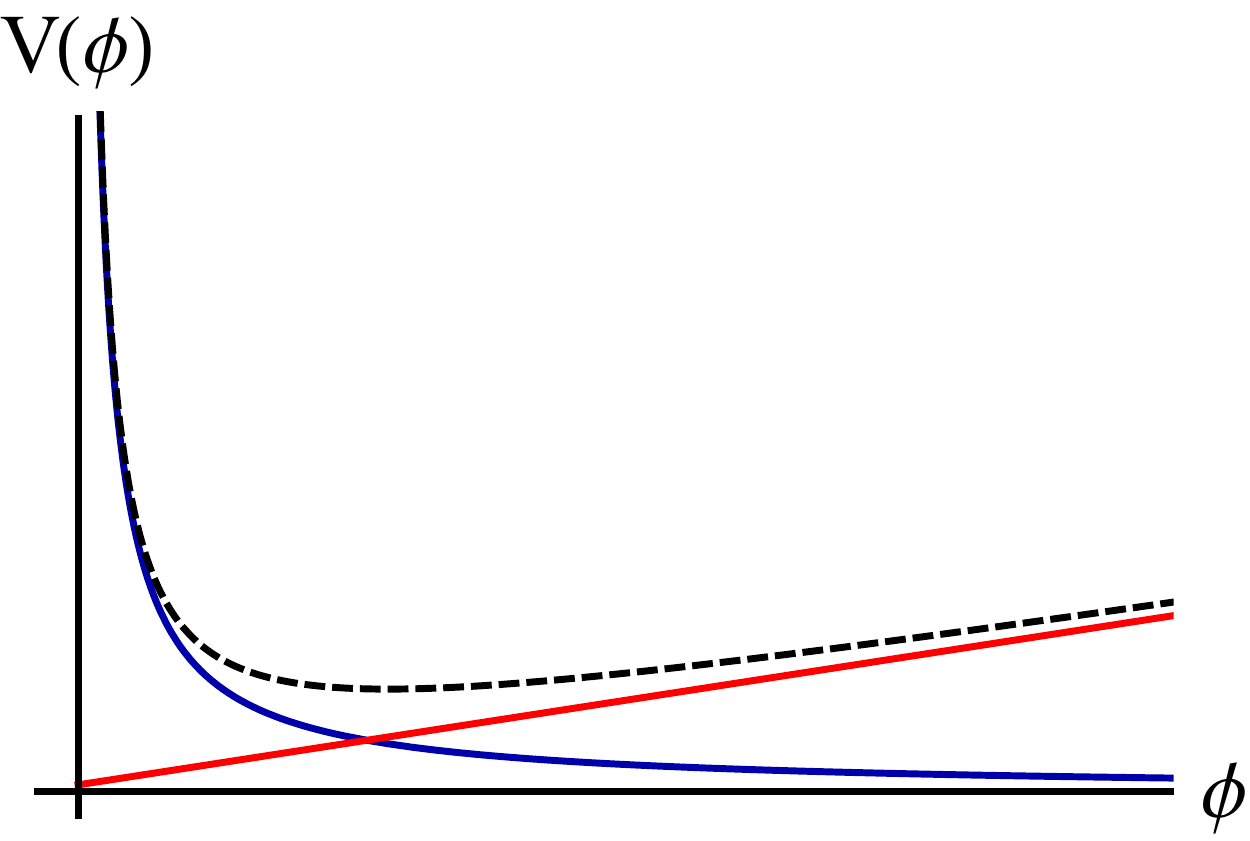}}}{$n>0$, low density}
\stackunder{
{\includegraphics[width=0.45\textwidth]{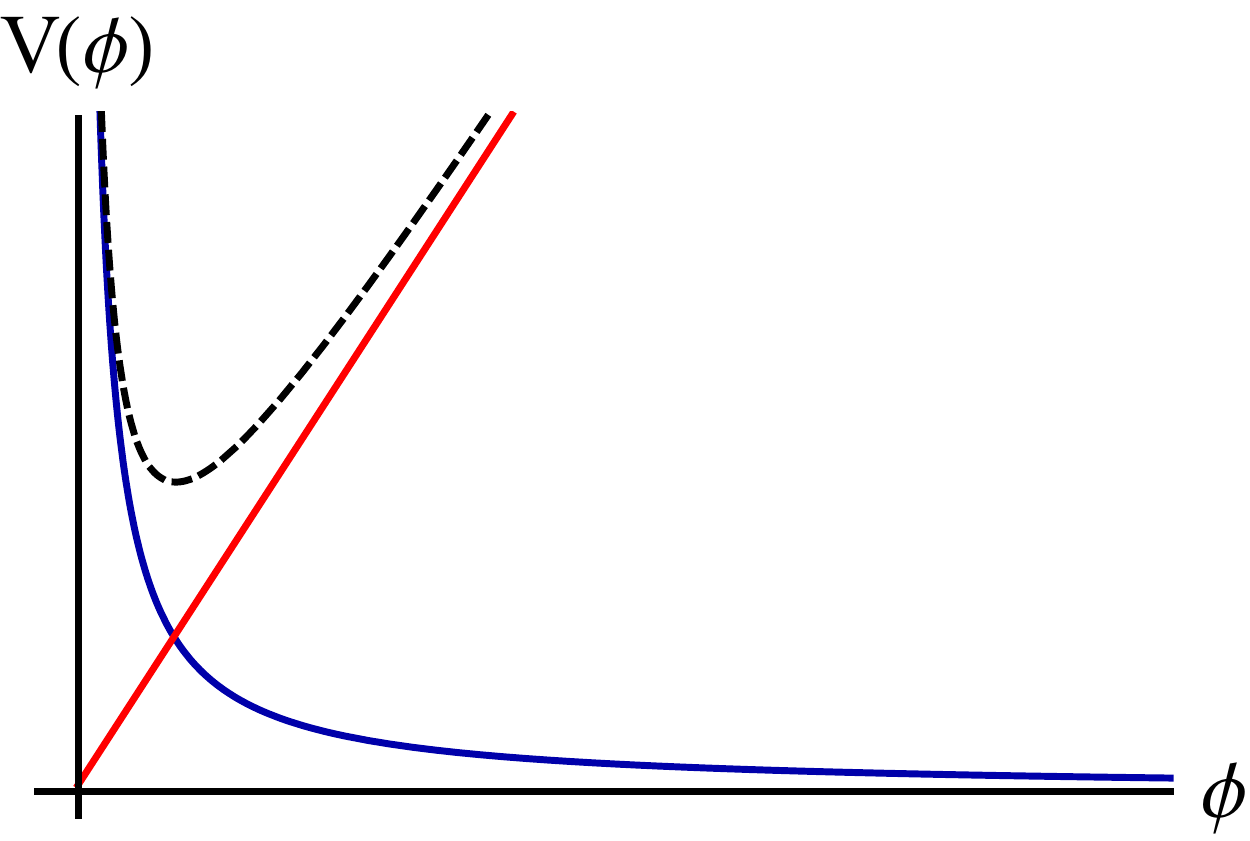}}}{$n>0$, high density}
\stackunder{
{\includegraphics[width=0.45\textwidth]{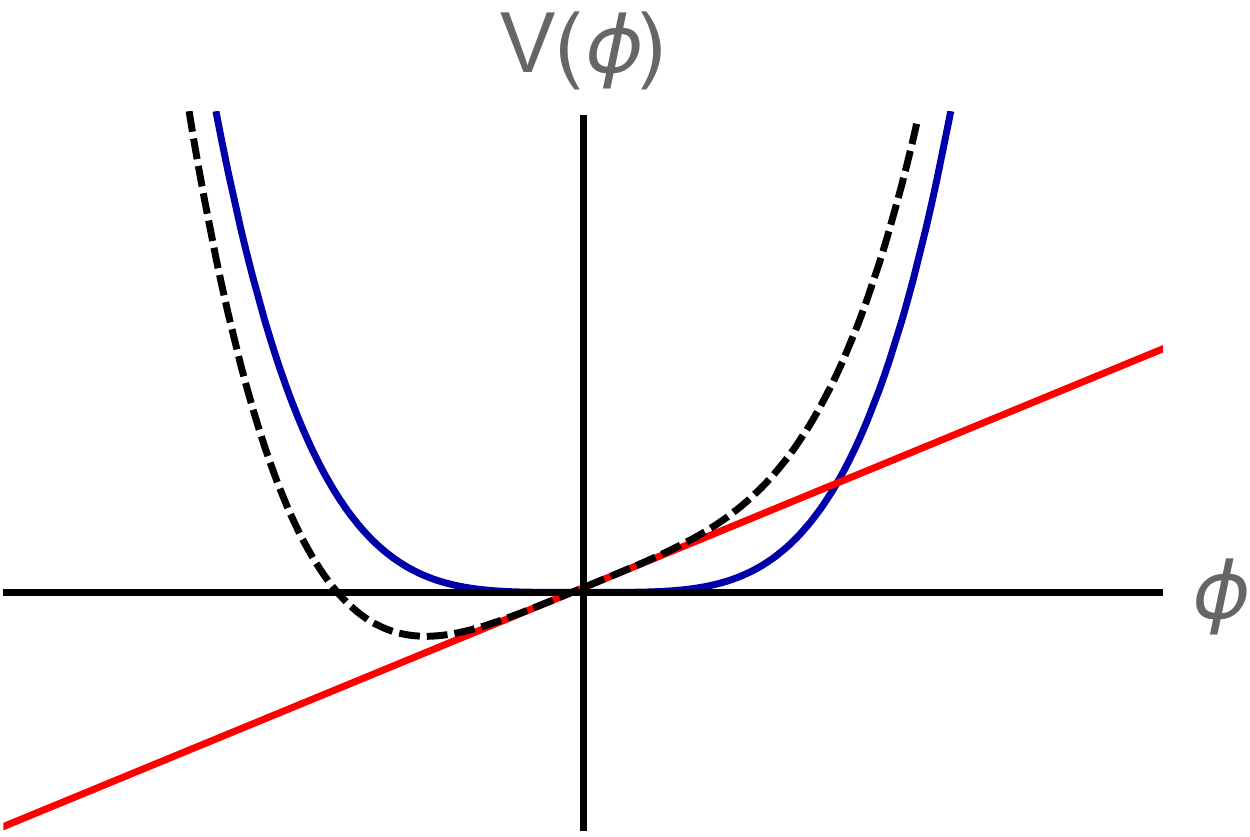}}}{$n<0$, low density}
\stackunder{
{\includegraphics[width=0.45\textwidth]{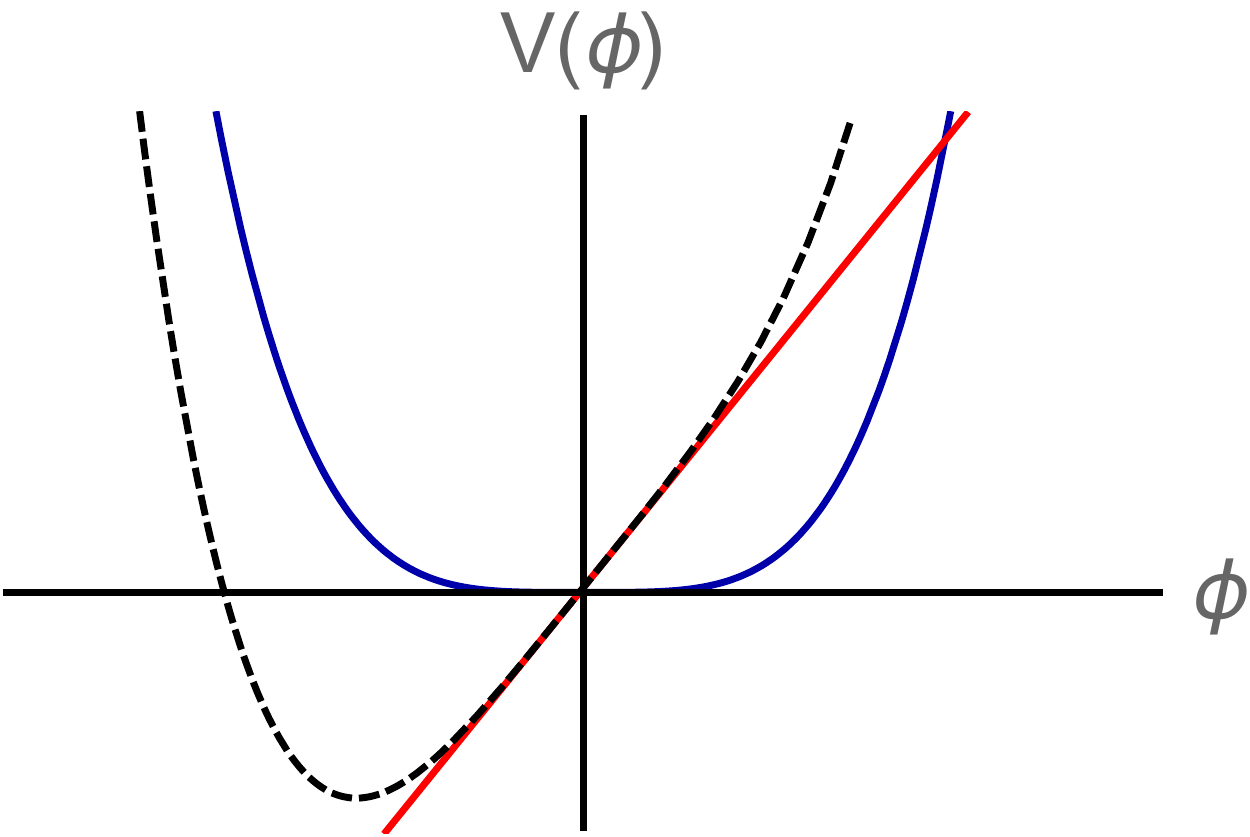}}}{$n<0$, high density}
\caption{Sketches of the chameleon effective potential for positive $n$ (upper panels) and negative $n$ (lower panels). The left and right panels show the cases of low and high density environments respectively. The blue lines show the bare potential and the red lines show the contribution from the coupling to matter. The black dashed lines show the resultant effective potential, which is the sum of the red and blue lines, and governs the dynamics of the scalar.  }\label{fig:cham_potential}
\end{figure}

The chameleon mechanism is illustrated in Figure \ref{fig:cham_potential}, which  sketches  the effective potential, as well as the separate contributions from the bare potential and the matter coupling, for positive and negative $n$ in both high and low densities. One can clearly see that the curvature around the high-density minimum is larger than around the low-density minimum, implying a larger mass for fluctuations. In practice, the difference can be several orders of magnitude, giving rise to very efficient screening. 

Since chameleon models are unable to self-accelerate cosmologically \cite{Wang:2012kj}, one typically adds a cosmological constant to the bare potential in order to account for dark energy. In this case, one has
\begin{equation}\label{eq:chamVgen}
V(\phi)= \Lambda_{\rm DE}^4 + \frac{\Lambda^{n+4}}{\phi^n}
\end{equation}
with $\Lambda_{\rm DE}=2.4$ meV. A common origin for the cosmological constant and the chameleon is an enticing scenario, for example one could have $V(\phi)=\Lambda^4\exp(\Lambda^4/\phi^n)$ \cite{Brax:2004qh}, and so special attention is often paid to the case $\Lambda=\Lambda_{\rm DE}=2.4$ meV. 

Another important model is the case $n=-4$. In this case, the mass-scale $\Lambda$ that governs that chameleon self-interactions is absent and one instead has the renormalizable potential
\begin{equation}\label{eq:Vchamphi4}
V(\phi)=\Lambda_{\rm DE}^4+\lambda_{\rm c}\phi^4.
\end{equation}
One generally expects $\lc\sim \oo(1)$ to be natural since values larger than this can give rise to  large quantum corrections to the potential  and smaller values typically require some degree of fine-tuning. Comparing with the form of the potential when $n\ne4$ one has $\lc=(\Lambda/\Lambda_{\rm DE})^4$. Even with this choice of renormalisable potential, the full chameleon model itself is non-renormalisable because the coupling to matter introduces higher-order operators of the form
\begin{equation}
\mathcal{L}\supset T\ln[A(\phi)]\sim \left(\frac{\phi}{\mc}+\oo{\left(\frac{\phi^2}{\mc^2}\right)}+\cdots\right)T.
\end{equation}
We will discuss this further below.

\subsubsection{\texorpdfstring{$f(R)$}{TEXT} models}
\label{sec:f(R)}

Theories of gravity where one replaces the Einstein-Hilbert action by a generic function $R$, known as $f(R)$ theories (see \cite{DeFelice:2010aj} for more general reviews), can screen using the chameleon mechanism, indeed they need to possess a form of screening mechanism to be compatible with solar system constraints. The first example of such a model was that of Hu and Sawicki \cite{Hu:2007nk} 
\begin{equation}\label{eqSM:f(R)}
S=\frac{1}{16\pi G}\int\dd^4x\sqrt{-\tg}\left({R+f(R)}\right)+S_{\rm m}[\tg];\quad f(R)=-a\frac{\mu^2}{1+(R/\mu^2)^{-b}},
\end{equation}
where $a$ and $b$ and both positive and $R=R(\tg)$. Expanding this action for high curvatures ($R\gg\mu^2$) one finds that
\begin{equation}
f(R)=-a\mu^2+a\mu^2\left(\frac{R}{\mu^2}\right)^{-b}+\cdots
\end{equation}
so that the theory looks like a cosmological constant with small corrections to GR. Indeed, one can tune the constants $a$ and $b$ to match with the $\Lambda$CDM cosmological model and one is left with small deviations from GR at the level of cosmological perturbations. In the low-curvature regime ($R\ll\mu^2$) the theory behaves like inverse-power law models where $f(R)\sim (R/\mu)^{-b}$ so that deviations from GR are suppressed. One can see the chameleon mechanism in action using the equivalence between $f(R)$ and scalar-tensor theories \cite{Chiba:2003ir}. Introducing the auxiliary field $\varphi$, a classically-equivalent action to \eqref{eqSM:f(R)} is
\begin{equation}\label{eqSM:f(R)2}
S=\frac{1}{16\pi G}\int\dd^4x\sqrt{-\tg}\left(R+f(\varphi)+\frac{\dd f}{\dd\varphi}\left[R-\varphi\right]\right)+S_{\rm m}[\tg].
\end{equation}
One can verify this by varying with respect to $\varphi$ to find $\varphi=R$ on shell, thereby recovering the action \eqref{eqSM:f(R)}. Introducing the Weyl-rescaled Einstein frame metric
\begin{equation}\label{eqSM:Weylf(R)}
\tg_\nm=A^2(\phi)g_\nm;\quad A^2(\varphi)=1+\frac{\dd f}{\dd\varphi},
\end{equation}
the action \eqref{eqSM:f(R)2} can be recast into a scalar tensor theory of the form 
\begin{align}\label{eqSM:STF(R)}
S&=\int\dd^4x\sqrt{-g}\left[\frac{R}{16\pi G}-\frac{1}{2}\partial_\mu\phi\partial^\mu\phi-V(\phi)\right]+S\mmm[e^{\sqrt{\frac{2}{3}}\frac{\phi}{\mpl}}]\quad\textrm{with}\\
V(\phi)&=\frac{\mpl^2}{2}\frac{\phi f'(\phi)-f(\phi)}{(1+f'(\phi))^2},
\end{align}
where the canonically-normalised field
\begin{equation}
\phi=-\sqrt{\frac{3}{2}}\mpl\ln\left(1+f'(\varphi)\right).
\end{equation}
The theory is then a chameleon with $M_c = \sqrt{6}\mpl$. The Hu-Sawicki model corresponds to a chameleon with $n=-b/(1+b)$ so that only a narrow range in the chameleon theory space is covered i.e $-1<n<-1/2$. The most well-studied models are $b=1$ ($n=-1/2$) and $b=3$ ($n=-3/4$), although, typically, results are only quoted for $n=1$, and so we will only focus on this model here.

\subsubsection{UV Properties}

Screening relies on the presence of non-linear self-interactions of the scalar field, and on coupling the scalar to the matter energy momentum tensor.  Written in the Einstein frame, this necessarily introduces  non-renormalisible operators, meaning that additional physics is required in order to UV complete the model \cite{Joyce:2014kja}.  Additionally, we might worry that integrating out physics in the UV changes the form of the low energy theory, either rescaling the coefficients, or introducing new terms into the Lagrangian. 

For the theory to be fully predictive it is important to understand whether the low energy theory we study is protected from corrections coming from UV physics. One commonly used way to estimate the size of these effects is to compute the Coleman-Weinberg \cite{Coleman:1973jx} corrections to the scalar mass \cite{Upadhye:2012vh}. To do this one computes the corrections to the scalar mass from scalar fields running in loops, these loop corrections arise precisely because the scalar field has non-trivial self interactions in its potential.  The Coleman-Weinberg corrections are found to be at least logarithmically divergent with scale. Even if these corrections to the mass are assumed to be small at some scale, they may become important at another scale, or in another environment. 

In \cite{Upadhye:2012vh} the relevance of these corrections for the E\"{o}t-Wash experiment was computed.  With some simple assumptions about the scale at which the logarithmic terms become important it was shown that the current constraints from these experiments are computed in a regime in which the quantum corrections are indeed under control.  However, as the experimental sensitivity improves these corrections will become more relevant. 

Keeping track of the quantum corrections is also important in order to understand the behaviour of the chameleon in the early universe. In \cite{Erickcek:2013dea,Erickcek:2013oma} it was shown that, with the exception of gravitationally coupled chameleons,  it is not possible to evolve the chameleon through the radiation dominated era without knowing the UV completion of the model.  This is because the decoupling of standard model particles during this epoch give a large impulse to the otherwise slowly rolling chameleon field \cite{Brax:2004qh}.  This causes the chameleon scalar to rapidly roll to the part of the potential where the field's self interactions are large, and so high energy quantum fluctuations of the field are excited. It is possible that some non-perturbative physics could resolve this, but in the absence of a proof of this, we do not know how to evolve the chameleon model from the early universe to late times in a predictive way. {One model that can evade this problem is the case $n=-4$ due to the absence of a low mass scale (that is problematic in the early Universe when energies are typically high) \cite{Miller:2016xpq}.  }

The most reliable way to compute UV corrections to the low energy chameleon model would be to know exactly what the UV-completion of the theory is. A number of attempts have been made to embed the chameleon mechanism within string theory \cite{Brax:2004ym,Conlon:2010jq,Hinterbichler:2010wu,Nastase:2013los,Nastase:2013ik}, within supersymmetry \cite{Brax:2012mq,Brax:2013yja}, and using non-canonical kinetic terms \cite{Padilla:2015wlv}, but, as yet, no complete theory exists.

\subsection{Symmetron Screening}

\begin{figure}[ht]\centering
\includegraphics[width=0.5\textwidth]{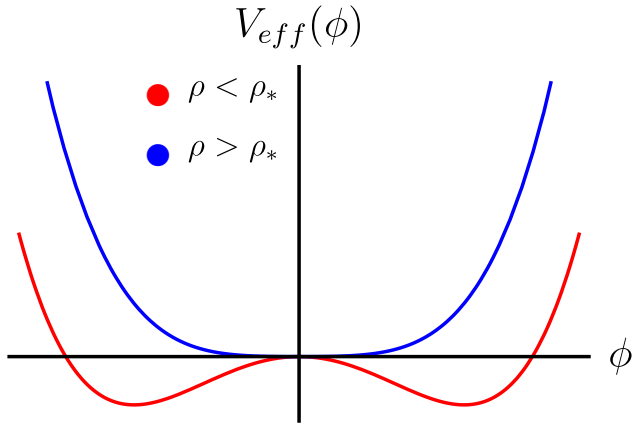}
\caption{The effective potential for the symmetron when $\rho<\rho_\star$ (red, lower) and when $\rho>\rho_\star$ (blue, upper).}\label{fig:symmetron_veff}
\end{figure}

The symmetron model does not rely on varying mass, 
%is constructed to have a similar mass in all environments, 
instead, the screening works by suppressing the coupling to matter in high-density regions. This is accomplished using $\mathbb{Z}_2$ symmetry restoration. The bare potential and coupling function are
\begin{equation}
V(\phi)=-\frac{1}{2}\mu^2\phi^2+\frac{\lambda}{4}\phi^4;\quad A(\phi)=1+\frac{\phi^2}{2\ms^2}
\end{equation}
so that the effective potential is
\begin{equation}
\label{eq:veff_symmetron}
V\eff(\phi)=\frac{1}{2}\mu^2\left(1-\frac{\rho}{\mu^2\ms^2}\right)\phi^2+\frac{\lambda}{4}\phi^4.
\end{equation}
This is $\mathbb{Z}_2$ ($\phi\rightarrow-\phi$) symmetric (as are $V(\phi)$ and $A(\phi)$ individually). The coefficient of the quadratic term can be either positive or negative depending on the density and, in particular, there is a critical density
\begin{equation}
\rst \equiv \mu^2 \ms^2
\end{equation}
where the sign changes. The screening mechanism is best exemplified by examining the shape of the effective potential sketched in Figure \ref{fig:symmetron_veff}. When $\rho<\rho_\star$ there are two degenerate minima at 
\begin{eqnarray}
\pmi^{\pm}&=&\pm\frac{\mu}{\sqrt{\lambda}}\sqrt{1-\frac{\rho}{\mu^2\ms^2}}\\
&\approx& \pm\frac{\mu}{\sqrt{\lambda}}, \;\;\;\;\;\mbox{ if } \rho \ll \rst
\end{eqnarray}
%where the second approximation holds when $\rho\ll\rst$. 
In this case, the $\mathbb{Z}_2$ symmetry is spontaneously broken and the coupling to matter is
\begin{equation}
\beta(\phi_0)=\left\vert\frac{\mpl \pmi^\pm}{\ms^2}\right\vert\approx \frac{\mu\mpl}{\lambda\ms^2},
\end{equation}
giving rise to a fifth-force potentially stronger than gravity. When $\rho>\rst$ the only minimum is at $\phi=0$ so that the symmetry is restored and the coupling $\beta(\phi_0)=0$. In which case no fifth force can be sourced.  One can tune the parameters $\mu$, and $\lambda$ in terms of $\ms$ to ensure that $\rst$ coincides with the present day cosmological density, or so that the fifth-force is of gravitational strength \cite{Hinterbichler:2010es,Hinterbichler:2011ca}, but we shall not do so here since we are considering a range of different experimental tests that constrain the parameters in very different environments and on many different scales.

\subsubsection{Generalized Symmetrons}

The symmetron screening mechanism is by no means reliant on the specific form of the effective potential \eqref{eq:veff_symmetron}. Indeed, clearly any effective potential of the form
\begin{equation}
V\eff(\phi)= -\mu^4\left(1-\frac{\rho}{\mu^{4-2n}\ms^{2n}}\right)\frac{\phi^{2n}}{\mu^{2n}}+\frac{\phi^{2m}}{\Lambda_{\rm s}^{2m-4}}
\end{equation}
with $n<m$ and $n,\, m\in \mathbb{Z}^+$ exhibits qualitatively similar features to the canonical symmetron. Such an effective potential can arise through the bare potential and coupling functions
\begin{equation}
V(\phi)=-\frac{\phi^{2n}}{\mu^{2n-4}}+\frac{\phi^{2m}}{\Lambda_{\rm s}^{2m-4}}; \;\;\;\;\;A(\phi) = 1 + \frac{\phi^{2n}}{\ms^{2n}}.
\end{equation}
First discovered by \cite{Brax:2011aw,Brax:2012gr} using tomographic methods, there has been little investigation of these models at the present time and so we do not consider them further here. 

\subsubsection{Radiativly-Stable Symmetrons}
The symmetron model, as described here, suffers the same UV stability properties as the chameleon. In particular that Coleman-Weinberg corrections could dramatically alter the shape of the potential needed for the symmetron mechanism to work. In this case, however, the one-loop corrections can also be exploited to give rise to the screening in a radiatively stable way \cite{Burrage:2016xzz}.

The Coleman-Weinberg model \cite{Coleman:1973jx} was originally discussed as a way of using radiative  corrections to generate a spontaneous symmetry breaking transition. The classical model is scale invariant, but the one-loop corrections generate a scale through dimensional transmutation of the logarithmic divergences.  In the one field model higher order loop corrections become important in the spontaneously broken vacuum, but in a multi-field model these can be kept under control \cite{Garbrecht:2015yza}, and the one loop potential can undergo a symmetry breaking transition whilst the higher order loop corrections remain small. 

The radiatively stable symmetron has a different bare potential to that discussed above
\begin{equation}
V(\phi) = \left(\frac{\lambda}{16 \pi}\right)^2 \phi^4\left(\ln \frac{\phi^2}{m^2}-\frac{17}{6}\right)
\end{equation}
however overall the phenomenology this gives rise to is very similar to that of the standard symmetron.

\subsection{Coupling to Photons}
\label{sec:photoncoup}

A conformally coupled scalar field does not have a classical coupling to photons.  This is because the scalar couples to the trace of the energy momentum tensor of the matter fields, and photons, being relativistic, have a traceless energy momentum tensor. This is not the end of the story, however, as quantum effects make it easy to generate  such a coupling.  One way to do this is to assume the presence of a new heavy fermion which has an electromagnetic charge. Then an interaction between one conformally coupled scalar, and two photons can be mediated by a triangle-loop of the heavy fermion.   If the fermion is sufficiently heavy that it can be integrated out, to leave the Standard Model plus the chameleon as a low energy effective theory, then the low energy theory has a contact interaction between the chameleon and two photons \cite{Brax:2009ey}. Such heavy, charged fermions are ubiquitous in theories of physics beyond that Standard Model, including, string theory, supersymmetry and GUTs.
It can also be shown that the Weyl rescaling that allows us to change from Jordan to Einstein frame, gives rise to a coupling to photons after quantisation of these fields, this was shown for the chameleon in \cite{Brax:2010uq}, following earlier work by Kaplunovsky and Louis in the context of supersymmetry \cite{Kaplunovsky:1994fg}. 

The coupling to photons that arises in all of these cases is that of a scalar axion-like particle
\begin{equation}
\mathcal{L}\supset \frac{\phi}{M_{\gamma}} F_{\mu\nu}F^{\mu\nu} \;.
\label{eq:photon}
\end{equation}
(For a symmetron model with $\mathbb{Z}_2$ symmetry the leading coupling would instead be quadratic in $\phi$.) Here $M_{\gamma}$ is the energy scale that controls the coupling to photons, this is not necessarily the scale at which the chameleon couples to other matter particles $\mc$.
The coupling in Equation (\ref{eq:photon}) means that existing constraints on axion-like particles can be applied to the chameleon, although some care must be taken in doing this as standard axion-like particles have  fixed mass and couplings, and so constraints from environments of vastly different density can be easily combined. This  is not the case for a screened scalar.  
This axion-like coupling is not necessary for a screening mechanism to work, however it is difficult to forbid such a coupling in a truly quantum theory.  Including the coupling opens new avenues for detecting the chameleon, as high precision searches for axions and axion-like particles can be exploited to detect or constrain the chameleon. For example, the interaction in equation (\ref{eq:photon}) means that chameleons can be produced through the Primakov effect as photons propagate through a magnetic field.  This underlies a range of different experimental search strategies. 

%Screening - JS: included this to split up astro and laboratory screening from the experiments section

\section{Screening}
\label{sec:screening}

In this section we discuss screening mechanisms in the context of astrophysical objects and typical laboratory configurations, and discuss some salient features that are specific to screening mechanisms.

\subsection{Astrophysical Screening: The Thin Shell Effect}

Astrophysical objects are typically spherical and so in this section we consider the screening of a non-relativistic, static, spherically symmetric object of mass $\mathcal{M}$, radius $R$, and density $\delta\rho(r)$ immersed in a much larger medium with density $\bar{\rho}$. The total density is $\rho=\brho+\delta\rho$. This could represent a star inside a galaxy or a galaxy/dark matter halo/cluster embedded in the cosmological background, in which case $\brho$ is the mean cosmic density. We follow the method of \cite{Hui:2009kc,Davis:2011qf,Sakstein:2013pda,Sakstein:2015oqa,Burrage:2016bwy}. (Other derivations using slightly different procedures recover the same results \cite{Brax:2012gr} but the current astrophysical constraints have been derived using the method we present here.) Far away from the object, the field minimizes the effective potential so that one has $\phi(r)\rightarrow\bar{\phi}\equiv \pmi(\brho)$. Near the object, the equation of motion in Eq. \eqref{eq:EOMgen} becomes (in spherical coordinates)
\begin{equation}
\nabla^2\phi=\frac{1}{r^2}\frac{\dd}{\dd r}\left(r^2\frac{\dd \phi}{\dd r}\right)=\frac{\dd V}{\dd\phi} +\frac{\beta(\phi)\rho}{\mpl}.\label{eq:astroscreen_gen}
\end{equation}
One can then envisage two regimes. If the field can reach the minimum of the effective potential inside the object then one has $V\eff'(\phi)=0$ and the right hand side of \eqref{eq:astroscreen_gen} is unsourced so that $\phi=\pmi(\rho)$ is constant and there is no fifth-force. If instead the field remains close to $\bar{\phi}$ we can linearise $\phi=\bar{\phi}+\varphi$ to find
\begin{equation}
\label{eq:astroscreen_linear}
\frac{1}{r^2}\frac{\dd}{\dd r}\left(r^2\frac{\dd \phi}{\dd r}\right) = m_0^2\varphi+\frac{\beta(\phi_0)}{\mpl}\delta\rho,
\end{equation}
where $m_0^2=V''(\phi)$. $V(\phi)$ is typically chosen so that $\phi$ is cosmologically relevant i.e. $m_0 R\ll1$ and one can ignore the first term on the right hand side of \eqref{eq:astroscreen_linear}, in which case one is left with a Poisson equation
\begin{equation}
\frac{1}{r^2}\frac{\dd}{\dd r}\left(r^2\frac{\dd \phi}{\dd r}\right) = \frac{\beta(\bar{\phi})}{\mpl}\delta\rho.\label{eq:astroscreen_Poisson}
\end{equation}

\begin{figure}
\centering
\includegraphics[width=0.5\textwidth]{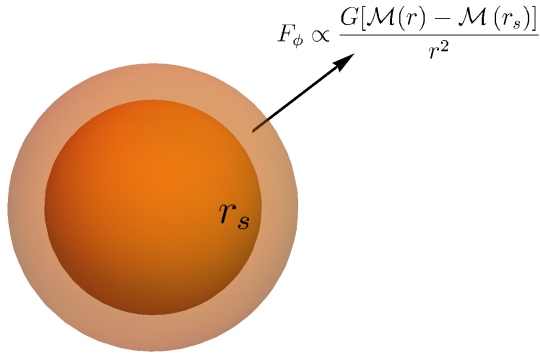}
\caption{The thin-shell effect. The fifth-force only receives a contribution from the mass in the thin-shell $\rs<r<R$.}\label{fig:thinn_shell}
\end{figure}

In practice, we expect a hybrid of these two cases where the field sits close to the minimum of the effective potential at the centre of the object and remains there up to some radius $\rs$ at which it begins to roll towards its asymptotic value and enter the second regime. There is therefore no fifth-force interior to $\rs$; for this reason we will refer to $\rs$ as the \emph{screening radius}. Outside the screening radius one can integrate \eqref{eq:astroscreen_Poisson} once to find
\begin{equation}
\label{eq:phi_prime_thin_shell}
\frac{\dd\phi}{\dd r}=\frac{\beta(\bar{\phi})\left(\mathcal{M}(r)-\mathcal{M}(\rs)\right)}{4\pi\mpl r^2},
\end{equation}
where
\begin{equation}\label{eq:continuity}
\mathcal{M}(r)=\int_0^r 4\pi r'^2\delta\rho(r')\dd r';\quad \mathcal{M}\equiv M(R).
\end{equation}
The fifth-force \eqref{eq:F5gen} is then 
\begin{equation}
F_5=\frac{2\beta^2(\bar{\phi})G\left[\mathcal{M}(r)-\mathcal{M}(\rs)\right]}{r^2}.\label{eq:F5_spherical}
\end{equation}
The field equation is only sourced by the density outside the screening radius and so only the mass exterior to this contributes to the fifth-force. Objects that have $\rs\ll R$ have
\begin{equation}
\frac{F_5}{F_{\rm N}}\approx 2\beta^2(\bar{\phi})
\end{equation}
and are hence unscreened whereas those for which $\rs\approx R$ have $F_5/F_N\ll1$ and are hence screened. In this case, the fifth-force only receives contributions from the mass in a very thin shell outside the screening radius. This phenomena has been dubbed \emph{the thin-shell effect}; we depict this in Figure \ref{fig:thinn_shell}. Outside the object the mass term in 
\eqref{eq:phi_prime_thin_shell} is more important than the density and one has 
\begin{equation}
F_5=\frac{2\beta^2(\bar{\phi})G\left[\mathcal{M}-\mathcal{M}(\rs)\right]}{r^2}e^{-m\eff(r-R)}.
\end{equation}

In order to determine whether an object is screened we must calculate $\rs$. This can be accomplished by integrating \eqref{eq:phi_prime_thin_shell} from $\rs$ (where $\phi=\phi_{\rm s}\approx\pmi(\rho)$) to $\infty$ to find
\begin{equation}
\label{eq:screening_condition_1}
\bar{\phi}-\phi_{\rm s}=\frac{\beta(\bar{\phi})\mathcal{M}(\rs)}{4\pi \mpl\rs}+\int_{\rs}^\infty \frac{\beta(\bar{\phi})\mathcal{M}(r')}{4\pi r'^2}\dd r'.
\end{equation}
Performing integration by parts and using Equation \eqref{eq:continuity} one finds an implicit relation for $\rs$
\begin{equation}\label{eq:screening_radius_1}
\chi\equiv\frac{\bar{\phi}}{2\beta(\bar{\phi})\mpl}=4\pi G\int_{\rs}^\infty r'\delta\rho(r')\dd r',
\end{equation}
where we have ignored $\phi_{\rm s}$ since the screening mechanisms always act to push $\phi$ to smaller values inside dense objects. Alternatively, one can use the relation $\dd\pn/\dd r= G\mathcal{M}(r)/r$ to write \eqref{eq:screening_condition_1} as
\begin{equation}
\label{eq:screening_radius_2}
\chi=-\pn(\rs)-\rs\pn'(\rs).
\end{equation}
If \eqref{eq:screening_radius_2} (or, equivalently, \eqref{eq:screening_radius_1}) has no solution then $\rs=0$ and the object is fully unscreened. Given that $\pn<0$ whilst $\pn'>0$ there can be no solution when $\chi>\pn=G\mathcal{M}/R$. Hence, only objects where $\chi<G\mathcal{M}/R$ can be partially (or fully for $\chi\ll G\mathcal{M}/R$) screened. 

The screening criteria above is particularly useful for determining which astrophysical objects will be partially unscreened and for which range of parameters; one simply needs to calculate the Newtonian potential. Commonly studied examples are given in Table \ref{tab:astro_screen_chi}. In the case of main sequence stars one can find the Newtonian potential using the mass-radius relation
\begin{equation}
\frac{M}{M_\odot}=\left(\frac{R}{R_\odot}\right)^{\nu},
\end{equation}
where $\nu$ depends on the type of star in question. In the case of galaxies one can use the Virial theorem to calculate the Newtonian potential from the circular velocity:
\begin{equation}
v^2=\frac{G\mathcal{M}}{R}.
\end{equation}
Dwarf galaxies are particularly good probes due to their low Newtonian potentials. Indeed, many of the astrophysical tests we will discuss below use either dwarf galaxies themselves or their constituent objects. Setting $\bar{\phi}=\phi_0$ the parameter of interest is 
\begin{equation}
\chi_0\equiv\frac{\phi_0}{2\beta(\phi_0)\mpl}.
\end{equation}
Unscreened dwarf galaxies can then, in theory, test $\chi_0\gsim 10^{-8}$. 

\begin{table}
\centering
\begin{tabular}{c | c }
Object & $\pn$\\\hline
Earth & $10^{-9}$\\
Moon & $10^{-11}$\\
Main-sequence stars ($M_\odot$) & $10^{-6}$\\
post Main-sequence stars ($1$--$10 M_\odot$) & $10^{-7}$--$10^{-8}$\\
Spiral and elliptical galaxies & $10^{-6}$\\
Dwarf galaxies & $10^{-8}$
\end{tabular}\caption{Astrophysical objects of interest and their Newtonian potentials.}\label{tab:astro_screen_chi}
\end{table}

In practice, one also needs to worry about environmental screening. So far, we have only considered the screening of a single object embedded in a larger background, but real astrophysical objects are typically not isolated; galaxies are found in clusters and stars come in pairs or groups. The non-linear nature of the field equations means that we cannot simply superimpose solutions sourced by different objects to obtain a new solution. This implies that an object's environment can affect whether it is screened or not. The most important example of this is the screening of dwarf galaxies. Taken in isolation, the Newtonian potential for a dwarf galaxy is $\oo(10^{-8})$ but the typical potential associated with clusters of galaxies is $\oo(10^{-4})$ so that only values of $\chi_0$ larger than this can be tested. The ideal probes are therefore dwarf galaxies located in voids that do not suffer from environmental screening. There has been a great effort towards determining the criteria for environmental screening \cite{Li:2011pj,Lombriser:2012nn,Lombriser:2013wta,Cai:2014fma}. Most of these rely on numerical N-body simulations, whose description lies outside the scope of this review, but the end result is a \emph{screening map} \cite{Cabre:2012tq} of the local universe that classifies galaxies as either screened, partially screened, or unscreened. To date, all astrophysical tests using dwarf galaxies have been taken from this screening map.

Astrophysical tests ultimately end up constraining regions in the $\chi_0$--$\beta(\phi_0)$ plane. For our models of interest, one has
\begin{equation}
\label{eq:astro_beta}
\beta(\phi_0)=\begin{cases}
    \frac{\mpl}{\mc}     & \quad \text{Chameleons }\\
    \frac{\mu\mpl}{\sqrt{\lambda}\ms^2}  & \quad \text{Symmetrons}\\
  \end{cases},
\end{equation}
and
\begin{equation}
\label{eq:astro_chi}
\chi_0=\begin{cases}
    \frac{1}{2}\left(\frac{\mc}{\mpl^2}\right)^{\frac{n+2}{n+1}}\left(\frac{n\Lambda^{n+4}}{3\Omega\mmm H_0^2}\right)^{\frac{1}{n+1}}     & \quad \text{Chameleons }\\
    \frac{1}{2}\left(\frac{\ms}{\mpl}\right)^{2}  & \quad \text{Symmetrons}\\
  \end{cases},
\end{equation}
where we have replaced the cosmic density in $\pmi(\rho)$ with $3\Omega\mmm\mpl^2H_0^2$.

\subsubsection{Screening in \texorpdfstring{$f(R)$}{TEXT} Theories}

Given that $f(R)$ models only cover a restricted range of $n$ and have a fixed value of $M_c$, it is not particularly enlightening to constrain $f(R)$ theories in terms of $\Lambda$ and $n$, even more so since the cosmological constant is fixed by tuning the parameters so that  $\Lambda=2.4\times10^{-3}$ eV does not have any special significance. (In this sense, $f(R)$ theories should be thought of as describing deviations from the $\Lambda$CDM model). Instead, constraints are often quoted in terms of the parameter $f_{R0}=f'(R_0)$, the first-derivative of $f(R)$ evaluated at the present time in the cosmological background. The significance of this parameter can be seen by examining the screening in the $f(R)$ formalism. Consider an object of density $\rho_0$ embedded in the cosmological background where the Ricci scalar curvature is $R_0$ and the density is $\rho_0$. If one embeds an object with density $\delta\rho$ into this background then it will source a Newtonian potential ($g_{00}=-a^2(1+2\Phi)$) and perturb $R=R_0+\delta R$, $f_R=f_{R0}+\delta f_R$ \cite{Schmidt:2010jr} such that
\begin{align}
\nabla^2\Phi&=\frac{16\pi G}{3}\rho-\frac{1}{6}\delta R(f_{R0})\label{eqSM:frscreen1}\\\label{eqSM:frscreen2}
\nabla^2 \delta f_R&=\frac{1}{3}\left(\delta R(f_{R0})-8\pi G\delta \rho\right).
\end{align}
In the limit where $\delta f_R\ll f_{R0}$ there can be no source for $\delta f_R$ and one has $\delta R(f_{R0})=8\pi G\delta \rho$ so that \eqref{eqSM:frscreen1} becomes $\nabla^2\Phi=4\pi G\delta\rho$. Therefore, in this limit we recover the Poisson equation and there are no deviations from GR; the fifth force is  screened. In the opposite limit where $\delta f_R\gg f_{R0}$ we can expand $\delta R (f_R)\approx \delta f_R/f_{RR}$ so that equation \eqref{eqSM:frscreen2} becomes
\begin{align}
\nabla^2 \delta f_R&=m_f^2\delta f_R-\frac{8\pi G}{3}\delta \rho,\quad m_f^2=\frac{1}{3 f_{RR}}\label{eqSM:frscreen3},
\end{align}
which is the equation of motion for a massive scalar with mass $m_f$. On scales shorter than $m_f^{-1}$ the mass can be ignored and one finds, using \eqref{eqSM:frscreen1}, $\nabla^2\Phi=16\pi G\delta\rho/3$ so that the Newtonian potential is enhanced by a factor of $4/3$; the force is fully unscreened. Note that, in this limit, Equation \eqref{eqSM:frscreen2} gives $|\delta f_R|=2\Phi/3$ but the maximal value of $\delta f_R$ is $f_{R0}$ so we conclude that objects must be partially screened if $f_{R0}<2\Phi/3$. Thus we see that $f_{R0}$ is the $f(R)$ equivalent of the $\chi$.

\subsubsection{Gravitational Lensing: Dynamical vs. Lensing masses}
\label{sec:lening_mass}

Conformal transformations leave null geodesics unchanged \cite{padmanabhan2010gravitation} ($\tg_\nm\dot{x}^\mu\dot{x}^\nu=A^2(\phi)g_\nm\dot{x}^\mu\dot{x}^\nu=0$) so that photons move on geodesics of both $\tg_\nm$ and $g_\nm$. This has some novel implications for gravitational lensing by massive bodies. Expanding the Einstein frame metric in the Newtonian gauge:
\begin{equation}\label{eq:WFEF}
\dd s^2=(-1+2\pn)\dd t^2 + (1+2\Psi_{\rm N})\dd x^2,
\end{equation}
the Jordan frame metric is
\begin{equation}
\dd \tilde{s}^2=\left(-1+2\pn-2\beta(\bar{\phi})\frac{\phi}{\mpl}\right)\dd t^2 + \left(1+2\Psi_{\rm N}+2\beta(\bar{\phi})\frac{\phi}{\mpl}\right)\dd x^2,
\end{equation}
where we have set $\phi\rightarrow\bar{\phi}+\phi$ and have absorbed factors of $A(\phi_0)^2$ into $t$ and $x^i$ (see Section \ref{sec:SStests}). We can thus identify the Jordan frame potentials
\begin{align}
\tilde{\Phi}_{\rm N}&=\pn-\beta(\bar{\phi})\frac{\phi}{\mpl}\quad
\tilde{\Psi}=\Psi+\beta(\bar{\phi})\frac{\phi}{\mpl}.
\end{align}
The Newtonian potential, which governs the motion of non-relativistic particles, therefore depends on $\phi$ whereas the lensing potential, $\Psi$, which governs the motion of photons is
\begin{equation}
\tilde{\Psi}_{\rm L}=\frac{1}{2}\left(\tilde{\Phi}_{\rm N}+\tilde{\Psi}\right)=\pn,
\end{equation}
where we have used the relationship $\Psi_{\rm N}=\pn$, which is a result of working in the Einstein frame. For an extended object of mass $\mathcal{M}$, the mass inferred from lensing
%---the \emph lensing mass---
is the true mass $\mathcal{M}$ because the Einstein frame potentials satisfy the Poisson equation. Conversely, the potential governing the motion of non-relativistic objects satisfies
\begin{equation}
\label{eq:dynamicalmass}
\tilde{\Phi}_{\rm N}'=\frac{G\mathcal{M}}{r^2}A^2(\bar{\phi})\left[1+2\beta^2(\bar{\phi})\left(1-\frac{\mathcal{M}(\rs)}{\mathcal{M}}\right)\right]=\frac{G\mathcal{M}_{\rm dyn}}{r^2},
\end{equation}
which defines a dynamical mass $\mathcal{M}_{\rm dyn}\ge\mathcal{M}$ with equality for fully screened objects. The difference between the lensing and dynamical masses is in stark contrast to GR, and is a particularly useful feature for testing modified gravity using astrophysical observations.

\subsection{Solar System Tests}
\label{sec:SStests}
Classical tests of GR use the PPN formalism applied to solar system objects and so in this Section we will  illustrate how these tests apply to screened modified gravity, and why they yield only weak constraints.

\subsubsection{PPN Parameters}
\label{sec:PPN}

The PPN metric is both an ansatz (for the possible potentials that could appear in the metric sourced by a massive body) and a gauge choice. There are 10 parameters that can be calculated and compared with observations, but only two are relevant for conformal scalar-tensor theories (disformal theories involve four parameters \cite{Ip:2015qsa}). The PPN metric with these two parameters is (for a spherically symmetric object of mass $\mathcal{M}$)
\begin{equation}
g_{00}=-1+2\frac{G \mathcal{M}}{r}
-2\beta\left(\frac{G\mathcal{M}}{r}\right)^2,
\quad g_{0i}=0,\quad\textrm{and} \quad g_{ij}=\left(1+2\gamma\frac{G\mathcal{M}}{r}\right)\delta_{ij}.
\end{equation}
The parameter $\gamma$ ($=1$ in GR) sets the amount of light-bending by massive objects, and the Shapiro time-delay effect; and the parameter $\beta$ ($=1$ in GR) measures the amount of non-linearity in the field equations. The term proportional to $\beta$ is responsible for the precession of the perihelion of Mercury. Note that the first term in $g_{00}$ is not free to vary, this is a gauge choice that implies that $G$ is Newton's constant as measured in Cavendish-type experiments.

General expressions for $\gamma$ and $\beta$ in screened scalar-tensor theories can be found in \cite{Hees:2011mu,Zhang:2016njn}.
%but this is not particularly enlightening. Instead, 
It is more instructive, however,  to consider the solution for the fifth-force profile of a static object derived in \eqref{eq:F5_spherical}. We will ignore the mass of the scalar for simplicity but including it does not change any of what follows. The calculation of the fifth-force was performed in the Einstein frame but the PPN metric is defined in the Jordan frame since it is the metric that controls the geodesics of matter and so our task is to calculate the Jordan frame metric given $\phi$ to $\oo(v^2/c^2)$ to find $\gamma$. The calculation of $\beta$ is analogous except one continues to $\oo(v^4/c^4)$; this calculation is long and tedious, and one does not gain any additional insight. For this reason, we will only calculate $\gamma$.

To begin, we summarize our Einstein frame solution. This is
\begin{align}
g_{00}=-1+2\frac{G\mathcal{M}}{r}
,\quad g_{0i}=0,\quad g_{ij}=\left(1+2\frac{G\mathcal{M}}{r}\right)\delta_{ij},\quad\textrm{and}\quad\phi=\bar{\phi}-\beta(\bar{\phi})\frac{\mathcal{M}-\mathcal{M}(\rs)}{4\pi \mpl r},
\end{align}
where we have used the fact that $F_5=2\beta(\bar{\phi})\phi'$ to find the field profile. Next, we can expand the metric as
\begin{equation}
\tg_\nm=A^2(\phi)g_\nm\approx A^2(\bar{\phi})(1+2\beta(\bar{\phi})\varphi) g_\nm,
\end{equation}
where $\varphi=\phi-\bar{\phi}$. The factor of $A^2(\bar{\phi})$ is usually ignored claiming
%by making statements such as 
``$A(\bar{\phi})\approx1$,'' but a more correct treatment is to rescale the coordinates such that $t\rightarrow t/A(\bar{\phi})$ and $r\rightarrow r/A(\bar{\phi})$. We also need to rescale the mass $\mathcal{M}$ since this was defined using Einstein frame coordinates, and Einstein frame densities. Note that one has $\tilde{T}^{\nm}=A^6T^\nm$, which implies $\tilde{\rho}=\tg_\nm\tilde{T}^\nm=A^4\rho$. The mass then needs to be rescaled as $\mathcal{M}\rightarrow A(\bar{\phi})\mathcal{M}$. Rescaling the mass and the coordinates, the Jordan frame metric is
\begin{align}
\tg_{00}&=-1+2\frac{A^2(\bar{\phi})G\mathcal{M}}{r}\left(1+2\beta(\bar{\phi})^2\left[1-\frac{\mathcal{M}(\rs)}{\mathcal{M}}\right]\right),\quad \tg_{0i}=0,\quad\textrm{and}\\ g_{ij}&=\left[1+2\frac{A^2(\bar{\phi})G\mathcal{M}}{r}\left(1-2\beta(\bar{\phi})^2\left[1-\frac{\mathcal{M}(\rs)}{\mathcal{M}}\right]\right)\right]\delta_{ij},
\end{align}
where the weak-field limit implies we ignore all higher-order polynomials involving $\phi$. More correctly, the PPN counting scheme assumes $\phi\le GM/r\sim v^2/c^2$ and higher-power terms, and cross terms, are therefore higher-order. 

The Jordan frame metric is not yet in the PPN gauge; we need to rescale
\begin{equation}
G\rightarrow G_{\rm N}\equiv A^2(\bar{\phi})\left(1+2\beta^2(\bar{\phi})\left[1-\frac{\mathcal{M}(\rs)}{\mathcal{M}}\right]\right).
\end{equation}
This defines Newton's constant as measured in laboratory experiments. The distinction between $G$ and $\GN$ is not overly important for screened modified gravity because these experiments are performed deep in the screened regime and $G\approx\GN$ but is crucial for theories without screening mechanisms. Performing this rescaling, one finds a metric in precisely the PPN form with \cite{Saaidi:2011zza,Hees:2011mu,Scharer:2014kya,Zhang:2016njn,Sakstein:2017pqi}
\begin{equation}
\gamma=\left[1-2\beta(\bar{\phi})^2\left(1-\frac{\mathcal{M}(\rs)}{\mathcal{M}}\right)\right]\left[1+2\beta(\bar{\phi})^2\left(1-\frac{\mathcal{M}(\rs)}{\mathcal{M}}\right)\right]^{-1}\approx 1-4\beta(\bar{\phi})^2\left(1-\frac{\mathcal{M}(\rs)}{\mathcal{M}}\right).
\end{equation}
Note that throughout this derivation we have not made use of any screening mechanisms directly, we could have taken any conformal field theory and applied the same procedure. The novel aspect of screening mechanisms is the non-linearity in the field equations, which means that instead of having $|\gamma-1|\propto 2\beta^2(\bar{\phi})$, one instead has $|\gamma-1|\propto 2\beta^2(\bar{\phi})(1-\mathcal{M}(\rs)/\mathcal{M})\ll2\beta^2(\bar{\phi})$ in the screened regime. Without screening mechanisms, we would have to tune $\beta^2(\bar{\phi})<10^{-5}$ in order to satisfy the Cassini bound  $|\gamma-1|<(2.1\pm2.3)\times10^{-5}$ \cite{Fomalont:2009zg}. With screening mechanisms, this bound can be  automatically satisfied for screened objects ($\mathcal{M}(\rs)\approx\mathcal{M}(r)$) without the need to perform any tunings.

\subsubsection{Lensing Revisited}

The careful reader will now be puzzled by a conundrum. We have already argued in section \ref{sec:lening_mass} that screened modified gravity (in fact, our derivation above applies equally to all conformal scalar-tensor theories) does not affect the lensing of light. We have also argued in this section that the PPN parameter $\gamma\ne1$ so that light bending by the Sun is different than in GR, which implies that the scalar does affect lensing. In fact both of these statements are compatible, the difference is merely a choice of coordinates. 

In section \ref{sec:lening_mass}, we did not fix to the PPN gauge, and so what we called $G$ is not the same as $\GN$, the value measured in laboratory experiments (although these should be approximately the same since we live in a screened environment). In fact, we could equivalently write equation \eqref{eq:dynamicalmass} as
\begin{equation}
\tilde{\Phi}_{\rm N}'=\frac{\GN \mathcal{M}}{r^2}.
\end{equation}
This relation is typically tested using kinematics, i.e. by equating it to $v_c^2/r$, where $v_c$ is the circular velocity. Such a test does not determine the mass, but rather, the product $\GN\mathcal{M}=G\mathcal{M}_{\rm dyn}$. If one chooses to set $G=\GN$ then this measurement determines $\mathcal{M}_{\rm dyn}$, and one finds that this is larger than $\mathcal{M}$. Alternatively, one could remove $G$ completely by measuring $\tilde{\Psi}=G\mathcal{M}_{\rm lens}/r$ and take the ratio $\tilde{\psi}/\tilde{\Phi}_{\rm N}=\mathcal{M}_{\rm lens}/\mathcal{M}_{\rm dyn}=\gamma$. Only the ratio of the two metric potentials is relevant physically, that is to say, the amount of gravitational lensing relevant to the force felt by non-relativistic objects. Whether or not $\phi$ directly affects lensing or not is completely a matter of coordinates, and how one chooses to interpret them.

\subsection{Equivalence Principle Violations}

One important feature of screened modified gravity models is that they do not satisfy the equivalence principle. By this, we mean that extended objects with identical masses but differing compositions will not fall at the same rate in externally applied gravitational (Newtonian + scalar) fields\footnote{Note that point particles do satisfy the equivalence principle because every matter species appearing in the action \eqref{eq:STtheoriesgeneral} is universally coupled to the Jordan frame metric and thus follow the same geodesics. The motion of extended objects is governed by energy-momentum conservation and it is here that the difference arises. See \cite{Hui:2009kc} for an extended discussion of this.}. This can be quantified by considering the Newtonian equation of motion for an extended object in external fields $\pn^{\rm ext}$ and $\phi^{\rm ext}$ (defined in the Einstein frame) respectively 
\begin{equation}
\label{eq:EP_violation_1}
\mathcal{M}\ddot{\vec{r}}=-\mathcal{M}\nabla\pn^{\rm ext}-Q\nabla\phi^{\rm ext}.
\end{equation}
The mass on the left hand side is the inertial mass of the object whereas the mass on the right hand side is the gravitational mass, which can be thought of as a gravitational charge (analogous to the electric charge) for the object. Since we are working in the Einstein frame, these two are equal. The quantity $Q$ is the object's scalar charge, which describes its response to the externally applied scalar gradient; one can show that \cite{Hui:2009kc} 
\begin{equation}
\label{eq:scalar_charge}
Q=\beta(\phi_0)\left(\mathcal{M}-\mathcal{M}(\rs)\right).
\end{equation}
This implies that the motion of the object depends on the screening radius, which in turn depends on the objects internal structure. The equivalence principle is thus violated for all objects except those that are completely screened (because $Q=0$) or fully unscreened (because $\rs=0$ and $Q=\mathcal{M}$). This equivalence principle violation allows for several novel tests that we will discuss below.

\subsection{Laboratory Screening}

Laboratory searches for screened fifth forces, and the particles that mediate them, are typically performed in a vacuum chamber.  Inside this chamber the position of the minimum of the effective potential can be different to the minimum of the effective potential in the walls of the vacuum chamber and its environment.  This is the key difference between screening in the laboratory, and screening in other astrophysical environments;  in a vacuum chamber there is a region of low density surrounded by a region of higher density. 

The behaviour of the field in the experimental apparatus depends on its mass, as the corresponding Compton wavelength sets the scale over which the field can vary its value.  The field can only change its value from the exterior of the experiment to the interior of the walls of the vacuum chamber if its Compton wavelength in the walls is of order the thickness of the walls or smaller.  Similarly, the field can only vary its value from the walls to the vacuum at the center of the chamber if its Compton wavelength in the chamber is comparable to, or smaller, than the diameter of the chamber. 

The chameleon field can vary its mass much more easily than the symmetron, and as a result laboratory tests constrain a much broader range of models for the chameleon.  If the symmetron mass is too small it will not be able to vary its VEV over the scale of the experiment. In this case there are no field gradients in the  experiment, and no resulting fifth forces, so no constraints can be placed. As the symmetron mass increases the vev starts to vary within the experiment, and a fifth force is present, however this fifth force may then be exponentially suppressed by the Yukawa term $e^{-m r}$, where $m$ is the mass of the symmetron in the vacuum. In general, therefore, laboratory experiments will only constrain a small range of symmetron masses \cite{Upadhye:2012rc,Burrage:2016rkv,Brax:2016wjk}. 

The chameleon field can vary  more easily in a laboratory vacuum, and therefore is much more amenable to laboratory constraints.  Over a wide range of the chameleon parameter space the chameleon will not be able to reach the value which minimises its potential in the interior of the vacuum chamber, and instead it will evolve to the value which sets its mass to be of order the size of the chamber. Once the corresponding Compton wavelength becomes smaller than the size of the chamber the field is able to reach the minimum of its effective potential. 

If the experiment is performed in a sufficiently small region at the center of the vacuum chamber then we can assume that the background value due to the vacuum chamber is constant.  Then, the screening condition simplifies. A sphere at the center of the vacuum chamber will be screened if there is a solution for the screening radius $\rs>0$ to 
\begin{equation}
1-\frac{r_S^2}{R^2}=\left(\frac{\mc}{M_P}\right)^2\frac{8\pi M_P^2 R}{M_{\rm obj}}\left(\frac{\phi_{\rm vac}-\phi_{\rm min}(\rho_{\rm obj})}{M}\right)
\end{equation}
where $M_{\rm obj}$ is the mass of the sphere, $R$ its radius  and $\rho_{\rm obj}$ its density. $\phi_{\rm vac}$ is the background chameleon value due to the vacuum chamber. The right hand side of this can be viewed as the ratio of the chameleon to Newtonian potentials at the surface of the object; this relation can be found by evaluating equation \eqref{eq:screening_radius_1} for a sphere of constant density. 

Clearly determining both the background value of the scalar field and the condition for screening become more complicated for non-spherical geometries, and in these cases numerics are needed to place definitive constraints. However the principles described here will still guide the shape of the field profile, and the conditions for screening. 

Laboratory searches for fifth forces are performed with both classical and quantum experiments. To determine the condition for screening in a quantum experiment requires a little more thought.  If the experiment is sufficiently low energy that the internal structure of the source is not disrupted, it must still be checked how the chameleon screening condition is affected by the delocalisation of the object's center of mass \cite{Burrage:2014oza}. The chameleon can respond to changes in the position of the source on timescales on the order of $1/ m_{\rm eff}(\phi_{\rm vac})$, and a delocalised source can be considered to fluctuate around with a time-scale $R_{\rm trap}/v$ where $R_{\rm trap}$ is the spatial extent of the trapping potential, and $v$ is the velocity of the particle. If $(v/R_{\rm trap})<m_{\rm vac}$ the chameleon field can respond to the quantum fluctuations of the object and therefore it is the object's density and size which determine whether the object is screened, regardless of the uncertainty on its center of mass position. Otherwise the chameleon cannot respond to the fluctuations in the position of the source, and the relevant density in the screening condition is $\bar{\psi}_{\rm obj}\psi_{\rm obj}$, where $\psi_{\rm obj}$ is the wavefunction of the object \cite{Burrage:2014oza}.

\subsection{Screening in the Jordan Frame}

{

In this review we will work exclusively in the Einstein frame but, for completeness, and because it has received little attention in the literature, we will discuss how screening works in the Jordan frame. We will follow the notation of \cite{Hui:2009kc}, who have provided the most comprehensive treatment to date\footnote{Note that our conventions differ from theirs. They use tildes to refer to Einstein frame quantities whereas we use them to refer to Jordan frame quantities and their function $\Omega(\phi)$ is related to our coupling function via $\Omega(\phi)=A^{-1}(\phi)$.}, although we will not perform the full Einstein-Infeld-Hoffmann approach for extended objects, instead we will work with the one-body problem to be consistent with our analyses above. 
Written in the Jordan frame, the action \eqref{eq:STtheoriesgeneral} is
\begin{equation}
\label{eq:JFaction}
S=\int\dd^4 x\sqrt{-\tg}\left[\frac{\mpl^2 }{2A^2(\phi)}\tilde{R}(\tg)-\frac{k(\phi)}{2}\partial_\mu\phi\partial^\mu\phi-\frac{V(\phi)}{A^4(\phi)}\right]+S_{\rm m}[\tilde{g}_{\mu\nu}],
\end{equation}
where
\begin{equation}
k(\phi)=\frac{1}{A^{2}(\phi)}\left[1+6\mpl^2\left(\frac{\dd\ln A}{\dd\phi}\right)^2\right].
\end{equation}
In the Jordan frame, the matter is minimally coupled to $\tg_\nm$ but the scalar has a non-canonical kinetic term, is non-minimally coupled to $R$, and the scalar potential is $V_{\rm J}(\phi)=V(\phi)/A^4(\phi)$. The scalar equation of motion is
\begin{equation}
\label{eq:seomJF}
k(\phi)\Box\phi+\frac{\dd k}{\dd\phi}\partial_\mu\phi\partial^\mu\phi-\frac{\dd V_{\rm J}}{\dd\phi}+\frac{1}{2}\frac{\dd A^{-2}(\phi)}{\dd \phi}\tilde{R}=0.
\end{equation}
Since the Ricci scalar appears in this equation, we also need the Einstein equations, which are
\begin{equation}
\label{eq:EEJF}
G_\nm=\frac{A^2(\phi)}{\mpl^2}\left[ \tilde{T}_{{\rm m}\,\,\nm}+k(\phi)\partial_\mu\phi\partial_\nu\phi-g_\nm\left(\frac{k}{2}\nabla_\alpha\phi\nabla^\alpha\phi+V_{\rm J}(\phi)\right)+\left(\nabla_\mu\nabla_\nu -g_\nm\Box\right)A^{-2}\right].
\end{equation}
Taking the trace of this, one finds
\begin{equation}
\tilde{R}=-\frac{A^2(\phi)}{\mpl^2}\left[\tilde{T}_{\rm m}-k\partial_\alpha\phi\partial^\alpha\phi+V_{\rm J}+3\Box A^{-2}(\phi)\right],
\end{equation}
which can be used to eliminate $\tilde{R}$ in equation \eqref{eq:seomJF}. These equations are complicated, but they simplify significantly in the Newtonian (weak-field) limit. As discussed by \cite{Will:2004nx,Ip:2015qsa}, the expansion parameter in the Newtonian limit is $v^2/c^2$ (or $GM/R$, the Newtonian potential) and one should take $\phi\sim v^2/c^2$ or smaller. In this case, one has 
\begin{align}\label{eq:WF}
A^n(\phi)\approx 1+\frac{n\beta(\phi_0)\phi}{\mpl},\,V_J(\phi)\approx V(\phi),\,\partial_\alpha\phi\partial^\alpha\phi\sim\mathcal{O}\left(\frac{v^4}{c^4}\right),\textrm{ and } \tilde{T}_{\rm m}\approx -\tilde{\rho},
\end{align}
where we have neglected terms at higher-order than $v^2/c^2$ and possible time-derivatives of the asymptotic field. We remind the reader that $\tilde{\rho}\sim v^2/c^2$ is the Jordan frame density. In the weak-field limit, we can therefore ignore all factors of $k(\phi)$ since they multiply terms that are higher-order than $v^2/c^2$\footnote{Technically one does have an $\mathcal{O}(1)$ contribution to $k(\phi)\approx 1+6\mpl^2/M^2$ which can be $\gg1$ for some values of $M$ considered here. In fact, it should be the canonically normalized field, $\varphi=\sqrt{k(\phi)}\phi\sim\mathcal{O}(v^2/c^2)$ (at this order), which is why we can neglect this contribution.}. We may ignore this contribution. With these approximations, one has $\tilde{R}\approx -\tilde{T}_{\rm m}/\mpl\approx \tilde{\rho}/\mpl^2$ so that equation \eqref{eq:seomJF} becomes (sending $\Box\rightarrow\nabla^2$ as time-derivatives are of order $v/c$ in the Newtonian limit)
\begin{equation}
\nabla^2\phi=\frac{\dd V(\phi)}{\dd \phi}+\frac{\beta(\phi_0)\tilde{\rho}}{\mpl}.
\end{equation}
This is none other than equation \eqref{eq:EOM_rho} (the Einstein frame scalar equation of motion) with the Einstein frame density replaces by the Jordan frame density. In fact, since $\tilde{T}_{\rm m}^\nm=A^{-6}T^{\nm}_{\rm m}$ one has $\tilde{T}_{\rm m} = A^{-4} T_{\rm m}$ so that $\tilde{\rho}=\rho+\mathcal{O}(v^4/c^4)$. The equation of motion for the scalar is therefore identical in both frames in the weak-field limit. Non-relativistic screening, which is all we are concerned with in this review, therefore works identically in both frames. 

In order to find the fifth-force, one can perform the Weyl-rescaling $\tilde{g}_\nm=A^2(\phi)g_\nm$ (taking the weak-field limit \eqref{eq:WF}) on equation \eqref{eq:WFEF} 
to find
\begin{equation}
\dd\tilde{s}=\left(-1+2\Phi+2\frac{\beta(\phi_0)\phi}{\mpl}\right)\dd t^2 +\left(1+2\Psi-2\frac{\beta(\phi_0)\phi}{\mpl}\right)\delta_{ij}\dd x^i\dd x^j
\end{equation}
so that the Jordan frame potentials are
\begin{align}
\tilde{\Phi}&=\Phi+\frac{\beta(\phi_0)\phi}{\mpl}\textrm{ and}\\
\tilde{\Psi}&=\Psi-\frac{\beta(\phi_0)\phi}{\mpl}.
\end{align}
In the weak-field limit the force is
\begin{equation}
F= -\vec{\nabla}\tilde{\Phi}=-\nabla\Phi-\frac{\beta(\phi_0)}{\mpl}\vec{\nabla}\phi.
\end{equation}
The second term is the fifth-force, which is identical to the total force calculated in the Einstein frame. 

}

% Experimental tests
\section{Experimental Tests}
\label{sec:experiments}

In this section we summarize the present experimental tests of chameleon and symmetron screening, which range from particle collider and precision laboratory experiments to astrophysical tests using stars and galaxies. %We first discuss screening in the context of typical laboratory set-ups and astrophysical sources before proceeding to discuss the state of the art experimental tests.

\subsection{Fifth-Force Searches}

Fifth-force searches aim to directly measure the force between two objects and search for deviations from Newton's law. The experiment is performed inside a vacuum chamber to reduce noise, and the geometry of the experiment is designed to minimize the Newtonian force. Recently, some experiments have been designed specifically for the task of searching for chameleons, either by adapting the geometry to maximize the chameleon force, or by varying the density inside the vacuum chamber. Typically, scales of order $\mu$m or greater are probed.

\subsubsection{Torsion Balance Experiments}

Torsion balance experiments typically consist of one mass that acts as a pendulum suspended above a second that sources a gravitational field and acts as an attractor. The two masses are arranged in a manner that cancels the inverse-square contribution to the total force so that the experiment is sensitive to any deviations.

The state-of-the-art in torsion balance tests is the E\"{o}t-Wash experiment \cite{Adelberger:2003zx,Kapner:2006si,Lambrecht:2005km}, which uses two circular disks as test-masses. The disks have holes bored into them which act as missing masses, giving rise to a net torque due to dipole (and higher-order multipole) moments. The upper disk is rotated at an angular velocity such that the contribution from any inverse-square forces to the torque is zero, and therefore any residual force is non-Newtonian. The absence of any such forces places strong constraints on non-inverse-square law modifications of gravity. This includes any scalar-tensor theory where the field is massive, including Yukawa interactions, and chameleons. 

In order to reduce electromagnetic noise, the pendulum and attractor are coated in gold and a beryllium-copper membrane is placed between them. This poses no additional problems for linear theories such as Yuakawa forces, but does present several technical complications for chameleon theories. The  membrane may or may not have a thin shell depending on the parameters under study, and the highly non-linear nature of the field equations make the theoretical modelling of this non-symmetric system difficult. Over time, several works have appeared with the aim of improving the accuracy of the theoretical calculation of the chameleon torque \cite{Brax:2008hh,Adelberger:2006dh,Mota:2006ed,Mota:2006fz,Upadhye:2012fz}, the most recent being the work of Upadhye \cite{Upadhye:2012qu}, which uses the so-called \emph{one-dimensional plane-parallel} approximation to include the effects of the missing masses on the chameleon force profile. A similar effort has been undertaken for symmetron models, with the most stringent constraints presented in \cite{Upadhye:2012rc}.  

\subsubsection{Casimir Force Tests}

The Casimir force (or Casimir-Polder force) is a prediction of quantum electrodynamics. Classically, two uncharged parallel plates placed in a vacuum would source no electromagnetic fields and therefore would feel no force; quantum mechanically, they interact with virtual photons of the vacuum resulting in a net force that can be interpreted as being due to the zero-point energy of the field between the plates. This force scales as $d^{-4}$ ($d$ is the distance between the plates) and is hence sub-dominant to the Newtonian force except at small separations.

This intriguing force has inspired several experiments to measure it, many of which operate at sub-mm (and even sub-micron) distances \cite{Lamoreaux:2005zza,Lambrecht:2011qm}. A chameleon force (per unit area) between the two plates would scale as \cite{Mota:2006fz,Brax:2007vm,Brax:2014zta}
\begin{equation}
\frac{F_{\rm cham}}{A}\propto d^{-\frac{2n}{n+2}},
\end{equation}
which always scales with a power $\ge-4$ (the bound is saturated when $n=-4$). This would dominate over the Casimir force at large separations and therefore the absence of any deviation from the Casimir prediction can constrain chameleon models.

In practice, it is difficult to keep the plates perfectly parallel, and very smooth plates are required for high-precision results. A more convenient scenario is the case where one of the plates is replaced by a sphere whose radius is larger compared with the separation. In this case, the Casimir force scales as $d^{-3}$ and the chameleon force would scale as 
\begin{equation}
\frac{F_{\rm cham}}{A}\propto d^{\frac{2-n}{n+2}}.
\end{equation}
Again, this power is always $\ge-3$.

The current generation of Casimir force experiments place strong constraints on $n=-4$ and $n=-6$ chameleon models when $\Lambda_c$ is fixed to the dark energy scale. The constraints on other models are not presently competitive with other experiments discussed in this review. The next generation of experiments will use larger separations where the chameleon force is more pronounced \cite{Lambrecht:2005km,Lamoreaux:2005zza} so more stringent constraints on a broader class of models are expected. 

Interestingly, experiments such as these can be adapted to the chameleon's unique properties because one can vary the density of the partial vacuum inside the chamber where the experiment operates. By changing the pressure of the ambient gas, one can look for a density-dependent change in the force, which would be a smoking gun of chameleon models \cite{Brax:2010xx,Almasi:2015zpa}.

At the present time, Casimir force experiments have not been applied to symmetron models, mainly due to the lack of any theoretical calculations of the symmetron force between objects of different geometries.

\subsubsection{Levitated Microspheres}

A recent addition to the fifth-force hunter's arsenal, optically-levitated microspheres are capable of probing forces $\lsim\oo(10^{-8}\textrm{ N})$ \cite{Geraci:2010ft}. The spheres have radii of $\oo(\mu\textrm{m})$ and, in the context of chameleon models, they would therefore be unscreened when $\Lambda_c\ge4.6$ meV (a factor of two above the dark energy scale). The spheres are held in an upward pointing laser beam trap by virtue of radiation pressure so as to counteract the Earth's gravity; any anomalous motion would then be due to non-gravitational interactions. In the case of chameleon models, a microsphere held in a chameleon gradient would experience a additional force given by 
\begin{equation}
F=\lambda\left(\frac{\rho}{\mc}\right)\int_{\rm sphere}\dd^3\vec{x}\frac{\partial\phi}{\partial z},
\end{equation}
where $z$ is the vertical direction and the sphere's density $\rho$ is assumed to be constant. The parameter $\lambda$ is the scalar charge of the sphere. When the sphere is unscreened, which is the case for $\mc\lsim10^{10}$ TeV, the chameleon force is unsuppressed and $\lambda=1$. When the sphere has a thin shell one has $\lambda<1$ and the constraints are not as stringent in this regime.

An experiment measuring forces using levitated microspheres has recently been applied to chameleon models resulting in new constraints on $n=1$ models \cite{Rider:2016xaq}; other models have yet to be considered. %Similarly the effectiveness of this technique to constrain symmetron models has yet to be explored.
Constraints on symmetron models are not currently competitive with other experiments \cite{Burrage:2016rkv}.

\subsection{Precision Atomic Tests}

Precision atomic tests search for corrections to the structure of hydrogenic atoms by looking for non-standard perturbations to the Hamiltonian. In the case of chameleons, electrons would feel a chameleon potential in addition to the Coulomb potential given by
\begin{equation}
\delta H=\frac{m_e}{\mc}\phi_{\rm N},
\end{equation}
where $\phi_{\rm N}$ is the chameleon field sourced by the nucleus. Since the vacuum chamber shields the experiment from the effects of the external field, chameleons with strong couplings to matter 
%(that would otherwise be screened by the galactic density) 
can be probed by looking for the shifts in the atomic energy levels due to this perturbation. In particular, this shielding implies that the nucleus is fully unscreened so that the shifts to the lowest energy levels are \cite{Brax:2010gp}
\begin{align}
\Delta E_{\rm 1s}&=-\frac{Zm_Nm_e}{4\pi a_0\mc^2}\\
\Delta E_{\rm 2s}=\Delta E_{\rm 2p}&=-\frac{Zm_Nm_e}{16\pi a_0\mc^2},
\end{align}
where $Z$ is the atomic number, $m_N$ is the nucleon mass, and $a_0$ is the Bohr radius. The potential coupling of the chameleons to photons will break the degeneracy between the 2S and 2P levels.

Presently, the 1S-2S transition in atomic hydrogen is the best constrained, having a total uncertainty of $10^{-9}$ eV (at $1\sigma$) \cite{Jaeckel:2010xx,Schwob:1999zz,Simon:1980hu}. The excellent agreement with standard atomic theory constrains the chameleon coupling
\begin{equation}
\mc\gsim 10\textrm{ TeV}.
\end{equation}
The effects of symmetron models on atomic transitions has yet to be investigated, although the $\mathbb{Z}_2$ means that the effective interaction with nucleons and electrons is higher-order i.e.
\begin{equation}
\mathcal{L}\supset m_e\frac{\phi^2}{2\ms^2}\bar{e}e,
\end{equation}
so that one would not expect this test to be as constraining.

\subsection{Atom Interferometry}

Atom interferometry is a hybridization of classical interferometric experiments and quantum mechanical double slit experiments. Atoms can be put into a superposition of two states, which travel along different paths and hence act like the arms of an interferometer. The two paths can be recombined later to produce an interference pattern that can be measured.

The atoms can be moved within the interferometer by shining laser light on them. If an atom absorbs a photon, it will be excited into a higher energy state and acquire the photon's momentum, resulting in some linear motion. In the absence of any observation, the atom is in a superposition of the ground state (where it is stationary) and an excited state (where it is in motion). The atom can be put into a superposition of states that travel along different paths by repeating this process several times.

The probability of measuring the atom in an excited state at the output of the interferometer is a function of the difference in phases accumulated by the wave functions on the two paths. If the atom is moving in an external force field that causes some constant acceleration $a$ then this probability is
%of measuring it in an excited state is given by
\begin{equation}
P=\propto\cos^2\left[\frac{a k T^2}{\hbar}\right],
\end{equation}
where $k$ is the photon momentum and $T$ is the duration of the experiment. 

A massive object placed inside the vacuum chamber will source a gravitational field that contributes to $a$. If, in addition to this, the object sources a chameleon field then this too contributes and the probability of measuring excited atoms is sensitive to it. Since atoms placed in vacuum chambers are unscreened over a large range of the parameter space, this experiment is incredibly sensitive to chameleon and symmetron forces \cite{Burrage:2014oza,Burrage:2015lya,Elder:2016yxm}. Indeed, the first generation of atom interferometry experiments designed to test screened modified gravity was able to constrain any anomalous acceleration down to levels of $10^{-6}g$ ($g\equiv GM_\Earth/R_{\Earth}$ is the gravitational acceleration at the surface of the  Earth), placing new constraints on chameleons and symmetrons that vastly reduced the viable parameter space \cite{Hamilton:2015zga,Burrage:2016rkv}. The current generation of experiments has constrained this further to $\lesssim 10^{-8}g$, reducing the parameter space further \cite{Jaffe:2016fsh}. 
%The references in this section have reported chameleon bounds for $n=1$ models. In this review we will translate their constraints into more general models. 

\subsection{Precision Neutron Tests}

Neutrons are perfect objects for testing short-range gravitational physics because they are electrically neutral, and are therefore not sensitive to electromagnetic noise such as background fields and van de Walls forces\footnote{Atoms are neutral as well but one advantage of neutrons is that their polarizability is 15 orders of magnitude smaller, making Van de Waals forces less of a background. We are grateful to Tobias Jenke for pointing this out to us.}.
%that plague charged particle experiments. 
This has motivated a recent interest in using neutrons to test chameleon models, which we summarize below. At the present time, all of the constraints derived using neutron experiments fix $\Lambda_c$ to the dark energy scale.   

\subsubsection{Ultra-cold Neutrons}

It is possible to arrange for neutrons produced in nuclear reactors to bounce above a mirror. These neutrons interact with the Newtonian potential of the Earth leading to a quantized energy spectrum. The mirror itself could source a chameleon field, which would act as a perturbation to the neutron Hamiltonian given by \cite{Brax:2011hb,Ivanov:2012cb}
\begin{equation}\label{eq:bouncing_neutrons}
\Delta H=\frac{m_N}{\mc}\phi=\frac{2.2\textrm{ keV}^2}{\mc}\left(\frac{z}{82\,\mu\textrm{m}}\right),
\end{equation}
where $z$ is the distance above the mirror. If this perturbation were large enough, new bound states would appear in the spectrum. No such states have been observed by a {qBounce experiment at the Institut Laue-Langevin in Grenoble}, which immediately places a new constraint \cite{Brax:2011hb}
\begin{equation}
\mc>10^4\textrm{ TeV}.
\end{equation}
Away from this regime, the perturbation \eqref{eq:bouncing_neutrons} leads to a shift in the energy levels. This can be probed using resonance spectroscopy, the most constraining transition being $|3\rangle\rightarrow|1\rangle$. The absence of any observed shift leads to the stronger constraint \cite{Jenke:2014yel}
\begin{equation}
\mc>1.7\times10^6\textrm{ TeV}
\end{equation}
for $n=1$. {In this review, we use the most up to date (at the time of writing) constraints given in \cite{Cronenberg:2015bol}\footnote{We thank Tobias Jenke for providing us with the numeric values.}.}

Bouncing neutron techniques have not yet been applied to symmetron models. The effective interaction for these models would be
\begin{equation}
m_N\frac{\phi^2}{\ms^2}\bar{n}n,
\end{equation}
and so one may expect a similar issue to testing symmetrons using precision atomic tests i.e. the higher-order nature of the interaction means that it would be naturally suppressed, leading to weaker constraints than chameleons. 

\subsubsection{Neutron Interferometry}

In an analogous manner to optical interferometry, a coherent beam of neutrons can be split and later recombined to produce interesting interference patterns \cite{Pokotilovski:2013pra,Brax:2013cfa}. A mono-silicone crystal plate can be used for this purpose. 

The proposal for testing chameleons using this technique is to introduce a cell composed of two parallel plates into the path one of the beams. A chameleon profile will develop between the two plates leading to a phase shift for the neutrons given by \cite{Brax:2013cfa,Brax:2014gja}
\begin{equation}
\delta\varphi=\frac{m_N^2}{\hbar^2k\mc}\int_{-d}^d\phi(x)\dd x,
\end{equation}
where $x$ is the horizontal direction and the plates are located at $x=\pm d$. This phase shift is maximum if the plates are in vacuum (or, rather, a partial vacuum) but diminishes if one were to inject gas at a higher density due to the suppression of the chameleon field. Such an experiment has been performed by two groups \cite{Lemmel:2015kwa,Li:2016tux}, who report consistent bounds in the range
\begin{equation}
M>10^7\textrm{--}10^8 \textrm{ TeV}
\end{equation}
for models with $1\le n\le 6$, with stronger bounds being obtained for lower $n$.

\subsection{Astrophysical Tests}

In this section we describe tests of chameleon and symmetron models using astrophysical objects. In many cases, the constraints are phrased in terms of $\chi_0$ and $\beta(\phi_0)$ and so the specific model is not important. {We will not include bounds from binary pulsars since they are uncompetitive and subject to astrophysical uncertainties to do with the screening level of the Milky Way \cite{Brax:2013uh,Zhang:2017srh}.}

\subsubsection{Distance Indicator Tests}

Determining the distance to astrophysical objects is a notoriously difficult task because only the flux of emitted photons, can be measured. Since this depends on both the distance and the absolute luminosity of the source via
\begin{equation}
F=\frac{L}{4\pi d^2}
\end{equation}
some knowledge of the luminosity $L$ is needed to infer the distance. Distance indicators are objects with some intrinsic or empirical relation between their luminosity and other observable properties. One famous example are type-Ia supernovae, where the luminosity can be found by fitting their light curve, making them standard candles. 

In the context of modified gravity, it is possible that the relation used to determine the luminosity is sensitive to gravitational physics. If the relation has been calculated using general relativity, or has been determined empirically using local (screened) observations, then it will give incorrect distances when applied to unscreened galaxies. In contrast, relations that are insensitive to the theory of gravity will always give the correct distance. Comparing how well different distance estimates to theoretically unscreened galaxies agree can therefore yield new constraints.

One robust distance indicator that is not sensitive to screened modified gravity is the tip of the red giant branch (TRGB). Low mass post-main-sequence stars ($M_\odot\lsim M\lsim 2M_\odot$) in the process of ascending the red giant branch (RGB) consist of an isothermal helium core surrounded by a thin hydrogen-burning shell. The hydrogen in this shell is continually processed into helium that is deposited onto the core, causing it's temperature to rise steadily as the RGB is ascended. When the temperature is sufficiently high, the triple-$\alpha$ process (core helium burning) can proceed efficiently, at which point the star moves to the asymptotic giant branch in a very short time-scale. This leaves a visible discontinuity in the I-band. The discontinuity occurs at fixed luminosity ($I=4.0\pm0.1$, the error is due to a very weak metallicity dependence \cite{1999IAUS..183...48S,Freedman:2010xv,Beaton:2016nsw}), making the TRGB a standard candle. Importantly, the physics of the helium flash is set by nuclear physics and is non-gravitational in origin, elucidating our earlier assertion that this distance indicator is insensitive to modified gravity\footnote{Technically, this is only the case when $\chi\lsim10^{-6}$, corresponding to parameters where the hydrogen burning shell becomes unscreened. When this happens, the core temperature increases at a faster rate leading to a reduction of the tip luminosity because the star has less time time to ascend the RGB. We will see shortly that $\chi>10^{-6}$ can be ruled out by other, independent means and so we will not dwell on this too much here. }.

Cepheid variable stars are distance indicators that are sensitive to modified gravity. With masses between $4$ and $10M_\odot$, these stars enter a phase where their structure is dominated by semi-convection---a convective process driven by inverse-gradients in the chemical composition---shortly after ascending the RGB, resulting in large temperature increases with a relatively small change in luminosity. This results in so called \emph{blue loops} in the Hertzprung-Russell (or color-magnitude) diagram. Whilst traversing the blue loop, the star crosses the instability strip where it is unstable to pulsations due to the presence of a layer of doubly-ionized helium\footnote{This has the result that small compressions result in an increased opacity that in turn causes an increase in the energy absorbed. The energy dammed up by this compression drives the pulsations. This is known as the $\kappa$-mechanism}. Cepheids pulsate with a well-measured period-luminosity relation (PLR) where the period $\Pi\propto\sqrt{R^3/G\mathcal{M}}$. This relation is therefore different in unscreened galaxies, and, in particular, if one applies the locally measured formula to an unscreened galaxy one under-estimates the distance by a factor
\begin{equation}
\frac{\Delta d}{d}\approx -0.3\frac{\Delta G}{G}.
\end{equation}

The screening mechanisms above can therefore be tested by comparing TRGB and Cepheid distances to unscreened galaxies. Reference \cite{Jain:2012tn} have done precisely this for a sample of 25 galaxies taken from the screening map \cite{Cabre:2012tq}. They also compared distances to a similar sample of screened galaxies as a control set. They found a similar agreement and scatter in both cases, and a $\chi^2$-fit to both GR and modified gravity models yielded constraints\footnote{Metallicity and other corrections produce a positive $\Delta d/d$, which makes the constraints even stronger.} in the $\chi_0$--$\beta(\phi_0)$ plane that we translate into chameleon, symmetron, and $f(R)$ parameters in section \ref{sec:constraints}.

\subsubsection{Rotation Curve Tests}
\label{sec:rot_tests}

The circular velocity of objects orbiting the center of galaxies is given by
\begin{equation}
v_c^2=\frac{G\mathcal{M}_{\rm gal}(r)}{r^2}\left(1+2\beta(\phi_0)\frac{Q}{\mathcal{M}}\right),
\end{equation}
where the scalar charge $Q$ is defined in equation \eqref{eq:scalar_charge} and $\mathcal{M}_{\rm gal}(r)$ is the galactic mass enclosed by $r$. If $10^{-8}\lsim \chi_0\lsim 10^{-6}$ then dwarf galaxies are unscreened but their constituent stars are not because their Newtonian potential allows them to self screen (see Table \ref{tab:astro_screen_chi}). Stars in unscreened dwarf galaxies therefore have $Q/\mathcal{M}=0$. In contrast, diffuse hydrogen gas with $\pn\sim 10^{-11}$--$10^{-12}$ cannot self-screen and has $Q/\mathcal{M}=\beta(\phi_0)$. Assuming that the galaxy is completely unscreened, the ratio of the circular velocity of stars and gas is then
\begin{equation}
\label{eq:vc_ratio_rotation_curve}
\frac{v_{c,\,{\rm gas}}}{v_{c,\,\star}}=\sqrt{1+2\beta^2(\phi_0)}
\end{equation}
implying that the galactic rotation curve measured using stellar observations will disagree with the rotation curve measured using observations of the interstellar gas. This is a direct consequence of the equivalence principle violation (i.e. $Q\ne\mathcal{M}$).

Measurements of the galactic rotation curves typically use either H$\alpha$ emission or the 21 cm line, both of which probe the gaseous component. An alternate but less prevalent method involves measuring the Mgb triplet lines, which are due to absorption in the atmosphere of K- and G-stars ($0.45M_\odot\lsim M\lsim 1.2M_\odot$). At present, the screening map contains six unscreened dwarf galaxies for which both Mgb and either H$\alpha$ or 21 cm data (or both) are available. Using this, \cite{Vikram:2014uza} have reconstructed both the gaseous and stellar rotation curves, and have used them to test the prediction \eqref{eq:vc_ratio_rotation_curve} using a separate $\chi^2$ fit for each galaxy. The has placed new constraints in the $\chi_0$--$\beta(\phi_0)$ plane which are comparable with the Cepheid bounds. 

\subsubsection{Galaxy Clusters}

The predicted difference between the dynamical and lensing masses discussed in Section \ref{sec:lening_mass} can be tested using observations of galaxy clusters, for which there is a wealth of X-ray and weak lensing data available. The X-ray brightness temperature is a measure of the mass of the hot gas in the intra-cluster medium, which is in hydrostatic equilibrium and hence satisfies\footnote{This assumes that the gas entirely supported by thermal pressure. In practice, one expects a small amount of non-thermal pressure but N-body simulations of chameleon theories have shown this to be negligible \cite{Wilcox:2016guw}. }
\begin{equation}
\label{eq:hydrostatic_mass}
\frac{\dd P}{\dd r} = -\frac{G\mathcal{M}_{\rm dyn}\rho}{r^2}.
\end{equation}
X-ray observations therefore probe the dynamical mass whereas weak lensing probes the lensing mass, so comparing the two places new constraints on screening. This was first done by \cite{Terukina:2013eqa} using observations of the Coma cluster to find the new constraint $f_{R0}<6\times10^{-5}$. Reference \cite{Wilcox:2016guw} subsequently applied the same methodology to a sample of 58 clusters using X-ray data from the XMM Cluster Survey and weak lensing data from CFHTLenS to obtain further constraints on more general chameleon models.

\subsection{\texorpdfstring{$f(R)$}{TEXT} Specific Tests}

In this section we will briefly summarize tests that have been specifically designed to test the Hu \& Sawicki \cite{Hu:2007nk} $f(R)$ theories discussed in section \ref{sec:f(R)}. Note that, since these theories correspond to chameleons with $-1<n<-1/2$, many of these tests are unconstraining for more general chameleon models. Similarly, specific tests are needed to target this parameter range. Note also, that $f(R)$ models are designed to be cosmologically relevant, and so the majority of the tests discussed here are astrophysical in nature. In what follows, we will only focus on $b=1$ ($n=-1/2$) models because the majority of tests have reported constraints for this model only. Larger values of $b$ are more readily screened and so one would expect the constraints to be weaker. Note that some tests mentioned above report bounds on $f_{R0}$. We will not repeat that discussion here. A full list of constraints on $f_{R0}$ can be found in \cite{Lombriser:2014dua} Table \ref{tab:astro_screen_chi}.

\subsubsection{Solar System Bounds}

One can solve the field equations sourced by the Sun to find a bound on the the value of $f_R^{\rm gal}=\dd f(R)/\dd R(\rho^{\rm gal})$ (defined as $\dd f(R)/\dd R$ at the Milky Way density) \cite{Hu:2007nk}
\begin{equation}
f_R^{\rm gal}=(\gamma-1)\frac{G\mathcal{M}_\odot}{R_\odot}\lsim 4.9\times 10^{-11},
\end{equation}
where $\gamma$ is the Eddington light-bending parameter in the PPN formalism. Relating the galactic density to the cosmological density ($\rho^{\rm gal}=10^{-24}$ g cm$^{-3}$) one finds
\begin{equation}
f_{R0}<74(1.23\times 10^6)^{b-1}\left[\frac{R_0}{\mu^2}\frac{\Omega_mh^2}{0.13}\right]^{-(b+1)},
\end{equation}
which gives $f_{R0}\lsim 0.03$ for $b=1$.

\subsubsection{Strong Gravitational Lensing}

Another method to probe the predicted discrepancy between the dynamical and lensing mass of an object is to use strong lensing by individual galaxies. In this case one can use the stellar dispersion relation to calculate the dynamical mass. Reference \cite{Smith:2009fn} has performed such a test for a sample of galaxies from the Sloan Lens ACS (SLACS) survey and find a constraint $f_{R0}<2.5\times10^{-6}$.

\subsubsection{Cluster Density Profiles}

N-body simulations of $f(R)$ gravity have repeatedly predicted an enhancement in the dark matter halo density profiles around the virial radius compared with GR \cite{Schmidt:2008tn,Schmidt:2009sv}. This is an artefact of the late-time unscreening in $f(R)$ models. The center of the galaxy is largely unaffected because it is both screened and formed earlier when the screening was more efficient. In contrast, there is a pile-up of mass in the outer regions, which form at later times, due to the weaker screening. Reference \cite{Lombriser:2011zw} has used weak lensing data for the Max-BCG galaxy cluster sample from the SDSS to probe this potential novel feature, finding a constraint $f_{R0}<3.5\times10^{-3}$.  

\subsubsection{Cluster Abundances}

The statistics of galaxy clusters is very sensitive to the theory of gravity. For $f(R)$ theories, the enhanced gravitational force results in a higher abundance of rare massive clusters compared with GR \cite{Schmidt:2008tn} meaning the halo mass function is modified. Making quantitative theoretical predictions for this requires knowledge of physics deep within the non-linear cosmological regime and so N-body simulations and spherical collapse halo models calibrated on them are required in order to make quantitative predictions. 

The first bound obtained by looking at cluster abundances yielded $f_{R0}<1.2\times10^{-4}$ \cite{Schmidt:2009am}. This was obtained by using X-ray inferred clusters in combination with a variety of different cosmological datasets available at the time. A stronger bound $f_{R0}<1.6\times10^{-5}$ has subsequently been obtained by \cite{Cataneo:2014kaa} using a full MCMC analysis of the cluster likelihood function for updated datasets from more recent cosmological surveys.

\subsubsection{Cosmic Microwave Background}

Modifications of GR change the structure of the equations describing linear cosmological perturbations, and can hence effect the cosmic microwave background (CMB) \cite{Zhang:2005vt,Song:2007da,Dossett:2014oia}. Updating various CMB codes to include the effects of $f(R)$ gravity, several groups have all obtained a similar bound $f_{R0}<10^{-3}$ \cite{Song:2007da,Dossett:2014oia,Raveri:2014cka,Cataneo:2014kaa}.

\subsubsection{Scalar Radiation}

As was first pointed out by \cite{Silvestri:2011ch}, pulsating stars should source scalar radiation and hence lose energy over time. If too much scalar monopole radiation (which is absent in GR) is emitted then the pulsations may quench. This was investigated by \cite{Upadhye:2013nfa}, who found that the energy loss to monopole radiation is too weak to place any meaningful bounds. They identified another scenario whereby the scalar radiation sourced by an expanding type II supernovae could drain the kinetic energy of the expanding matter and significantly impede the expansion. This places the weak constraint $f_{R0}<10^{-2}$.

\subsubsection{Redshift-space Distortions}

The clustering of matter can be greatly modified in $f(R)$ cosmologies compared with GR, and this can be particularly pronounced in redshift space \cite{Jennings:2012pt,Bose:2016qun,Bose:2017dtl}. The possibility of testing this was first investigated by \cite{Yamamoto:2010ie}, who examined a sample of luminous red galaxies (LRGs) from the SDSS to find a bound $f_{R0}<10^{-4}$. A more recent study combining redshift-space distortion observations with other cosmological datasets found the stronger bound $f_{R0}<2.6\times10^{-6}$ \cite{Xu:2014wda}.

\subsection{Tests of the Coupling to Photons}

In this section we summarize experimental tests of the coupling to photons discussed in Section \ref{sec:photoncoup}. We will restrict our attention to chameleon models, for which the coupling to photons has been widely studied. Extending these constraints to other models with screening remains a topic for future work. 

\subsubsection{PVLAS}
The PVLAS experiment \cite{Zavattini:2005tm} studied the polarisation of light propagating through a magnetic field.  The presence of an axion, or axion-like particle coupled as in Equation (\ref{eq:photon}) would mean that, in the presence of a magnetic field, one polarisation of the propagating photon can convert into the scalar particle and vice versa.  The second polarisation will propagate through unimpeded \cite{Raffelt:1987im}.  This induces rotation and ellipticity into the polarisation of the incoming laser beam.  The PVLAS experiment bounded the induced rotation to be less than  $1.2\times 10^{−8}$ rad at 5 T and $1.0\times10^{−8}$ rad at 2.3 T, and the induced ellipticity to be less than $1.4 \times 10^{−8}$ at 2.3 T. This constraints the coupling strength $M_{\gamma}$ of a light axion-like particle. 

In such experiments chameleon particles behave very differently to standard axion-like particles, precisely because of their density dependent mass. If standard axion-like particles were produced in PVLAS they would pass through the walls at the end of the vacuum chamber without interacting and so leave the experiment. For a chameleon to pass through the wall, the chameleon particle must have enough energy that it can adjust its mass to the higher value needed for it to exist inside the wall.  If it does not have this energy it is instead reflected from the wall and back into the vacuum chamber \cite{Brax:2007ak,Brax:2007hi}. This leads to a large ratio of the rotation to the ellipticity of the polarisation which is a unique signal of chameleon models. For a chameleon with a potential $V(\phi)=(2.3 \times 10^{-3} \mbox{ eV})^5/\phi$, and assuming the coupling to photons is the same as the coupling to other matter fields, the results of the PVLAS experiment constrain $\mc=M_{\gamma}> 2\times 10^6 \mbox{ GeV}$.

\subsubsection{GammeV-CHASE}
A second commonly used experimental design to look for axion-like particles, light-shining-through-walls, also needs to be modified in order to search for chameleon particles. Experiments searching for standard ALPs rely on the ability of ALPs to pass through walls which are impermeable to photons.  Light is shone into a cavity across which a magnetic field is applied. A wall is then placed in this cavity, in the absence of ALPs no light would be seen on the far side of the wall. But if a photon converts into an ALP before hitting the wall this ALP can pass through and then may reconvert into a photon on the far side of the wall. 

As discussed in the previous subsection, chameleon ALPs cannot pass through walls in the way that standard ALPs do, and so light-shining through walls experiments cannot constrain chameleons.  However this inability to pass through walls can be developed into a new type of experiment specifically designed to look for chameleons; these are known as after-glow experiments \cite{Gies:2007su,Ahlers:2007st}.  The basic design of the experiment is to shine a laser beam into a vacuum chamber across which a magnetic field is applied.  If there is a non-zero probability of the photons converting into chameleons then the number of chameleons trapped  inside the chamber (because they cannot pass through the walls) will increase the longer the laser beam is on.  The laser is then turned off, but the magnetic field is left on.  Then the chameleons can reconvert into photons, leading to a detection of light, after the laser has been turned off. 

This experiment was successfully performed by the GammeV collaboration, and was known as GammeV-CHASE (GammeV CHameleon Afterglow SEarch) \cite{Upadhye:2009iv}. Constraints were placed on values of the chameleon coupling to photons, as a function of the effective chameleon mass in the chamber \cite{Chou:2008gr}.  This mass depends on the choice of the chameleon potential and the strength of the coupling to other matter fields.  For the lightest chameleons inside the vacuum chamber, GammeV-CHASE constrains the coupling to photons to be $M_{\gamma}>3 \times 10^7 \mbox{ GeV}$ \cite{Steffen:2010ze,Upadhye:2012ar}.  The constraints weaken if the effective mass of the chameleon is above $10^{-3} \mbox{ eV}$. 

The modelling of how the chameleon behaves inside the experiment requires care.  Whilst a semi-classical approximation would predict that the chameleon bounces off the walls of the vacuum chamber unchanged, considering the chameleons as fluctuations in a quantum field opens up the possibility that the non-trivial self interactions of the chameleon field could allow a chameleon particle to fragment into a number of lower energy chameleons as it hits the wall. This was shown not to be a significant effect in the GammeV-CHASE experiment for the benchmark potentials $V(\phi)= \lambda \phi^4$ and $V(\phi)=\Lambda^5/\phi$ \cite{Brax:2013tsa}.  However, for steeper potentials this effect will start to become relevant.

\subsubsection{ADMX}
ADMX (Axion Dark Matter eXperiment), is another experiment aiming to detect axions and axion-like particles through the Primakov effect \cite{Asztalos:2009yp,Asztalos:2003px}.  However, in this case the axions come from outside the experiment, and are hypothesised to be responsible for the dark matter in our galaxy \cite{Sikivie:1983ip}.  This set up has been used to constrain chameleon theories using the same afterglow effect discussed above \cite{Rybka:2010ah}, but using microwave photons trapped in a cavity instead of laser light. The experiment excluded couplings $5 \times 10^3 \mbox{ GeV} < M_{\gamma}< 1 \times 10^9 \mbox{ GeV}$ for effective chameleon masses in the cavity $\sim 1.95 \mbox{ $\mu$eV}$.

\subsubsection{CAST}

The CAST (CERN Axion Solar Telescope) experiment searches for axions produced in the Sun, by looking for their reconversion into photons in the bore of a decommissioned LHC magnet \cite{Zioutas:2004hi}. Results from this search can be applied to chameleons, if they are also produced in the Sun.  At the particle level the processes which produce chameleons are the same as those that produce scalar axion-like particles, but determining the total flux of chameleons from the Sun requires taking into account  the added complication that the mass of the chameleon field varies with the density of the solar medium \cite{Brax:2010xq}. 

CAST has not yet detected a signal from the Sun, and so bounds can be placed on the chameleon couplings.  They exclude photon couplings $M_{\gamma} \geq 2.6 \times 10^7 \mbox{GeV}$, for a range of couplings to matter $10^{12} \mbox{ GeV} \leq \mc\leq 10^{18} \mbox{ GeV}$, assuming that the bare chameleon potential is $V(\phi) =(10^{-3}\mbox{ eV})^5/ \phi$ \cite{Anastassopoulos:2015yda}. 

There are also proposals by the CAST collaboration to detect solar chameleons using a novel force sensor \cite{Baum:2014rka}. While chameleons may be produced in the sun due to the coupling to photons, the detection mechanism itself does not rely on the coupling in Equation (\ref{eq:photon}).  The detection relies on having a force sensor sufficiently sensitive that it can measure the chameleon radiation pressure \cite{Karuza:2015sia}, which comes about as the chameleons emitted from the Sun bounce off the sensor, for the same reason that chameleons are reflected from the walls of vacuum chambers, the chameleon particle does not have enough energy to adjust its mass sufficiently to pass through the membrane of the sensor. 

\subsubsection{Collider Constraints}
The collider constraints on chameleon models can also be extended to include the coupling to photons in equation (\ref{eq:photon}).  This leads to additional loops, which should be inserted into the diagrams,  and allows for additional production and decay processes which should be included. Analysis of precision electro-weak data from LEP constrains $M_{\gamma}\gtrsim 10^3\mbox{ GeV}$ \cite{Brax:2009aw}. 

\subsubsection{Galactic and Extra-Galactic Constraints}

The effects of the chameleon on light propagating through magnetic fields, originating in the interaction of equation (\ref{eq:photon}), can also be relevant to astrophysical observations.  For many observations, light from distant sources has to propagate through galactic, intra-cluster, or extra-galactic magnetic fields in order to reach us. Whilst the magnetic fields strengths are much lower than those achievable in the laboratory they extend over much larger distances, meaning that the astrophysical constraints can in principle be more stringent that those achieved in the laboratory.  They do, however, come with much larger uncertainties around the initial luminosity of the source, the polarisation of the light it emits, and  over the structure of the magnetic fields. Astrophysical magnetic fields also display much more structure than the coherent magnetic fields used in laboratory, which adds to the complexity of the calculations. 

In \cite{Burrage:2008ii} it was shown that chameleons coupled to photons can induce both linear and circular polarisation into light from stars. As long as the chameleon mass is smaller than the local plasma density, then it can be neglected in these calculations, meaning that the constraints are largely model independent as long as the chameleon is light on astrophysical scales. Within the galaxy this requires $m_{\phi} <1.3 \times 10^{-11} \mbox{ eV}$. From measurements of the polarisation of galactic stars, expected to be largely unpolarized initially, the bound $M_\gamma> 1.1 \times 10^9 \mbox{ GeV}$ was derived. Assuming the magnetic field strength of the intergalactic medium is $B\approx 3 \mbox{ $\mu$G}$ and the coherrence length is $20 \mbox{ pc}$. The polarisation of light from the Crab nebula, type 1a supernova, high redshift quasars, gamma ray bursts and the CMB was also analysed but the bounds were weaker than those from observations of stars. 

Looking for the depletion in luminosity of astrophysical sources from photons converting into chameleons is difficult because there is generally no way of determining the intrinsic luminosity of the source. However, for some astrophysical objects, correlations have been observed between the luminosity of the source, and a second observable that should not be affected by the coupling to chameleons.  The best constraints of this form on chameleons come from looking at Active Galactic Nuclei (AGN) where the X-ray luminosity at $2 \mbox{ keV}$ is observed to be tightly correlated with the optical luminosity at $ 5 \mbox{ eV}$ \cite{Steffen:2006px,Young:2009jc}. Similar luminosity relations exist for blazars and gamma ray bursts, but these give rise to weaker constraints.  As the probability of a photon converting into a chameleon increases with the frequency of the photon, the effects of the chameleon on the X-ray luminosity of the AGN can be significant, whilst the effects on the optical luminosity remain small. Therefore the luminosity relation can be used to constrain the chameleon \cite{Burrage:2009mj}, with the current best constraint $M_{\gamma}\gtrsim 10^{11}\mbox{ GeV}$ assuming, again, that the chameleons are sufficiently light, $m_{\phi}<10^{-12} \mbox{ eV}$, on astrophysical distance scales that the effects of their mass are negligible \cite{Burrage:2009mj,Pettinari:2010ay}. 

The conversion of photons into chameleons also will increase the opacity of the universe at high frequencies. In \cite{Avgoustidis:2010ju} tests of the distance duality relation, which relates luminosity distance and angular diameter distance to sources, were used to derive constraints on cosmic opacity. This can be viewed as a test of chameleons because depletion of photons from the source will change the luminosity distance, whilst leaving the angular diameter distance unaffected. Constraints are currently not competitive with those from starlight polarisations, but should be expected to improve significantly with new data from upcoming cosmological surveys. 

Light from the cosmic microwave background also passes through magnetic fields on its way to us, although constraints from CMB intensity and polarisation data are difficult to apply because of our lack of knowledge about primordial magnetic fields \cite{Schelpe:2010he}. Knowledge of the magnetic fields of localised objects, such as the Coma cluster, mean that constraints can be obtained from measurements of the Sunyaev-Zel'dovich (SZ) effect. The SZ effect is the distortion of the CMB spectrum by inverse Compton scattering of high-energy electrons. The effect of converting photons into chameleons in the cluster's magnetic field, also depletes the expected photon number, but with a very different frequency dependence. Knowledge of the Coma cluster's magnetic fields leads to the constraint $1.1 \times 10^9 \mbox{ GeV} \lesssim M_{\gamma}$ \cite{Davis:2010nj}.

\subsection{Summary of Tests}

Here, we briefly summarize the tests that have been used to test screened modified gravity to date. The summary is given in Table \ref{tab:experiment_summary} and we do not include $f(R)$-specific tests because they do not carry over to more general models.

\begin{table}
\centering
\begin{tabular}{c | c| c }
Test & Chameleons & Symmetrons \\\hline
E\"{o}t-Wash & \cmark & \cmark    \\
Casimir force & \cmark & \xmark   \\
Microspheres & \cmark & \xmark    \\
Precision atomic tests & \cmark & \xmark   \\
Atom interferometry & \cmark & \cmark  \\
Cold neutrons & \cmark & \xmark   \\
Neutron interferometry & \cmark & \xmark \\
Distance indicators & \cmark & \cmark  \\
Rotation curves & \cmark & \cmark  \\
Cluster Lensing & \cmark & \xmark  \\
\end{tabular}
\caption{Summary of present tests of chameleon and symmetron theories.}\label{tab:experiment_summary}
\end{table}

% Compendium of constraints

\section{Constraints}\label{sec:constraints}

In this section we convert the constraints discussed in the previous section into a single and familiar parametrization and combine them to show the presently allowed parameter ranges. 

\subsection{Chameleon Constraints}

\begin{figure}
\includegraphics[width=0.85\textwidth]{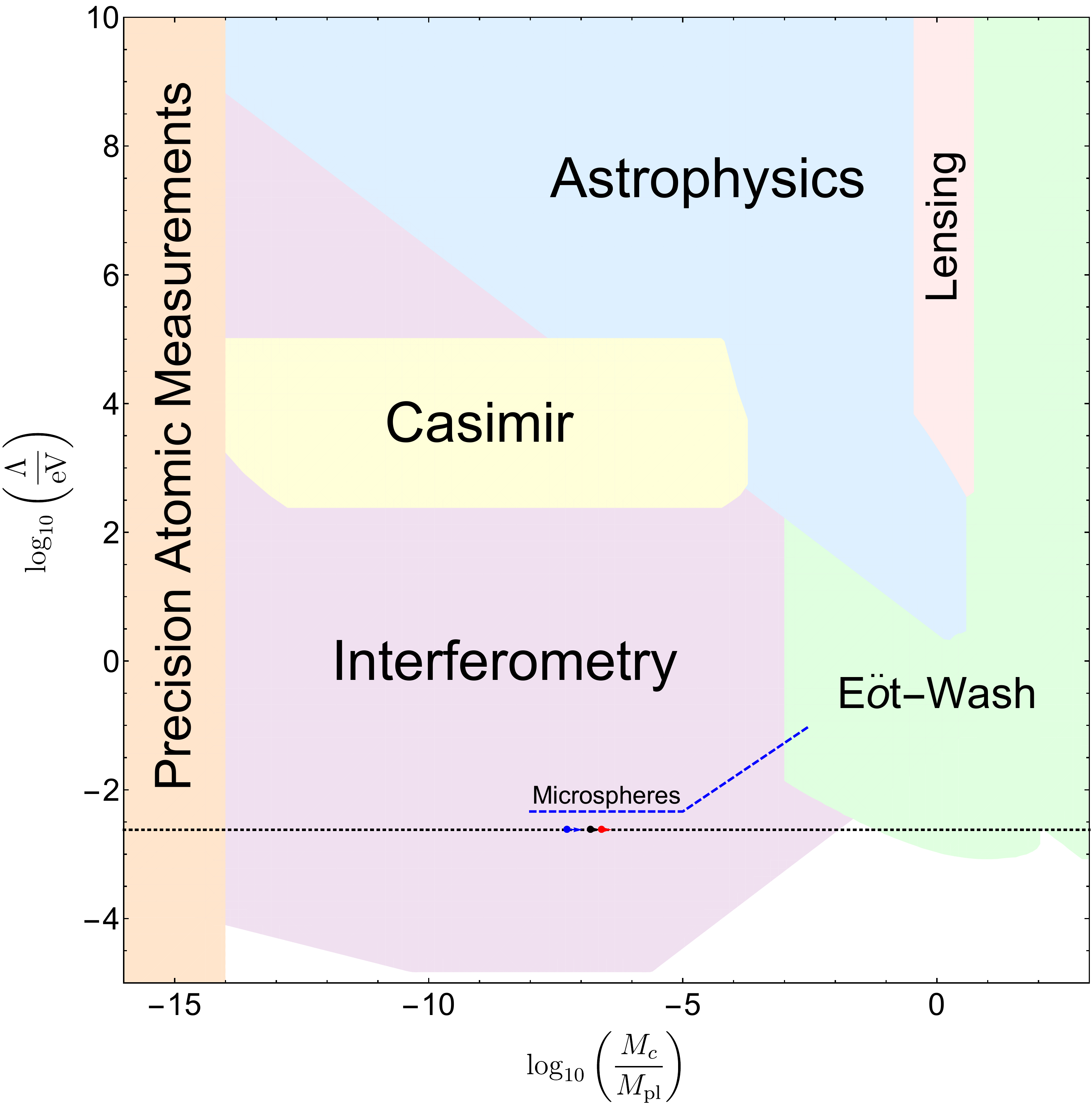}
\caption{Current bounds on the parameters $\mc$ and $\Lambda$ for $n=1$ chameleon models. The regions excluded by each specific test are indicated in the figure; the region labelled astrophysics contains the bounds from both Cepheid and rotation curve tests. The dashed line indicates the dark energy scale $\Lambda=2.4$ meV. The black, red, and blue arrows show the lower bound on $\mc$ coming from neutron bouncing and interferometry. The blue corresponds to the bounds of \cite{Lemmel:2015kwa} and the red to the bounds of \cite{Li:2016tux}.}
\label{fig:chameleon_n1}
\end{figure}

\begin{figure}
\includegraphics[width=0.85\textwidth]{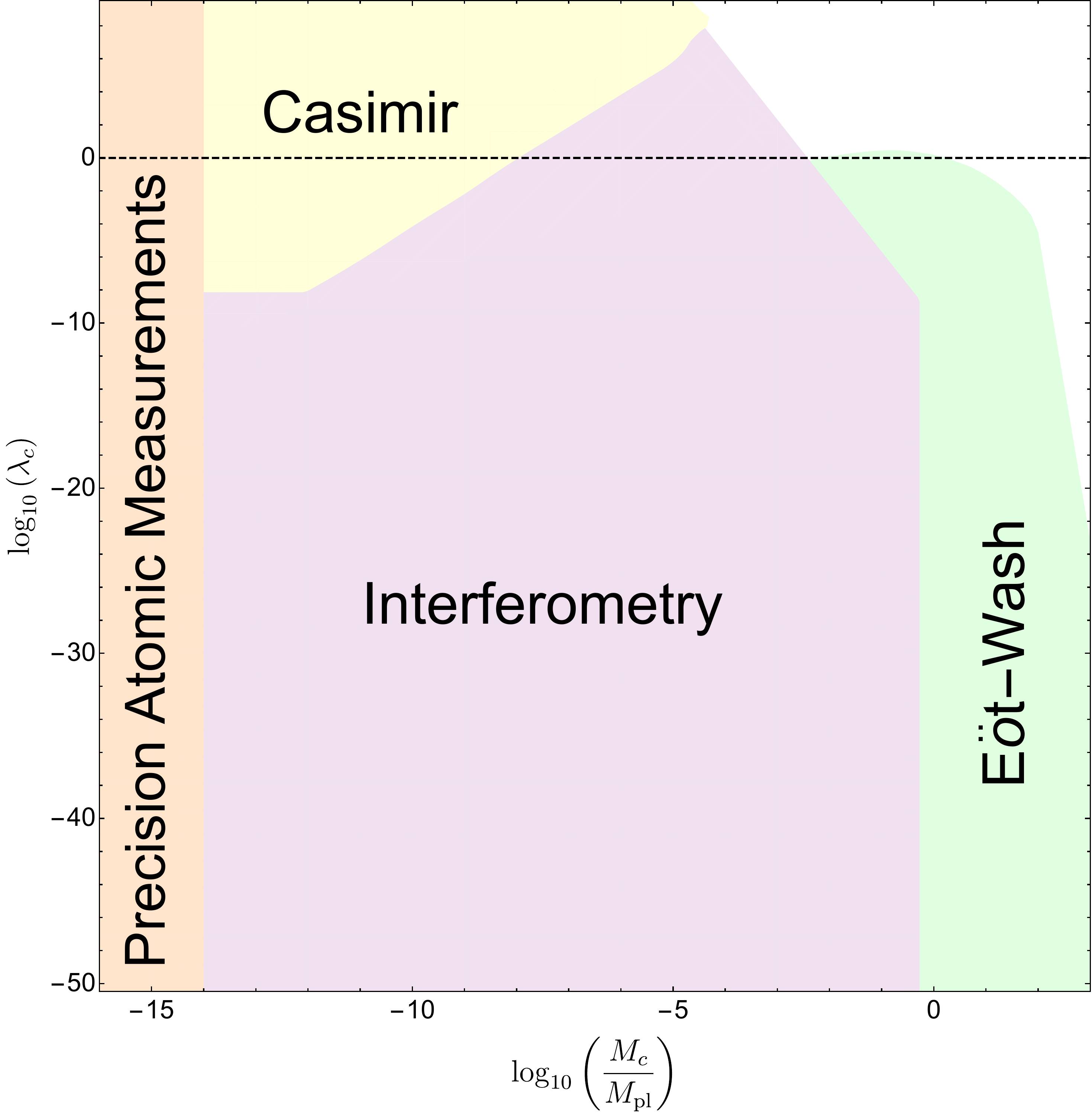}
\caption{Current bounds on the parameters $\mc$ and $\Lambda$ for $n=-4$ chameleon models. The regions excluded by each specific test are indicated in the figure. Comparing equation \ref{eq:chamVgen} with equation \ref{eq:Vchamphi4} reveals that $\lambda_{\rm c} = (\Lambda/\Lambda_{\rm DE})^4$ and so the values of $\lambda_{\rm c}$ plotted here cover the same range of $\Lambda$ as figure \ref{fig:chameleon_n1}. The black dashed line at $\lambda_{\rm c}=1$ therefore corresponds to the dark energy scale $\Lambda=\Lambda_{\rm DE}$.}
\label{fig:chameleon_nm4}
\end{figure}

\begin{figure}
\includegraphics[width=0.85\textwidth]{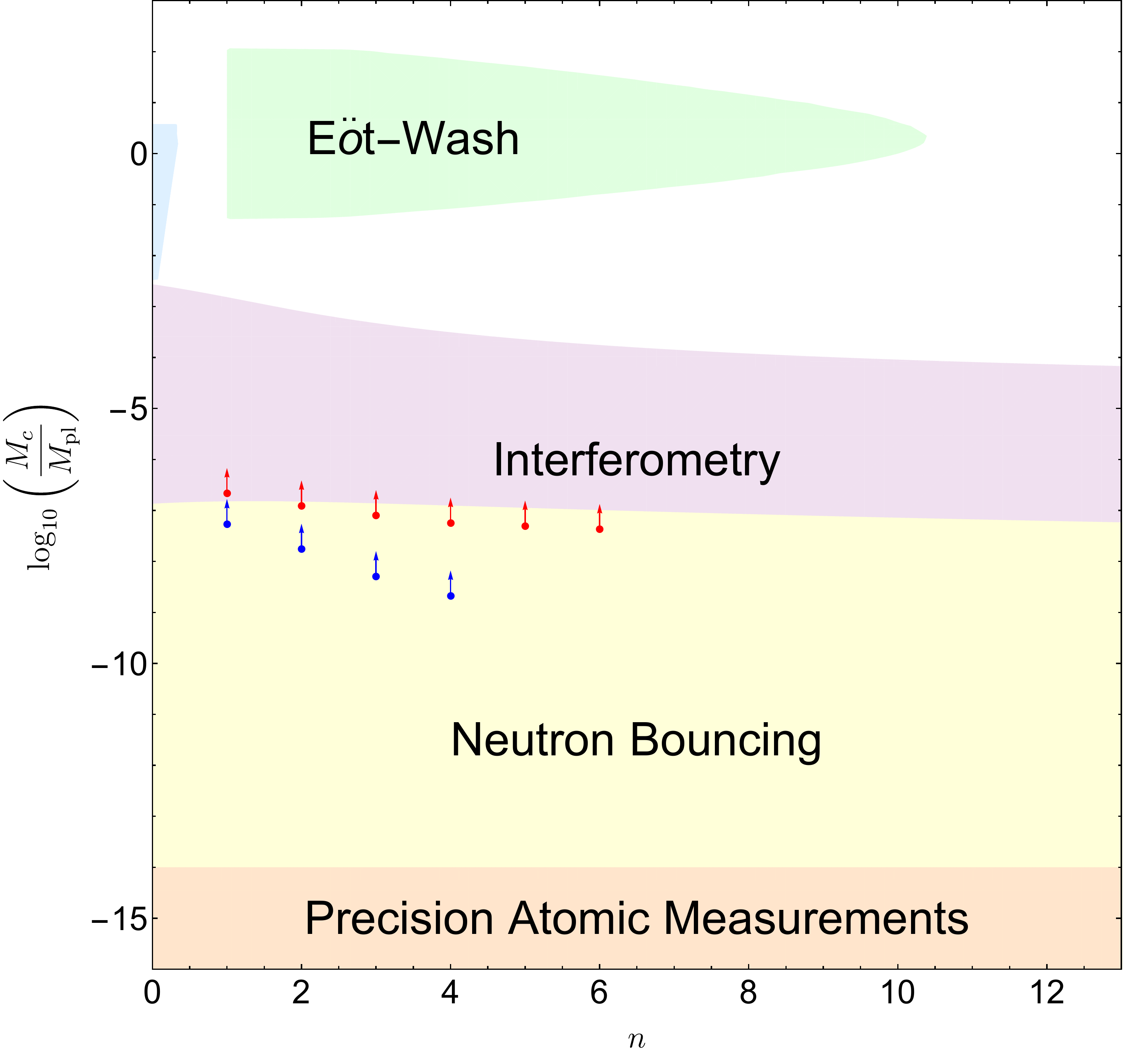}
\caption{Current bounds on the parameters $n$ and $\mc$ when $\Lambda$ is fixed to the dark energy scale $\Lambda_{\rm DE}$ and $n>0$. The regions excluded by each specific test are indicated in the figure. The blue region corresponds to astrophysical tests, which includes both Cepheid and rotation curve tests. The blue and red arrows indicate the lower bounds coming from the neutron interferometry experiments of \cite{Lemmel:2015kwa} and \cite{Li:2016tux} respectively.}
\label{fig:chameleon_ng0}
\end{figure}

\begin{figure}
\includegraphics[width=0.85\textwidth]{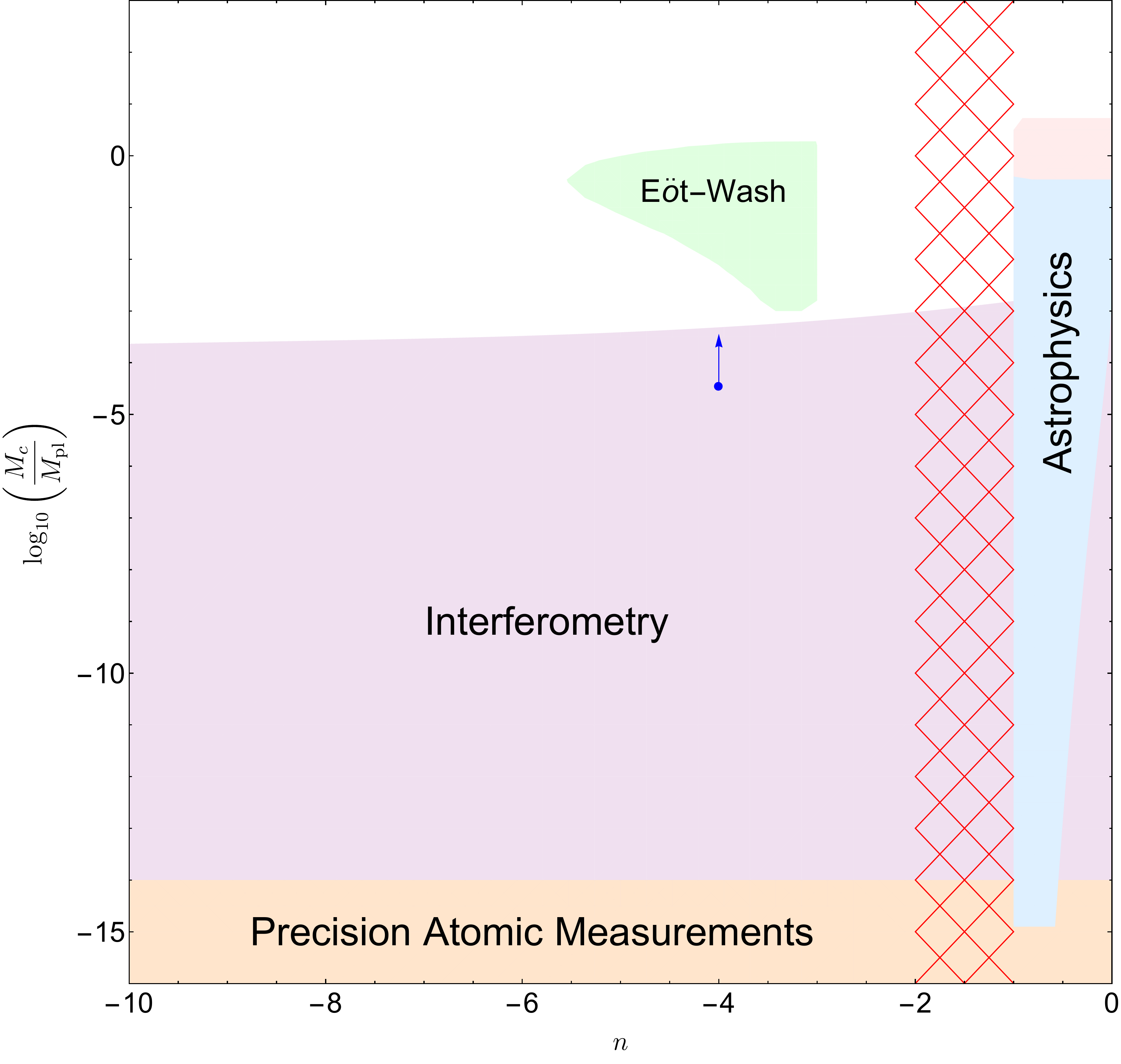}
\caption{Current bounds on the parameters $n$ and $\mc$ when $\Lambda$ is fixed to the dark energy scale $\Lambda_{\rm DE}$ and $n<0$. The red hashed region indicates values of $n$ where the model is not a chameleon, and the reader is reminded that only negative even integers are chameleons. The regions excluded by each specific test are indicated in the figure; the region labelled astrophysics contains the bounds from both Cepheid and rotation curve tests. The blue arrow indicates the lower bound coming from the neutron interferometry experiment of \cite{Lemmel:2015kwa}.  }
\label{fig:chameleon_nl0}
\end{figure}

The current bounds on chameleon models are shown below. We cover the two most commonly studied models $n=1$ (Figure \ref{fig:chameleon_n1}) and $n=-4$ (Figure \ref{fig:chameleon_nm4}). In these cases we plot $\Lambda$ vs. $\mc$. Furthermore, many experiments focus on the case $\Lambda=\Lambda_{\rm DE}=2.4$ meV (the dark energy scale) and so for this choice we plot $\mc$ vs. $n$ for both positive (Figure \ref{fig:chameleon_ng0}) and negative $n$ (Figure \ref{fig:chameleon_nl0}).


\subsubsection{\texorpdfstring{$f(R)$}{TEXT} Constraints}

\begin{figure}[ht]
\centering
\includegraphics[width=0.85\textwidth]{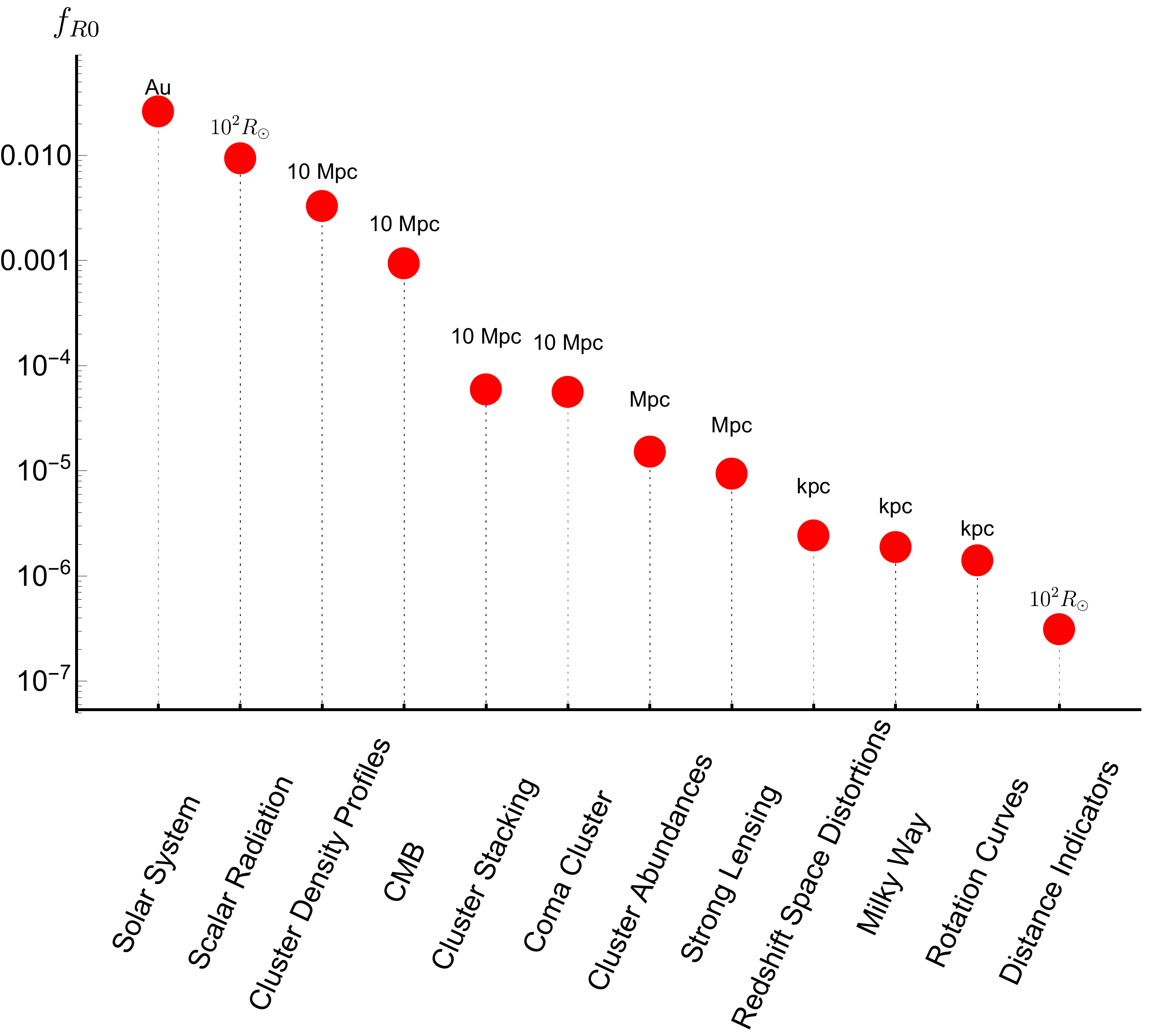}
\caption{Constraints on $f_{R0}$ for $b=1$ Hu \& Sawicki $f(R)$ models (see equation \eqref{eqSM:f(R)}). The red dots indicate the upper limit for the specific test given on the $x$-axis and the points are labelled by the typical distance scale associated with the relevant test.}\label{fig:frconstraints}
\end{figure}

We show the current constraints on the Hu \& Sawicki $b=1$ $f(R)$ model \eqref{eqSM:f(R)} in Figure \ref{fig:frconstraints}. The $x$-axis labels each specific test and the $y$-axis shows the resultant upper limit on $f_{R0}$. It is common to express constraints on $f_{R0}$ showing the length scale on which they were obtained (e.g. \cite{Lombriser:2014dua}). Whilst complementary tests on all scales are crucial consistency checks of the theory, it is important to note that this length is not a new parameter appearing in the theory, and that it is the same parameter $f_{R0}$ being constrained no matter the test or the length scale that it probes. For this reason, we have included the typical length scale for each test in the figure. 

The point labelled ``Milky Way" is not derived from any specific test and is simply the statement that the $f_{R0}$ should be smaller than the Newtonian potential of the Milky Way. One does not need to impose this \emph{a priori} because it is not clear whether or not the Milky Way is screened by the local group; we include it here for completeness, and to make contact with those parts of the literature that take this constraint as given.

\subsubsection{Constraints on the Coupling to Photons}

\begin{figure}
\includegraphics[width=0.85\textwidth]{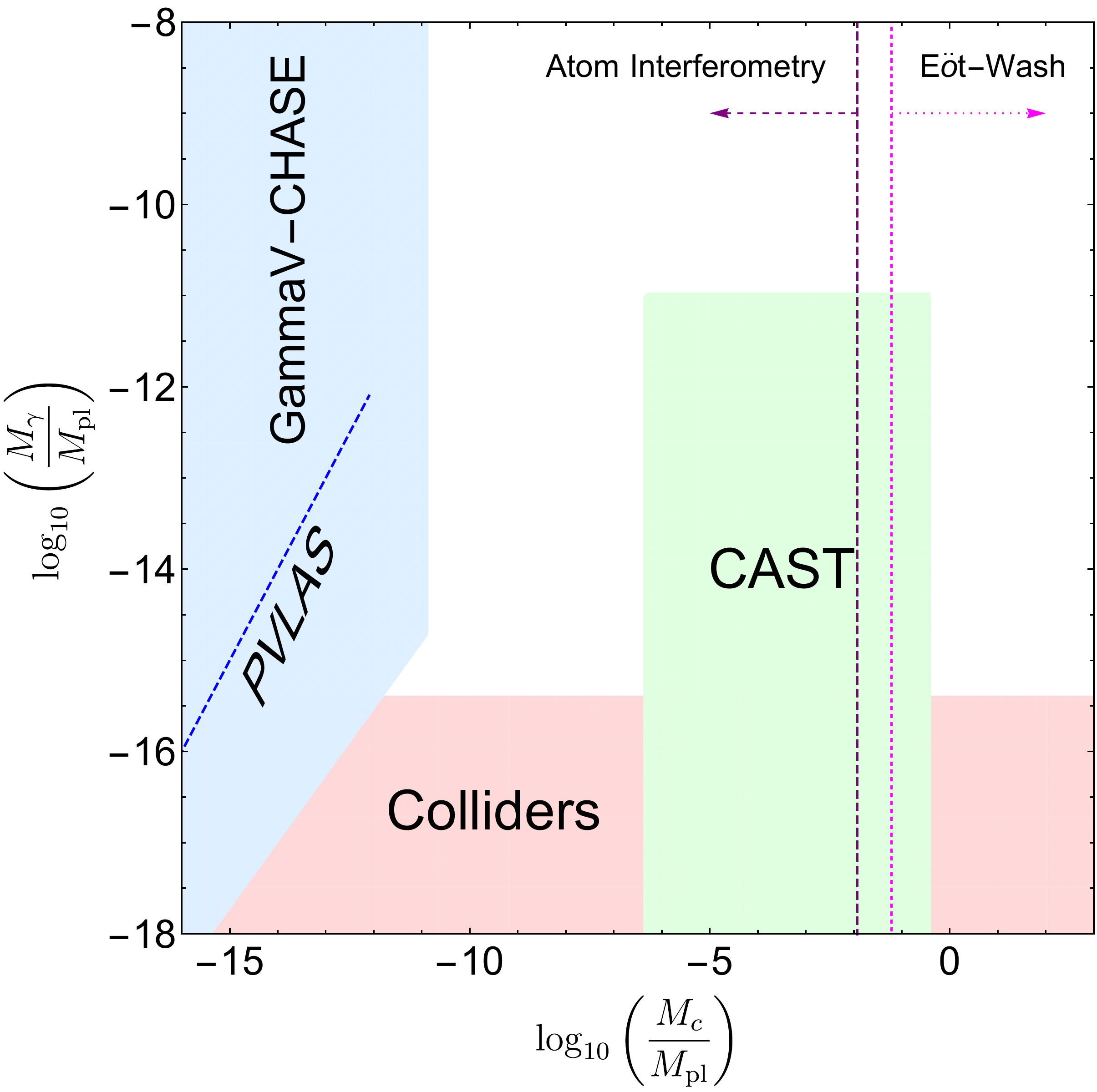}
\caption{Current constraints on the chameleon coupling to photons, $M_\gamma$, for $n=1$ models with $\Lambda$ set to the dark energy scale. The bounds coming from each specific test are indicated in the figure. }
\label{fig:photon_coupling}
\end{figure}

The constraints on the coupling to photons are shown  in Figure \ref{fig:photon_coupling}. We only show constraints for $n=1$ models since many experiments only report bounds for these models at the present time. Furthermore, many of the experiments restrict to the case $\Lambda=\Lambda_{\rm DE}=2.4$ meV and so we do the same here. The results from ADMX are not included since they are presented in terms of $m_{\rm eff}$ rather than the fundamental parameters. One could convert the constraints into the $\mc$--$M_\gamma$ plane, but this depends on the geometry and densities of the experimental apparatus, which are not sufficiently well known. Similarly, we do not include astrophysical bounds due to the need to make assumptions about the strength of magnetic fields and the value of the ambient density.

\subsection{Symmetron Constraints}

\begin{figure}
\includegraphics[width=0.85\textwidth]{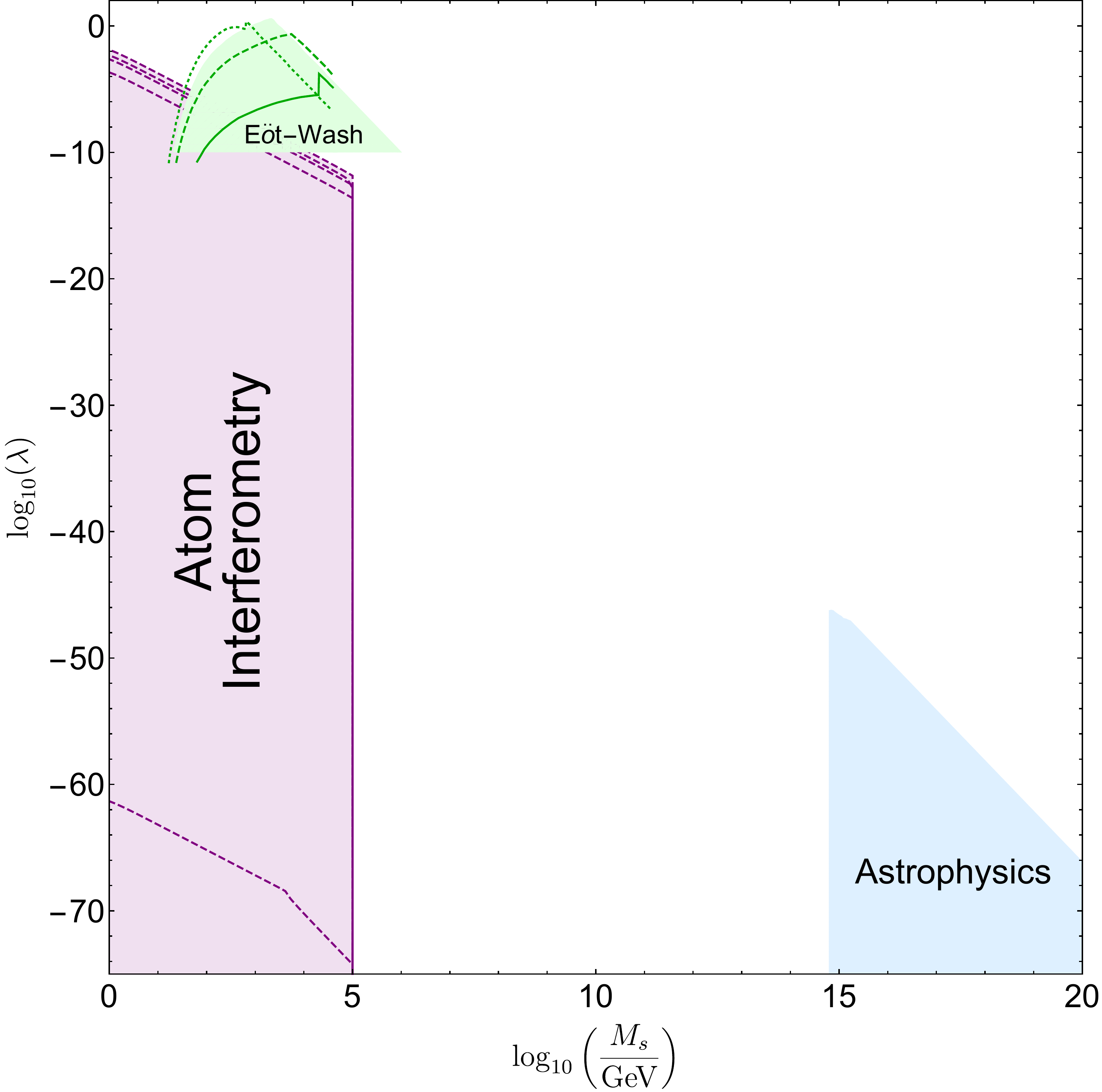}
\caption{The current bounds on the symmetron parameters $M_{\rm s}$ and $\lambda$. The region of parameter space excluded by each specific test is indicated in the figure. The E\"{o}t-Wash region corresponds to $\mu=2.4$ meV; the outlines for values $\mu=\{10^{-4},\,10^{-3},\,10^{-2}\}$ eV are shown by the solid, dashed, and dotted green lines respectively. The atom interferometry lines correspond to the regions excluded for $\mu=\{10^{-4},\,10^{-4.5},\,10^{-5},\,10^{-5},\, 2.4\times10^{-3}\}$ eV from top to bottom respectively, the latter value corresponding to the dark energy scale. The astrophysical bounds are insensitive to the value of $\mu$ for the values considered here.  }
\label{fig:symmetron_constraints}
\end{figure}

The current bounds on the symmetron parameters $M_{\rm s}$ and $\lambda$ are shown in Figure \ref{fig:symmetron_constraints} for some commonly studied values of $\mu$ indicated in the caption.

%Conclusions
\section{Conclusions and Outlook}
\label{sec:conclusions}

Chameleon and symmetron models have been a paragon for viable, interesting, and relevant infra-red modifications of general relativity for over a decade. The screening mechanism has resulted in theories of gravity that are perfectly consistent with general relativity's predictions in the solar system but are yet falsifiable using novel approaches such as astrophysical phenomena in distant galaxies, as well as specifically targeted laboratory searches. In many cases, these models may be relevant on linear (and non-linear) cosmological scales.

In this review, we have surveyed the omnibus of literature providing constraints and have translated them into a single parametrization in order to assess the current viability of the models. The main results are presented in Figures \ref{fig:chameleon_n1}--\ref{fig:symmetron_constraints}, which can be summarized as follows:
\begin{itemize}
\item $n=1$ and $n=-4$ chameleon models (two of the most commonly studied) are tightly constrained but there is a large parameter space remaining for $n>1$ and $n<-4$ when $\Lambda$ is fixed to the dark energy scale. Away from this, the constraints are not as strong. In many cases, this is because bounds on other models are not reported.
\item Symmetron models are well-constrained by astrophysical probes and atom interferometry but there is a lack of theoretical work translating the bounds from existing experimental results into symmetron constraints. This has resulted in a desert separating astrophysical and laboratory tests (this could be filled in partially by constraints from future space-based tests of relativistic gravitation \cite{Sakstein:2017pqi}).
\item The coupling of chameleons to photons for $n=1$ models is tightly constrained and there is only a narrow window remaining. The coupling of symmetrons to photons and chameleon models with $n\ne1$ has yet to be explored.
\item Hu \& Sawicki $f(R)$ models \cite{Hu:2007nk} are well-constrained for $b=1$ but, presently, there are not enough reported bounds on larger values to make a meaningful comparison. For $b=1$ the bounds on $f_{R0}$ are at the $10^{-7}$ level. In theory, $10^{-8}$ would be achievable with better statistics; below this, dwarf galaxies begin to become screened and higher-precision tests are necessary.
\item At the present time, the environment-dependent dilaton, which screens in a distinct manner from chameleon and symmetron models, has not been studied sufficiently in the context of laboratory and astrophysical tests to produce any meaningful constraints.

\subsection{Prospects for Future Bounds}

We end by discussing the prospects for future tests of screened modified gravity.

\subsubsection{Laboratory Tests}
As new experimental techniques are been developed, and existing ones are improved we can expect bounds on chameleon and symmetron models of screening to continue to improve.  It is to be expected that this will be a combination of the reinterpretation of experimental results obtained when searching for other types of new physics, and a smaller number of experiments dedicated to directly searching for screening.  

It is difficult to imagine that a single experiment could cover all of the remaining chameleon and symmetron parameter space, and so ideally a combination of techniques and searches are needed in order to fully rule out the possibility that screened scalars exist in our universe.  

\subsubsection{Astrophysical Tests}

Astrophysical objects show strong deviations from GR when the Newtonian potential $\pn<\chi_0$ ($\sim f_{R0}$ for $f(R)$ theories). Given that current bounds place $\chi_0\lsim\oo(10^{-7})$, the only objects in the Universe with a low enough Newtonian potential to exhibit novel effects are dwarf galaxies located in voids, and several tests using such galaxies have been proposed.

The rotation curve test described in Section \ref{sec:rot_tests} suffers from a lack of unscreened galaxies, and a larger sample would improve the constraints. Future and upcoming data releases, in particular SDSS-MaNGA, can provide a larger sample size that would significantly improve the bounds. Additional tests, such as the warping of galactic disks due to equivalence principle violations have been proposed \cite{Jain:2011ji}, although a test using SDSS optical and ALFALFA radio observations did not yield any bounds on the model parameters \cite{Vikram:2013uba}. Future radio surveys such as VLT may be more fruitful. 

{Finally, N-body simulations are uncovering a variety of novel phenomena exhibited by chameleons on non-linear cosmological scales \cite{Jain:2013wgs}. Many of these are clear smoking-gun signals that could be measured with upcoming peculiar velocity and galaxy redshift surveys \cite{Hellwing:2014nma}.

}

\subsubsection{Tests of the Coupling to Photons}

The increase in interest in axions and axion-like particles as dark matter candidates has lead to a series of proposals and experiments aimed at further constraining these particles which, in many cases, focus on their interactions with photons. These experiments present an exciting opportunity for new constraints on theories with screening, but the details of how powerful these constraints can be remain to be worked out.

\end{itemize}

\bibliography{ref}

\providecommand{\href}[2]{#2}\begingroup\raggedright\begin{thebibliography}{100}

\bibitem{Abbott:2016blz}
{\bfseries Virgo, LIGO Scientific} Collaboration, B.~P. Abbott {\em et~al.},
  ``{Observation of Gravitational Waves from a Binary Black Hole Merger},''
  \href{http://dx.doi.org/10.1103/PhysRevLett.116.061102}{{\em Phys. Rev.
  Lett.} {\bfseries 116} no.~6, (2016) 061102},
\href{http://arxiv.org/abs/1602.03837}{{\ttfamily arXiv:1602.03837 [gr-qc]}}.
%%CITATION = ARXIV:1602.03837;%%.

\bibitem{TheLIGOScientific:2016src}
{\bfseries Virgo, LIGO Scientific} Collaboration, B.~P. Abbott {\em et~al.},
  ``{Tests of general relativity with GW150914},''
  \href{http://dx.doi.org/10.1103/PhysRevLett.116.221101}{{\em Phys. Rev.
  Lett.} {\bfseries 116} no.~22, (2016) 221101},
\href{http://arxiv.org/abs/1602.03841}{{\ttfamily arXiv:1602.03841 [gr-qc]}}.
%%CITATION = ARXIV:1602.03841;%%.

\bibitem{Clifton:2011jh}
T.~Clifton, P.~G. Ferreira, A.~Padilla, and C.~Skordis, ``{Modified Gravity and
  Cosmology},'' \href{http://dx.doi.org/10.1016/j.physrep.2012.01.001}{{\em
  Phys. Rept.} {\bfseries 513} (2012) 1--189},
\href{http://arxiv.org/abs/1106.2476}{{\ttfamily arXiv:1106.2476
  [astro-ph.CO]}}.
%%CITATION = ARXIV:1106.2476;%%.

\bibitem{Joyce:2014kja}
A.~Joyce, B.~Jain, J.~Khoury, and M.~Trodden, ``{Beyond the Cosmological
  Standard Model},''
  \href{http://dx.doi.org/10.1016/j.physrep.2014.12.002}{{\em Phys. Rept.}
  {\bfseries 568} (2015) 1--98},
\href{http://arxiv.org/abs/1407.0059}{{\ttfamily arXiv:1407.0059
  [astro-ph.CO]}}.
%%CITATION = ARXIV:1407.0059;%%.

\bibitem{Koyama:2015vza}
K.~Koyama, ``{Cosmological Tests of Modified Gravity},''
  \href{http://dx.doi.org/10.1088/0034-4885/79/4/046902}{{\em Rept. Prog.
  Phys.} {\bfseries 79} no.~4, (2016) 046902},
\href{http://arxiv.org/abs/1504.04623}{{\ttfamily arXiv:1504.04623
  [astro-ph.CO]}}.
%%CITATION = ARXIV:1504.04623;%%.

\bibitem{Bull:2015stt}
P.~Bull {\em et~al.}, ``{Beyond $\Lambda$CDM: Problems, solutions, and the road
  ahead},'' \href{http://dx.doi.org/10.1016/j.dark.2016.02.001}{{\em Phys. Dark
  Univ.} {\bfseries 12} (2016) 56--99},
\href{http://arxiv.org/abs/1512.05356}{{\ttfamily arXiv:1512.05356
  [astro-ph.CO]}}.
%%CITATION = ARXIV:1512.05356;%%.

\bibitem{Weinberg:1965rz}
S.~Weinberg, ``{Photons and gravitons in perturbation theory: Derivation of
  Maxwell's and Einstein's equations},''
\href{http://dx.doi.org/10.1103/PhysRev.138.B988}{{\em Phys. Rev.} {\bfseries
  138} (1965) B988--B1002}.
%%CITATION = PHRVA,138,B988;%%.

\bibitem{Blas:2014aca}
D.~Blas and E.~Lim, ``{Phenomenology of theories of gravity without Lorentz
  invariance: the preferred frame case},''
  \href{http://dx.doi.org/10.1142/S0218271814430093}{{\em Int. J. Mod. Phys.}
  {\bfseries D23} (2015) 1443009},
\href{http://arxiv.org/abs/1412.4828}{{\ttfamily arXiv:1412.4828 [gr-qc]}}.
%%CITATION = ARXIV:1412.4828;%%.

\bibitem{deRham:2014zqa}
C.~de~Rham, ``{Massive Gravity},''
  \href{http://dx.doi.org/10.12942/lrr-2014-7}{{\em Living Rev. Rel.}
  {\bfseries 17} (2014) 7},
\href{http://arxiv.org/abs/1401.4173}{{\ttfamily arXiv:1401.4173 [hep-th]}}.
%%CITATION = ARXIV:1401.4173;%%.

\bibitem{Will:2004nx}
C.~M. Will, ``{The confrontation between general relativity and experiment},''
\href{http://dx.doi.org/10.1007/BF02705195}{{\em Pramana} {\bfseries 63} (2004)
  731--740}.
%%CITATION = PRAMC,63,731;%%.

\bibitem{Williams:2004qba}
J.~G. Williams, S.~G. Turyshev, and D.~H. Boggs, ``{Progress in lunar laser
  ranging tests of relativistic gravity},''
  \href{http://dx.doi.org/10.1103/PhysRevLett.93.261101}{{\em Phys. Rev. Lett.}
  {\bfseries 93} (2004) 261101},
\href{http://arxiv.org/abs/gr-qc/0411113}{{\ttfamily arXiv:gr-qc/0411113
  [gr-qc]}}.
%%CITATION = GR-QC/0411113;%%.

\bibitem{Adelberger:2003zx}
E.~G. Adelberger, B.~R. Heckel, and A.~E. Nelson, ``{Tests of the gravitational
  inverse square law},''
  \href{http://dx.doi.org/10.1146/annurev.nucl.53.041002.110503}{{\em Ann. Rev.
  Nucl. Part. Sci.} {\bfseries 53} (2003) 77--121},
\href{http://arxiv.org/abs/hep-ph/0307284}{{\ttfamily arXiv:hep-ph/0307284
  [hep-ph]}}.
%%CITATION = HEP-PH/0307284;%%.

\bibitem{GBM:2017lvd}
{\bfseries GROND, SALT Group, OzGrav, DFN, INTEGRAL, Virgo, Insight-Hxmt, MAXI
  Team, Fermi-LAT, J-GEM, RATIR, IceCube, CAASTRO, LWA, ePESSTO, GRAWITA,
  RIMAS, SKA South Africa/MeerKAT, H.E.S.S., 1M2H Team, IKI-GW Follow-up, Fermi
  GBM, Pi of Sky, DWF (Deeper Wider Faster Program), Dark Energy Survey,
  MASTER, AstroSat Cadmium Zinc Telluride Imager Team, Swift, Pierre Auger,
  ASKAP, VINROUGE, JAGWAR, Chandra Team at McGill University, TTU-NRAO, GROWTH,
  AGILE Team, MWA, ATCA, AST3, TOROS, Pan-STARRS, NuSTAR, ATLAS Telescopes,
  BOOTES, CaltechNRAO, LIGO Scientific, High Time Resolution Universe Survey,
  Nordic Optical Telescope, Las Cumbres Observatory Group, TZAC Consortium,
  LOFAR, IPN, DLT40, Texas Tech University, HAWC, ANTARES, KU, Dark Energy
  Camera GW-EM, CALET, Euro VLBI Team, ALMA} Collaboration, B.~P. Abbott {\em
  et~al.}, ``{Multi-messenger Observations of a Binary Neutron Star Merger},''
  \href{http://dx.doi.org/10.3847/2041-8213/aa91c9}{{\em Astrophys. J.}
  {\bfseries 848} no.~2, (2017) L12},
\href{http://arxiv.org/abs/1710.05833}{{\ttfamily arXiv:1710.05833
  [astro-ph.HE]}}.
%%CITATION = ARXIV:1710.05833;%%.

\bibitem{TheLIGOScientific:2017qsa}
{\bfseries Virgo, LIGO Scientific} Collaboration, B.~Abbott {\em et~al.},
  ``{GW170817: Observation of Gravitational Waves from a Binary Neutron Star
  Inspiral},'' \href{http://dx.doi.org/10.1103/PhysRevLett.119.161101}{{\em
  Phys. Rev. Lett.} {\bfseries 119} no.~16, (2017) 161101},
\href{http://arxiv.org/abs/1710.05832}{{\ttfamily arXiv:1710.05832 [gr-qc]}}.
%%CITATION = ARXIV:1710.05832;%%.

\bibitem{Sakstein:2017xjx}
J.~Sakstein and B.~Jain, ``{Implications of the Neutron Star Merger GW170817
  for Cosmological Scalar-Tensor Theories},''
\href{http://arxiv.org/abs/1710.05893}{{\ttfamily arXiv:1710.05893
  [astro-ph.CO]}}.
%%CITATION = ARXIV:1710.05893;%%.

\bibitem{Ezquiaga:2017ekz}
J.~M. Ezquiaga and M.~Zumalacárregui, ``{Dark Energy after GW170817: dead ends
  and the road ahead},''
\href{http://arxiv.org/abs/1710.05901}{{\ttfamily arXiv:1710.05901
  [astro-ph.CO]}}.
%%CITATION = ARXIV:1710.05901;%%.

\bibitem{Creminelli:2017sry}
P.~Creminelli and F.~Vernizzi, ``{Dark Energy after GW170817},''
\href{http://arxiv.org/abs/1710.05877}{{\ttfamily arXiv:1710.05877
  [astro-ph.CO]}}.
%%CITATION = ARXIV:1710.05877;%%.

\bibitem{Baker:2017hug}
T.~Baker, E.~Bellini, P.~G. Ferreira, M.~Lagos, J.~Noller, and I.~Sawicki,
  ``{Strong constraints on cosmological gravity from GW170817 and GRB
  170817A},''
\href{http://arxiv.org/abs/1710.06394}{{\ttfamily arXiv:1710.06394
  [astro-ph.CO]}}.
%%CITATION = ARXIV:1710.06394;%%.

\bibitem{Crisostomi:2017lbg}
M.~Crisostomi and K.~Koyama, ``{Vainshtein mechanism after GW170817},''
\href{http://arxiv.org/abs/1711.06661}{{\ttfamily arXiv:1711.06661
  [astro-ph.CO]}}.
%%CITATION = ARXIV:1711.06661;%%.

\bibitem{Langlois:2017dyl}
D.~Langlois, R.~Saito, D.~Yamauchi, and K.~Noui, ``{Scalar-tensor theories and
  modified gravity in the wake of GW170817},''
\href{http://arxiv.org/abs/1711.07403}{{\ttfamily arXiv:1711.07403 [gr-qc]}}.
%%CITATION = ARXIV:1711.07403;%%.

\bibitem{Dima:2017pwp}
A.~Dima and F.~Vernizzi, ``{Vainshtein Screening in Scalar-Tensor Theories
  before and after GW170817: Constraints on Theories beyond Horndeski},''
\href{http://arxiv.org/abs/1712.04731}{{\ttfamily arXiv:1712.04731 [gr-qc]}}.
%%CITATION = ARXIV:1712.04731;%%.

\bibitem{Bartolo:2017ibw}
N.~Bartolo, P.~Karmakar, S.~Matarrese, and M.~Scomparin, ``{Cosmic structures
  and gravitational waves in ghost-free scalar-tensor theories of gravity},''
\href{http://arxiv.org/abs/1712.04002}{{\ttfamily arXiv:1712.04002 [gr-qc]}}.
%%CITATION = ARXIV:1712.04002;%%.

\bibitem{Bellini:2014fua}
E.~Bellini and I.~Sawicki, ``{Maximal freedom at minimum cost: linear
  large-scale structure in general modifications of gravity},''
  \href{http://dx.doi.org/10.1088/1475-7516/2014/07/050}{{\em JCAP} {\bfseries
  1407} (2014) 050},
\href{http://arxiv.org/abs/1404.3713}{{\ttfamily arXiv:1404.3713
  [astro-ph.CO]}}.
%%CITATION = ARXIV:1404.3713;%%.

\bibitem{Brax:2015dma}
P.~Brax, C.~Burrage, and A.-C. Davis, ``{The Speed of Galileon Gravity},''
  \href{http://dx.doi.org/10.1088/1475-7516/2016/03/004}{{\em JCAP} {\bfseries
  1603} no.~03, (2016) 004},
\href{http://arxiv.org/abs/1510.03701}{{\ttfamily arXiv:1510.03701 [gr-qc]}}.
%%CITATION = ARXIV:1510.03701;%%.

\bibitem{Khoury:2003rn}
J.~Khoury and A.~Weltman, ``{Chameleon cosmology},''
  \href{http://dx.doi.org/10.1103/PhysRevD.69.044026}{{\em Phys. Rev.}
  {\bfseries D69} (2004) 044026},
\href{http://arxiv.org/abs/astro-ph/0309411}{{\ttfamily arXiv:astro-ph/0309411
  [astro-ph]}}.
%%CITATION = ASTRO-PH/0309411;%%.

\bibitem{Khoury:2003aq}
J.~Khoury and A.~Weltman, ``{Chameleon fields: Awaiting surprises for tests of
  gravity in space},''
  \href{http://dx.doi.org/10.1103/PhysRevLett.93.171104}{{\em Phys. Rev. Lett.}
  {\bfseries 93} (2004) 171104},
\href{http://arxiv.org/abs/astro-ph/0309300}{{\ttfamily arXiv:astro-ph/0309300
  [astro-ph]}}.
%%CITATION = ASTRO-PH/0309300;%%.

\bibitem{Gessner:1992flm}
E.~Gessner, ``{A new scalar tensor theory for gravity and the flat rotation
  curves of spiral galaxies},''
\href{http://dx.doi.org/10.1007/BF00645239}{{\em Astrophys. Space Sci.}
  {\bfseries 196} no.~1, (1992) 29--43}.
%%CITATION = APSSB,196,29;%%.

\bibitem{Pietroni:2005pv}
M.~Pietroni, ``{Dark energy condensation},''
  \href{http://dx.doi.org/10.1103/PhysRevD.72.043535}{{\em Phys. Rev.}
  {\bfseries D72} (2005) 043535},
\href{http://arxiv.org/abs/astro-ph/0505615}{{\ttfamily arXiv:astro-ph/0505615
  [astro-ph]}}.
%%CITATION = ASTRO-PH/0505615;%%.

\bibitem{Olive:2007aj}
K.~A. Olive and M.~Pospelov, ``{Environmental dependence of masses and coupling
  constants},'' \href{http://dx.doi.org/10.1103/PhysRevD.77.043524}{{\em Phys.
  Rev.} {\bfseries D77} (2008) 043524},
\href{http://arxiv.org/abs/0709.3825}{{\ttfamily arXiv:0709.3825 [hep-ph]}}.
%%CITATION = ARXIV:0709.3825;%%.

\bibitem{Hinterbichler:2010es}
K.~Hinterbichler and J.~Khoury, ``{Symmetron Fields: Screening Long-Range
  Forces Through Local Symmetry Restoration},''
  \href{http://dx.doi.org/10.1103/PhysRevLett.104.231301}{{\em Phys. Rev.
  Lett.} {\bfseries 104} (2010) 231301},
\href{http://arxiv.org/abs/1001.4525}{{\ttfamily arXiv:1001.4525 [hep-th]}}.
%%CITATION = ARXIV:1001.4525;%%.

\bibitem{Hinterbichler:2011ca}
K.~Hinterbichler, J.~Khoury, A.~Levy, and A.~Matas, ``{Symmetron Cosmology},''
  \href{http://dx.doi.org/10.1103/PhysRevD.84.103521}{{\em Phys. Rev.}
  {\bfseries D84} (2011) 103521},
\href{http://arxiv.org/abs/1107.2112}{{\ttfamily arXiv:1107.2112
  [astro-ph.CO]}}.
%%CITATION = ARXIV:1107.2112;%%.

\bibitem{Damour:1994zq}
T.~Damour and A.~M. Polyakov, ``{The String dilaton and a least coupling
  principle},'' \href{http://dx.doi.org/10.1016/0550-3213(94)90143-0}{{\em
  Nucl. Phys.} {\bfseries B423} (1994) 532--558},
\href{http://arxiv.org/abs/hep-th/9401069}{{\ttfamily arXiv:hep-th/9401069
  [hep-th]}}.
%%CITATION = HEP-TH/9401069;%%.

\bibitem{Brax:2011ja}
P.~Brax, C.~van~de Bruck, A.-C. Davis, B.~Li, and D.~J. Shaw, ``{Nonlinear
  Structure Formation with the Environmentally Dependent Dilaton},''
  \href{http://dx.doi.org/10.1103/PhysRevD.83.104026}{{\em Phys. Rev.}
  {\bfseries D83} (2011) 104026},
\href{http://arxiv.org/abs/1102.3692}{{\ttfamily arXiv:1102.3692
  [astro-ph.CO]}}.
%%CITATION = ARXIV:1102.3692;%%.

\bibitem{Brax:2010gi}
P.~Brax, C.~van~de Bruck, A.-C. Davis, and D.~Shaw, ``{The Dilaton and Modified
  Gravity},'' \href{http://dx.doi.org/10.1103/PhysRevD.82.063519}{{\em Phys.
  Rev.} {\bfseries D82} (2010) 063519},
\href{http://arxiv.org/abs/1005.3735}{{\ttfamily arXiv:1005.3735
  [astro-ph.CO]}}.
%%CITATION = ARXIV:1005.3735;%%.

\bibitem{Babichev:2013usa}
E.~Babichev and C.~Deffayet, ``{An introduction to the Vainshtein mechanism},''
  \href{http://dx.doi.org/10.1088/0264-9381/30/18/184001}{{\em Class. Quant.
  Grav.} {\bfseries 30} (2013) 184001},
\href{http://arxiv.org/abs/1304.7240}{{\ttfamily arXiv:1304.7240 [gr-qc]}}.
%%CITATION = ARXIV:1304.7240;%%.

\bibitem{deRham:2010kj}
C.~de~Rham, G.~Gabadadze, and A.~J. Tolley, ``{Resummation of Massive
  Gravity},'' \href{http://dx.doi.org/10.1103/PhysRevLett.106.231101}{{\em
  Phys. Rev. Lett.} {\bfseries 106} (2011) 231101},
\href{http://arxiv.org/abs/1011.1232}{{\ttfamily arXiv:1011.1232 [hep-th]}}.
%%CITATION = ARXIV:1011.1232;%%.

\bibitem{Hinterbichler:2011tt}
K.~Hinterbichler, ``{Theoretical Aspects of Massive Gravity},''
  \href{http://dx.doi.org/10.1103/RevModPhys.84.671}{{\em Rev. Mod. Phys.}
  {\bfseries 84} (2012) 671--710},
\href{http://arxiv.org/abs/1105.3735}{{\ttfamily arXiv:1105.3735 [hep-th]}}.
%%CITATION = ARXIV:1105.3735;%%.

\bibitem{deRham:2016nuf}
C.~de~Rham, J.~T. Deskins, A.~J. Tolley, and S.-Y. Zhou, ``{Graviton Mass
  Bounds},'' \href{http://dx.doi.org/10.1103/RevModPhys.89.025004}{{\em Rev.
  Mod. Phys.} {\bfseries 89} no.~2, (2017) 025004},
\href{http://arxiv.org/abs/1606.08462}{{\ttfamily arXiv:1606.08462
  [astro-ph.CO]}}.
%%CITATION = ARXIV:1606.08462;%%.

\bibitem{Wang:2012kj}
J.~Wang, L.~Hui, and J.~Khoury, ``{No-Go Theorems for Generalized Chameleon
  Field Theories},''
  \href{http://dx.doi.org/10.1103/PhysRevLett.109.241301}{{\em Phys. Rev.
  Lett.} {\bfseries 109} (2012) 241301},
\href{http://arxiv.org/abs/1208.4612}{{\ttfamily arXiv:1208.4612
  [astro-ph.CO]}}.
%%CITATION = ARXIV:1208.4612;%%.

\bibitem{Copeland:2006wr}
E.~J. Copeland, M.~Sami, and S.~Tsujikawa, ``{Dynamics of dark energy},''
  \href{http://dx.doi.org/10.1142/S021827180600942X}{{\em Int. J. Mod. Phys.}
  {\bfseries D15} (2006) 1753--1936},
\href{http://arxiv.org/abs/hep-th/0603057}{{\ttfamily arXiv:hep-th/0603057
  [hep-th]}}.
%%CITATION = HEP-TH/0603057;%%.

\bibitem{Jain:2013wgs}
B.~Jain {\em et~al.}, ``{Novel Probes of Gravity and Dark Energy},''
\href{http://arxiv.org/abs/1309.5389}{{\ttfamily arXiv:1309.5389
  [astro-ph.CO]}}.
%%CITATION = ARXIV:1309.5389;%%.

\bibitem{Sakstein:2015oqa}
J.~Sakstein, ``{Astrophysical Tests of Modified Gravity},''
\href{http://arxiv.org/abs/1502.04503}{{\ttfamily arXiv:1502.04503
  [astro-ph.CO]}}.
%%CITATION = ARXIV:1502.04503;%%.

\bibitem{Bekenstein:1992pj}
J.~D. Bekenstein, ``{The Relation between physical and gravitational
  geometry},'' \href{http://dx.doi.org/10.1103/PhysRevD.48.3641}{{\em Phys.
  Rev.} {\bfseries D48} (1993) 3641--3647},
\href{http://arxiv.org/abs/gr-qc/9211017}{{\ttfamily arXiv:gr-qc/9211017
  [gr-qc]}}.
%%CITATION = GR-QC/9211017;%%.

\bibitem{Brax:2014vva}
P.~Brax and C.~Burrage, ``{Constraining Disformally Coupled Scalar Fields},''
  \href{http://dx.doi.org/10.1103/PhysRevD.90.104009}{{\em Phys. Rev.}
  {\bfseries D90} no.~10, (2014) 104009},
\href{http://arxiv.org/abs/1407.1861}{{\ttfamily arXiv:1407.1861
  [astro-ph.CO]}}.
%%CITATION = ARXIV:1407.1861;%%.

\bibitem{Sakstein:2014isa}
J.~Sakstein, ``{Disformal Theories of Gravity: From the Solar System to
  Cosmology},'' \href{http://dx.doi.org/10.1088/1475-7516/2014/12/012}{{\em
  JCAP} {\bfseries 1412} no.~12, (2014) 012},
\href{http://arxiv.org/abs/1409.1734}{{\ttfamily arXiv:1409.1734
  [astro-ph.CO]}}.
%%CITATION = ARXIV:1409.1734;%%.

\bibitem{Sakstein:2014aca}
J.~Sakstein, ``{Towards Viable Cosmological Models of Disformal Theories of
  Gravity},'' \href{http://dx.doi.org/10.1103/PhysRevD.91.024036}{{\em Phys.
  Rev.} {\bfseries D91} no.~2, (2015) 024036},
\href{http://arxiv.org/abs/1409.7296}{{\ttfamily arXiv:1409.7296
  [astro-ph.CO]}}.
%%CITATION = ARXIV:1409.7296;%%.

\bibitem{Ip:2015qsa}
H.~Y. Ip, J.~Sakstein, and F.~Schmidt, ``{Solar System Constraints on Disformal
  Gravity Theories},''
  \href{http://dx.doi.org/10.1088/1475-7516/2015/10/051}{{\em JCAP} {\bfseries
  1510} (2015) 051},
\href{http://arxiv.org/abs/1507.00568}{{\ttfamily arXiv:1507.00568 [gr-qc]}}.
%%CITATION = ARXIV:1507.00568;%%.

\bibitem{Sakstein:2015jca}
J.~Sakstein and S.~Verner, ``{Disformal Gravity Theories: A Jordan Frame
  Analysis},'' \href{http://dx.doi.org/10.1103/PhysRevD.92.123005}{{\em Phys.
  Rev.} {\bfseries D92} no.~12, (2015) 123005},
\href{http://arxiv.org/abs/1509.05679}{{\ttfamily arXiv:1509.05679 [gr-qc]}}.
%%CITATION = ARXIV:1509.05679;%%.

\bibitem{Burrage:2016bwy}
C.~Burrage and J.~Sakstein, ``{A Compendium of Chameleon Constraints},''
  \href{http://dx.doi.org/10.1088/1475-7516/2016/11/045}{{\em JCAP} {\bfseries
  1611} no.~11, (2016) 045},
\href{http://arxiv.org/abs/1609.01192}{{\ttfamily arXiv:1609.01192
  [astro-ph.CO]}}.
%%CITATION = ARXIV:1609.01192;%%.

\bibitem{wald2010general}
R.~Wald, {\em General Relativity}.
\newblock University of Chicago Press, 2010.
\newblock \url{https://books.google.co.uk/books?id=9S-hzg6-moYC}.

\bibitem{Zumalacarregui:2013pma}
M.~Zumalacárregui and J.~García-Bellido, ``{Transforming gravity: from
  derivative couplings to matter to second-order scalar-tensor theories beyond
  the Horndeski Lagrangian},''
  \href{http://dx.doi.org/10.1103/PhysRevD.89.064046}{{\em Phys. Rev.}
  {\bfseries D89} (2014) 064046},
\href{http://arxiv.org/abs/1308.4685}{{\ttfamily arXiv:1308.4685 [gr-qc]}}.
%%CITATION = ARXIV:1308.4685;%%.

\bibitem{Hui:2009kc}
L.~Hui, A.~Nicolis, and C.~Stubbs, ``{Equivalence Principle Implications of
  Modified Gravity Models},''
  \href{http://dx.doi.org/10.1103/PhysRevD.80.104002}{{\em Phys. Rev.}
  {\bfseries D80} (2009) 104002},
\href{http://arxiv.org/abs/0905.2966}{{\ttfamily arXiv:0905.2966
  [astro-ph.CO]}}.
%%CITATION = ARXIV:0905.2966;%%.

\bibitem{Brax:2011aw}
P.~Brax, A.-C. Davis, and B.~Li, ``{Modified Gravity Tomography},''
  \href{http://dx.doi.org/10.1016/j.physletb.2012.08.002}{{\em Phys. Lett.}
  {\bfseries B715} (2012) 38--43},
\href{http://arxiv.org/abs/1111.6613}{{\ttfamily arXiv:1111.6613
  [astro-ph.CO]}}.
%%CITATION = ARXIV:1111.6613;%%.

\bibitem{Brax:2011ta}
P.~Brax and A.-C. Davis, ``{Modified Gravity and the CMB},''
  \href{http://dx.doi.org/10.1103/PhysRevD.85.023513}{{\em Phys. Rev.}
  {\bfseries D85} (2012) 023513},
\href{http://arxiv.org/abs/1109.5862}{{\ttfamily arXiv:1109.5862
  [astro-ph.CO]}}.
%%CITATION = ARXIV:1109.5862;%%.

\bibitem{Brax:2012gr}
P.~Brax, A.-C. Davis, B.~Li, and H.~A. Winther, ``{A Unified Description of
  Screened Modified Gravity},''
\href{http://arxiv.org/abs/1203.4812}{{\ttfamily arXiv:1203.4812
  [astro-ph.CO]}}.
%%CITATION = ARXIV:1203.4812;%%.

\bibitem{Babichev:2009fi}
E.~Babichev and D.~Langlois, ``{Relativistic stars in f(R) and scalar-tensor
  theories},'' \href{http://dx.doi.org/10.1103/PhysRevD.81.124051}{{\em Phys.
  Rev.} {\bfseries D81} (2010) 124051},
\href{http://arxiv.org/abs/0911.1297}{{\ttfamily arXiv:0911.1297 [gr-qc]}}.
%%CITATION = ARXIV:0911.1297;%%.

\bibitem{Minamitsuji:2016hkk}
M.~Minamitsuji and H.~O. Silva, ``{Relativistic stars in scalar-tensor theories
  with disformal coupling},''
  \href{http://dx.doi.org/10.1103/PhysRevD.93.124041}{{\em Phys. Rev.}
  {\bfseries D93} no.~12, (2016) 124041},
\href{http://arxiv.org/abs/1604.07742}{{\ttfamily arXiv:1604.07742 [gr-qc]}}.
%%CITATION = ARXIV:1604.07742;%%.

\bibitem{Babichev:2016jom}
E.~Babichev, K.~Koyama, D.~Langlois, R.~Saito, and J.~Sakstein, ``{Relativistic
  Stars in Beyond Horndeski Theories},''
  \href{http://dx.doi.org/10.1088/0264-9381/33/23/235014}{{\em Class. Quant.
  Grav.} {\bfseries 33} no.~23, (2016) 235014},
\href{http://arxiv.org/abs/1606.06627}{{\ttfamily arXiv:1606.06627 [gr-qc]}}.
%%CITATION = ARXIV:1606.06627;%%.

\bibitem{Sakstein:2016oel}
J.~Sakstein, E.~Babichev, K.~Koyama, D.~Langlois, and R.~Saito, ``{Towards
  Strong Field Tests of Beyond Horndeski Gravity Theories},''
  \href{http://dx.doi.org/10.1103/PhysRevD.95.064013}{{\em Phys. Rev.}
  {\bfseries D95} no.~6, (2017) 064013},
\href{http://arxiv.org/abs/1612.04263}{{\ttfamily arXiv:1612.04263 [gr-qc]}}.
%%CITATION = ARXIV:1612.04263;%%.

\bibitem{Brax:2017wcj}
P.~Brax, A.-C. Davis, and R.~Jha, ``{Neutron Stars in Screened Modified
  Gravity: Chameleon vs Dilaton},''
  \href{http://dx.doi.org/10.1103/PhysRevD.95.083514}{{\em Phys. Rev.}
  {\bfseries D95} no.~8, (2017) 083514},
\href{http://arxiv.org/abs/1702.02983}{{\ttfamily arXiv:1702.02983 [gr-qc]}}.
%%CITATION = ARXIV:1702.02983;%%.

\bibitem{Brax:2004qh}
P.~Brax, C.~van~de Bruck, A.-C. Davis, J.~Khoury, and A.~Weltman, ``{Detecting
  dark energy in orbit - The Cosmological chameleon},''
  \href{http://dx.doi.org/10.1103/PhysRevD.70.123518}{{\em Phys. Rev.}
  {\bfseries D70} (2004) 123518},
\href{http://arxiv.org/abs/astro-ph/0408415}{{\ttfamily arXiv:astro-ph/0408415
  [astro-ph]}}.
%%CITATION = ASTRO-PH/0408415;%%.

\bibitem{DeFelice:2010aj}
A.~De~Felice and S.~Tsujikawa, ``{f(R) theories},''
  \href{http://dx.doi.org/10.12942/lrr-2010-3}{{\em Living Rev. Rel.}
  {\bfseries 13} (2010) 3},
\href{http://arxiv.org/abs/1002.4928}{{\ttfamily arXiv:1002.4928 [gr-qc]}}.
%%CITATION = ARXIV:1002.4928;%%.

\bibitem{Hu:2007nk}
W.~Hu and I.~Sawicki, ``{Models of f(R) Cosmic Acceleration that Evade
  Solar-System Tests},''
  \href{http://dx.doi.org/10.1103/PhysRevD.76.064004}{{\em Phys. Rev.}
  {\bfseries D76} (2007) 064004},
\href{http://arxiv.org/abs/0705.1158}{{\ttfamily arXiv:0705.1158 [astro-ph]}}.
%%CITATION = ARXIV:0705.1158;%%.

\bibitem{Chiba:2003ir}
T.~Chiba, ``{1/R gravity and scalar - tensor gravity},''
  \href{http://dx.doi.org/10.1016/j.physletb.2003.09.033}{{\em Phys. Lett.}
  {\bfseries B575} (2003) 1--3},
\href{http://arxiv.org/abs/astro-ph/0307338}{{\ttfamily arXiv:astro-ph/0307338
  [astro-ph]}}.
%%CITATION = ASTRO-PH/0307338;%%.

\bibitem{Coleman:1973jx}
S.~R. Coleman and E.~J. Weinberg, ``{Radiative Corrections as the Origin of
  Spontaneous Symmetry Breaking},''
\href{http://dx.doi.org/10.1103/PhysRevD.7.1888}{{\em Phys. Rev.} {\bfseries
  D7} (1973) 1888--1910}.
%%CITATION = PHRVA,D7,1888;%%.

\bibitem{Upadhye:2012vh}
A.~Upadhye, W.~Hu, and J.~Khoury, ``{Quantum Stability of Chameleon Field
  Theories},'' \href{http://dx.doi.org/10.1103/PhysRevLett.109.041301}{{\em
  Phys. Rev. Lett.} {\bfseries 109} (2012) 041301},
\href{http://arxiv.org/abs/1204.3906}{{\ttfamily arXiv:1204.3906 [hep-ph]}}.
%%CITATION = ARXIV:1204.3906;%%.

\bibitem{Erickcek:2013dea}
A.~L. Erickcek, N.~Barnaby, C.~Burrage, and Z.~Huang, ``{Chameleons in the
  Early Universe: Kicks, Rebounds, and Particle Production},''
  \href{http://dx.doi.org/10.1103/PhysRevD.89.084074}{{\em Phys. Rev.}
  {\bfseries D89} no.~8, (2014) 084074},
\href{http://arxiv.org/abs/1310.5149}{{\ttfamily arXiv:1310.5149
  [astro-ph.CO]}}.
%%CITATION = ARXIV:1310.5149;%%.

\bibitem{Erickcek:2013oma}
A.~L. Erickcek, N.~Barnaby, C.~Burrage, and Z.~Huang, ``{Catastrophic
  Consequences of Kicking the Chameleon},''
  \href{http://dx.doi.org/10.1103/PhysRevLett.110.171101}{{\em Phys. Rev.
  Lett.} {\bfseries 110} (2013) 171101},
\href{http://arxiv.org/abs/1304.0009}{{\ttfamily arXiv:1304.0009
  [astro-ph.CO]}}.
%%CITATION = ARXIV:1304.0009;%%.

\bibitem{Miller:2016xpq}
C.~Miller and A.~L. Erickcek, ``{Quartic Chameleons: Safely Scale-Free in the
  Early Universe},'' \href{http://dx.doi.org/10.1103/PhysRevD.94.104049}{{\em
  Phys. Rev.} {\bfseries D94} no.~10, (2016) 104049},
\href{http://arxiv.org/abs/1607.07877}{{\ttfamily arXiv:1607.07877
  [astro-ph.CO]}}.
%%CITATION = ARXIV:1607.07877;%%.

\bibitem{Brax:2004ym}
P.~Brax, C.~van~de Bruck, and A.~C. Davis, ``{Is the radion a chameleon?},''
  \href{http://dx.doi.org/10.1088/1475-7516/2004/11/004}{{\em JCAP} {\bfseries
  0411} (2004) 004},
\href{http://arxiv.org/abs/astro-ph/0408464}{{\ttfamily arXiv:astro-ph/0408464
  [astro-ph]}}.
%%CITATION = ASTRO-PH/0408464;%%.

\bibitem{Conlon:2010jq}
J.~P. Conlon and F.~G. Pedro, ``{Moduli-Induced Vacuum Destabilisation},''
  \href{http://dx.doi.org/10.1007/JHEP05(2011)079}{{\em JHEP} {\bfseries 05}
  (2011) 079},
\href{http://arxiv.org/abs/1010.2665}{{\ttfamily arXiv:1010.2665 [hep-th]}}.
%%CITATION = ARXIV:1010.2665;%%.

\bibitem{Hinterbichler:2010wu}
K.~Hinterbichler, J.~Khoury, and H.~Nastase, ``{Towards a UV Completion for
  Chameleon Scalar Theories},''
  \href{http://dx.doi.org/10.1007/JHEP06(2011)072,
  10.1007/JHEP03(2011)061}{{\em JHEP} {\bfseries 03} (2011) 061},
  \href{http://arxiv.org/abs/1012.4462}{{\ttfamily arXiv:1012.4462 [hep-th]}}.
[Erratum: JHEP06,072(2011)].
%%CITATION = ARXIV:1012.4462;%%.

\bibitem{Nastase:2013los}
H.~Nastase and A.~Weltman, ``{A natural cosmological constant from
  chameleons},'' \href{http://dx.doi.org/10.1016/j.physletb.2015.05.066}{{\em
  Phys. Lett.} {\bfseries B747} (2015) 200--204},
\href{http://arxiv.org/abs/1302.1748}{{\ttfamily arXiv:1302.1748 [hep-th]}}.
%%CITATION = ARXIV:1302.1748;%%.

\bibitem{Nastase:2013ik}
H.~Nastase and A.~Weltman, ``{Chameleons on the Racetrack},''
  \href{http://dx.doi.org/10.1007/JHEP08(2013)059}{{\em JHEP} {\bfseries 08}
  (2013) 059},
\href{http://arxiv.org/abs/1301.7120}{{\ttfamily arXiv:1301.7120 [hep-th]}}.
%%CITATION = ARXIV:1301.7120;%%.

\bibitem{Brax:2012mq}
P.~Brax, A.-C. Davis, and J.~Sakstein, ``{SUPER-Screening},''
  \href{http://dx.doi.org/10.1016/j.physletb.2013.01.044}{{\em Phys. Lett.}
  {\bfseries B719} (2013) 210--217},
\href{http://arxiv.org/abs/1212.4392}{{\ttfamily arXiv:1212.4392 [hep-th]}}.
%%CITATION = ARXIV:1212.4392;%%.

\bibitem{Brax:2013yja}
P.~Brax, A.-C. Davis, and J.~Sakstein, ``{Dynamics of Supersymmetric
  Chameleons},'' \href{http://dx.doi.org/10.1088/1475-7516/2013/10/007}{{\em
  JCAP} {\bfseries 1310} (2013) 007},
\href{http://arxiv.org/abs/1302.3080}{{\ttfamily arXiv:1302.3080
  [astro-ph.CO]}}.
%%CITATION = ARXIV:1302.3080;%%.

\bibitem{Padilla:2015wlv}
A.~Padilla, E.~Platts, D.~Stefanyszyn, A.~Walters, A.~Weltman, and T.~Wilson,
  ``{How to Avoid a Swift Kick in the Chameleons},''
  \href{http://dx.doi.org/10.1088/1475-7516/2016/03/058}{{\em JCAP} {\bfseries
  1603} no.~03, (2016) 058},
\href{http://arxiv.org/abs/1511.05761}{{\ttfamily arXiv:1511.05761 [hep-th]}}.
%%CITATION = ARXIV:1511.05761;%%.

\bibitem{Burrage:2016xzz}
C.~Burrage, E.~J. Copeland, and P.~Millington, ``{Radiative Screening of Fifth
  Forces},'' \href{http://dx.doi.org/10.1103/PhysRevLett.117.211102}{{\em Phys.
  Rev. Lett.} {\bfseries 117} no.~21, (2016) 211102},
\href{http://arxiv.org/abs/1604.06051}{{\ttfamily arXiv:1604.06051 [gr-qc]}}.
%%CITATION = ARXIV:1604.06051;%%.

\bibitem{Garbrecht:2015yza}
B.~Garbrecht and P.~Millington, ``{Self-consistent solitons for vacuum decay in
  radiatively generated potentials},''
  \href{http://dx.doi.org/10.1103/PhysRevD.92.125022}{{\em Phys. Rev.}
  {\bfseries D92} (2015) 125022},
\href{http://arxiv.org/abs/1509.08480}{{\ttfamily arXiv:1509.08480 [hep-ph]}}.
%%CITATION = ARXIV:1509.08480;%%.

\bibitem{Brax:2009ey}
P.~Brax, C.~Burrage, A.-C. Davis, D.~Seery, and A.~Weltman, ``{Higgs production
  as a probe of Chameleon Dark Energy},''
  \href{http://dx.doi.org/10.1103/PhysRevD.81.103524}{{\em Phys. Rev.}
  {\bfseries D81} (2010) 103524},
\href{http://arxiv.org/abs/0911.1267}{{\ttfamily arXiv:0911.1267 [hep-ph]}}.
%%CITATION = ARXIV:0911.1267;%%.

\bibitem{Brax:2010uq}
P.~Brax, C.~Burrage, A.-C. Davis, D.~Seery, and A.~Weltman, ``{Anomalous
  coupling of scalars to gauge fields},''
  \href{http://dx.doi.org/10.1016/j.physletb.2011.03.047}{{\em Phys. Lett.}
  {\bfseries B699} (2011) 5--9},
\href{http://arxiv.org/abs/1010.4536}{{\ttfamily arXiv:1010.4536 [hep-th]}}.
%%CITATION = ARXIV:1010.4536;%%.

\bibitem{Kaplunovsky:1994fg}
V.~Kaplunovsky and J.~Louis, ``{Field dependent gauge couplings in locally
  supersymmetric effective quantum field theories},''
  \href{http://dx.doi.org/10.1016/0550-3213(94)00150-2}{{\em Nucl. Phys.}
  {\bfseries B422} (1994) 57--124},
\href{http://arxiv.org/abs/hep-th/9402005}{{\ttfamily arXiv:hep-th/9402005
  [hep-th]}}.
%%CITATION = HEP-TH/9402005;%%.

\bibitem{Davis:2011qf}
A.-C. Davis, E.~A. Lim, J.~Sakstein, and D.~Shaw, ``{Modified Gravity Makes
  Galaxies Brighter},''
  \href{http://dx.doi.org/10.1103/PhysRevD.85.123006}{{\em Phys. Rev.}
  {\bfseries D85} (2012) 123006},
\href{http://arxiv.org/abs/1102.5278}{{\ttfamily arXiv:1102.5278
  [astro-ph.CO]}}.
%%CITATION = ARXIV:1102.5278;%%.

\bibitem{Sakstein:2013pda}
J.~Sakstein, ``{Stellar Oscillations in Modified Gravity},''
  \href{http://dx.doi.org/10.1103/PhysRevD.88.124013}{{\em Phys. Rev.}
  {\bfseries D88} no.~12, (2013) 124013},
\href{http://arxiv.org/abs/1309.0495}{{\ttfamily arXiv:1309.0495
  [astro-ph.CO]}}.
%%CITATION = ARXIV:1309.0495;%%.

\bibitem{Li:2011pj}
B.~Li, G.-B. Zhao, and K.~Koyama, ``{Halos and Voids in f(R) Gravity},''
  \href{http://dx.doi.org/10.1111/j.1365-2966.2012.20573.x}{{\em Mon. Not. Roy.
  Astron. Soc.} {\bfseries 421} (2012) 3481},
\href{http://arxiv.org/abs/1111.2602}{{\ttfamily arXiv:1111.2602
  [astro-ph.CO]}}.
%%CITATION = ARXIV:1111.2602;%%.

\bibitem{Lombriser:2012nn}
L.~Lombriser, K.~Koyama, G.-B. Zhao, and B.~Li, ``{Chameleon f(R) gravity in
  the virialized cluster},''
  \href{http://dx.doi.org/10.1103/PhysRevD.85.124054}{{\em Phys. Rev.}
  {\bfseries D85} (2012) 124054},
\href{http://arxiv.org/abs/1203.5125}{{\ttfamily arXiv:1203.5125
  [astro-ph.CO]}}.
%%CITATION = ARXIV:1203.5125;%%.

\bibitem{Lombriser:2013wta}
L.~Lombriser, B.~Li, K.~Koyama, and G.-B. Zhao, ``{Modeling halo mass functions
  in chameleon f(R) gravity},''
  \href{http://dx.doi.org/10.1103/PhysRevD.87.123511}{{\em Phys. Rev.}
  {\bfseries D87} no.~12, (2013) 123511},
\href{http://arxiv.org/abs/1304.6395}{{\ttfamily arXiv:1304.6395
  [astro-ph.CO]}}.
%%CITATION = ARXIV:1304.6395;%%.

\bibitem{Cai:2014fma}
Y.-C. Cai, N.~Padilla, and B.~Li, ``{Testing Gravity using Cosmic Voids},''
  \href{http://dx.doi.org/10.1093/mnras/stv777}{{\em Mon. Not. Roy. Astron.
  Soc.} {\bfseries 451} no.~1, (2015) 1036--1055},
\href{http://arxiv.org/abs/1410.1510}{{\ttfamily arXiv:1410.1510
  [astro-ph.CO]}}.
%%CITATION = ARXIV:1410.1510;%%.

\bibitem{Cabre:2012tq}
A.~Cabre, V.~Vikram, G.-B. Zhao, B.~Jain, and K.~Koyama, ``{Astrophysical Tests
  of Modified Gravity: A Screening Map of the Nearby Universe},''
  \href{http://dx.doi.org/10.1088/1475-7516/2012/07/034}{{\em JCAP} {\bfseries
  1207} (2012) 034},
\href{http://arxiv.org/abs/1204.6046}{{\ttfamily arXiv:1204.6046
  [astro-ph.CO]}}.
%%CITATION = ARXIV:1204.6046;%%.

\bibitem{Schmidt:2010jr}
F.~Schmidt, ``{Dynamical Masses in Modified Gravity},''
  \href{http://dx.doi.org/10.1103/PhysRevD.81.103002}{{\em Phys. Rev.}
  {\bfseries D81} (2010) 103002},
\href{http://arxiv.org/abs/1003.0409}{{\ttfamily arXiv:1003.0409
  [astro-ph.CO]}}.
%%CITATION = ARXIV:1003.0409;%%.

\bibitem{padmanabhan2010gravitation}
T.~Padmanabhan, {\em Gravitation: Foundations and Frontiers}.
\newblock Cambridge University Press, 2010.
\newblock \url{https://books.google.com/books?id=BSfe2MjbQ3gC}.

\bibitem{Hees:2011mu}
A.~Hees and A.~Fuzfa, ``{Combined cosmological and solar system constraints on
  chameleon mechanism},''
  \href{http://dx.doi.org/10.1103/PhysRevD.85.103005}{{\em Phys. Rev.}
  {\bfseries D85} (2012) 103005},
\href{http://arxiv.org/abs/1111.4784}{{\ttfamily arXiv:1111.4784 [gr-qc]}}.
%%CITATION = ARXIV:1111.4784;%%.

\bibitem{Zhang:2016njn}
X.~Zhang, W.~Zhao, H.~Huang, and Y.~Cai, ``{Post-Newtonian parameters and
  cosmological constant of screened modified gravity},''
  \href{http://dx.doi.org/10.1103/PhysRevD.93.124003}{{\em Phys. Rev.}
  {\bfseries D93} no.~12, (2016) 124003},
\href{http://arxiv.org/abs/1603.09450}{{\ttfamily arXiv:1603.09450 [gr-qc]}}.
%%CITATION = ARXIV:1603.09450;%%.

\bibitem{Saaidi:2011zza}
K.~Saaidi, A.~Mohammadi, and H.~Sheikhahmadi, ``{$\gamma$ Parameter and Solar
  System constraint in Chameleon Brans Dick theory},''
  \href{http://dx.doi.org/10.1103/PhysRevD.83.104019}{{\em Phys. Rev.}
  {\bfseries D83} (2011) 104019},
\href{http://arxiv.org/abs/1201.0271}{{\ttfamily arXiv:1201.0271 [gr-qc]}}.
%%CITATION = ARXIV:1201.0271;%%.

\bibitem{Scharer:2014kya}
A.~Schärer, R.~Angélil, R.~Bondarescu, P.~Jetzer, and A.~Lundgren, ``{Testing
  scalar-tensor theories and parametrized post-Newtonian parameters in Earth
  orbit},'' \href{http://dx.doi.org/10.1103/PhysRevD.90.123005}{{\em Phys.
  Rev.} {\bfseries D90} no.~12, (2014) 123005},
\href{http://arxiv.org/abs/1410.7914}{{\ttfamily arXiv:1410.7914 [gr-qc]}}.
%%CITATION = ARXIV:1410.7914;%%.

\bibitem{Sakstein:2017pqi}
J.~Sakstein, ``{Tests of Gravity with Future Space-Based Experiments},''
\href{http://arxiv.org/abs/1710.03156}{{\ttfamily arXiv:1710.03156
  [astro-ph.CO]}}.
%%CITATION = ARXIV:1710.03156;%%.

\bibitem{Fomalont:2009zg}
E.~Fomalont, S.~Kopeikin, G.~Lanyi, and J.~Benson, ``{Progress in Measurements
  of the Gravitational Bending of Radio Waves Using the VLBA},''
  \href{http://dx.doi.org/10.1088/0004-637X/699/2/1395}{{\em Astrophys. J.}
  {\bfseries 699} (2009) 1395--1402},
\href{http://arxiv.org/abs/0904.3992}{{\ttfamily arXiv:0904.3992
  [astro-ph.CO]}}.
%%CITATION = ARXIV:0904.3992;%%.

\bibitem{Upadhye:2012rc}
A.~Upadhye, ``{Symmetron dark energy in laboratory experiments},''
  \href{http://dx.doi.org/10.1103/PhysRevLett.110.031301}{{\em Phys. Rev.
  Lett.} {\bfseries 110} no.~3, (2013) 031301},
\href{http://arxiv.org/abs/1210.7804}{{\ttfamily arXiv:1210.7804 [hep-ph]}}.
%%CITATION = ARXIV:1210.7804;%%.

\bibitem{Burrage:2016rkv}
C.~Burrage, A.~Kuribayashi-Coleman, J.~Stevenson, and B.~Thrussell,
  ``{Constraining symmetron fields with atom interferometry},''
  \href{http://dx.doi.org/10.1088/1475-7516/2016/12/041}{{\em JCAP} {\bfseries
  1612} (2016) 041},
\href{http://arxiv.org/abs/1609.09275}{{\ttfamily arXiv:1609.09275
  [astro-ph.CO]}}.
%%CITATION = ARXIV:1609.09275;%%.

\bibitem{Brax:2016wjk}
P.~Brax and A.-C. Davis, ``{Atomic Interferometry Test of Dark Energy},''
  \href{http://dx.doi.org/10.1103/PhysRevD.94.104069}{{\em Phys. Rev.}
  {\bfseries D94} no.~10, (2016) 104069},
\href{http://arxiv.org/abs/1609.09242}{{\ttfamily arXiv:1609.09242
  [astro-ph.CO]}}.
%%CITATION = ARXIV:1609.09242;%%.

\bibitem{Burrage:2014oza}
C.~Burrage, E.~J. Copeland, and E.~A. Hinds, ``{Probing Dark Energy with Atom
  Interferometry},''
  \href{http://dx.doi.org/10.1088/1475-7516/2015/03/042}{{\em JCAP} {\bfseries
  1503} no.~03, (2015) 042},
\href{http://arxiv.org/abs/1408.1409}{{\ttfamily arXiv:1408.1409
  [astro-ph.CO]}}.
%%CITATION = ARXIV:1408.1409;%%.

\bibitem{Kapner:2006si}
D.~J. Kapner, T.~S. Cook, E.~G. Adelberger, J.~H. Gundlach, B.~R. Heckel, C.~D.
  Hoyle, and H.~E. Swanson, ``{Tests of the gravitational inverse-square law
  below the dark-energy length scale},''
  \href{http://dx.doi.org/10.1103/PhysRevLett.98.021101}{{\em Phys. Rev. Lett.}
  {\bfseries 98} (2007) 021101},
\href{http://arxiv.org/abs/hep-ph/0611184}{{\ttfamily arXiv:hep-ph/0611184
  [hep-ph]}}.
%%CITATION = HEP-PH/0611184;%%.

\bibitem{Lambrecht:2005km}
A.~Lambrecht, V.~V. Nesvizhevsky, R.~Onofrio, and S.~Reynaud, ``{Development of
  a high-sensitivity torsional balance for the study of the Casimir force in
  the 1-10 micrometre range},''
\href{http://dx.doi.org/10.1088/0264-9381/22/24/012}{{\em Class. Quant. Grav.}
  {\bfseries 22} (2005) 5397--5406}.
%%CITATION = CQGRD,22,5397;%%.

\bibitem{Brax:2008hh}
P.~Brax, C.~van~de Bruck, A.-C. Davis, and D.~J. Shaw, ``{f(R) Gravity and
  Chameleon Theories},''
  \href{http://dx.doi.org/10.1103/PhysRevD.78.104021}{{\em Phys. Rev.}
  {\bfseries D78} (2008) 104021},
\href{http://arxiv.org/abs/0806.3415}{{\ttfamily arXiv:0806.3415 [astro-ph]}}.
%%CITATION = ARXIV:0806.3415;%%.

\bibitem{Adelberger:2006dh}
E.~G. Adelberger, B.~R. Heckel, S.~A. Hoedl, C.~D. Hoyle, D.~J. Kapner, and
  A.~Upadhye, ``{Particle Physics Implications of a Recent Test of the
  Gravitational Inverse Sqaure Law},''
  \href{http://dx.doi.org/10.1103/PhysRevLett.98.131104}{{\em Phys. Rev. Lett.}
  {\bfseries 98} (2007) 131104},
\href{http://arxiv.org/abs/hep-ph/0611223}{{\ttfamily arXiv:hep-ph/0611223
  [hep-ph]}}.
%%CITATION = HEP-PH/0611223;%%.

\bibitem{Mota:2006ed}
D.~F. Mota and D.~J. Shaw, ``{Strongly coupled chameleon fields: New horizons
  in scalar field theory},''
  \href{http://dx.doi.org/10.1103/PhysRevLett.97.151102}{{\em Phys. Rev. Lett.}
  {\bfseries 97} (2006) 151102},
\href{http://arxiv.org/abs/hep-ph/0606204}{{\ttfamily arXiv:hep-ph/0606204
  [hep-ph]}}.
%%CITATION = HEP-PH/0606204;%%.

\bibitem{Mota:2006fz}
D.~F. Mota and D.~J. Shaw, ``{Evading Equivalence Principle Violations,
  Cosmological and other Experimental Constraints in Scalar Field Theories with
  a Strong Coupling to Matter},''
  \href{http://dx.doi.org/10.1103/PhysRevD.75.063501}{{\em Phys. Rev.}
  {\bfseries D75} (2007) 063501},
\href{http://arxiv.org/abs/hep-ph/0608078}{{\ttfamily arXiv:hep-ph/0608078
  [hep-ph]}}.
%%CITATION = HEP-PH/0608078;%%.

\bibitem{Upadhye:2012fz}
A.~Upadhye, ``{Particles and forces from chameleon dark energy},'' in {\em {8th
  Patras Workshop on Axions, WIMPs and WISPs (AXION-WIMP 2012) Chicago,
  Illinois, July 18-22, 2012}}.
\newblock 2012.
\newblock \href{http://arxiv.org/abs/1211.7066}{{\ttfamily arXiv:1211.7066
  [hep-ph]}}.
\newblock
\url{https://inspirehep.net/record/1204935/files/arXiv:1211.7066.pdf}.
\newblock
%%CITATION = ARXIV:1211.7066;%%.

\bibitem{Upadhye:2012qu}
A.~Upadhye, ``{Dark energy fifth forces in torsion pendulum experiments},''
  \href{http://dx.doi.org/10.1103/PhysRevD.86.102003}{{\em Phys. Rev.}
  {\bfseries D86} (2012) 102003},
\href{http://arxiv.org/abs/1209.0211}{{\ttfamily arXiv:1209.0211 [hep-ph]}}.
%%CITATION = ARXIV:1209.0211;%%.

\bibitem{Lamoreaux:2005zza}
S.~K. Lamoreaux and W.~T. Buttler, ``{Thermal noise limitations to force
  measurements with torsion pendulums: Applications to the measurement of the
  Casimir force and its thermal correction},''
\href{http://dx.doi.org/10.1103/PhysRevE.71.036109}{{\em Phys. Rev.} {\bfseries
  E71} (2005) 036109}.
%%CITATION = PHRVA,E71,036109;%%.

\bibitem{Lambrecht:2011qm}
A.~Lambrecht and S.~Reynaud, ``{Casimir and short-range gravity tests},'' in
  {\em {Experimental Gravity and Gravitational Waves, p.199-206 (Th\'e Gioi,
  2011)}}.
\newblock 2011.
\newblock \href{http://arxiv.org/abs/1106.3848}{{\ttfamily arXiv:1106.3848
  [quant-ph]}}.
\newblock
\url{https://inspirehep.net/record/914261/files/arXiv:1106.3848.pdf}.
\newblock
%%CITATION = ARXIV:1106.3848;%%.

\bibitem{Brax:2007vm}
P.~Brax, C.~van~de Bruck, A.-C. Davis, D.~F. Mota, and D.~J. Shaw, ``{Detecting
  chameleons through Casimir force measurements},''
  \href{http://dx.doi.org/10.1103/PhysRevD.76.124034}{{\em Phys. Rev.}
  {\bfseries D76} (2007) 124034},
\href{http://arxiv.org/abs/0709.2075}{{\ttfamily arXiv:0709.2075 [hep-ph]}}.
%%CITATION = ARXIV:0709.2075;%%.

\bibitem{Brax:2014zta}
P.~Brax and A.-C. Davis, ``{Casimir, Gravitational and Neutron Tests of Dark
  Energy},'' \href{http://dx.doi.org/10.1103/PhysRevD.91.063503}{{\em Phys.
  Rev.} {\bfseries D91} no.~6, (2015) 063503},
\href{http://arxiv.org/abs/1412.2080}{{\ttfamily arXiv:1412.2080 [hep-ph]}}.
%%CITATION = ARXIV:1412.2080;%%.

\bibitem{Brax:2010xx}
P.~Brax, C.~van~de Bruck, A.~C. Davis, D.~J. Shaw, and D.~Iannuzzi, ``{Tuning
  the Mass of Chameleon Fields in Casimir Force Experiments},''
  \href{http://dx.doi.org/10.1103/PhysRevLett.104.241101}{{\em Phys. Rev.
  Lett.} {\bfseries 104} (2010) 241101},
\href{http://arxiv.org/abs/1003.1605}{{\ttfamily arXiv:1003.1605 [quant-ph]}}.
%%CITATION = ARXIV:1003.1605;%%.

\bibitem{Almasi:2015zpa}
A.~Almasi, P.~Brax, D.~Iannuzzi, and R.~I. Sedmik, ``{Force sensor for
  chameleon and Casimir force experiments with parallel-plate configuration},''
  \href{http://dx.doi.org/10.1103/PhysRevD.91.102002}{{\em Phys. Rev.}
  {\bfseries D91} no.~10, (2015) 102002},
\href{http://arxiv.org/abs/1505.01763}{{\ttfamily arXiv:1505.01763
  [physics.ins-det]}}.
%%CITATION = ARXIV:1505.01763;%%.

\bibitem{Geraci:2010ft}
A.~A. Geraci, S.~B. Papp, and J.~Kitching, ``{Short-range force detection using
  optically-cooled levitated microspheres},''
  \href{http://dx.doi.org/10.1103/PhysRevLett.105.101101}{{\em Phys. Rev.
  Lett.} {\bfseries 105} (2010) 101101},
\href{http://arxiv.org/abs/1006.0261}{{\ttfamily arXiv:1006.0261 [hep-ph]}}.
%%CITATION = ARXIV:1006.0261;%%.

\bibitem{Rider:2016xaq}
A.~D. Rider, D.~C. Moore, C.~P. Blakemore, M.~Louis, M.~Lu, and G.~Gratta,
  ``{Search for Screened Interactions Below the Dark Energy Length Scale Using
  Optically Levitated Microspheres},''
\href{http://arxiv.org/abs/1604.04908}{{\ttfamily arXiv:1604.04908 [hep-ex]}}.
%%CITATION = ARXIV:1604.04908;%%.

\bibitem{Brax:2010gp}
P.~Brax and C.~Burrage, ``{Atomic Precision Tests and Light Scalar
  Couplings},'' \href{http://dx.doi.org/10.1103/PhysRevD.83.035020}{{\em Phys.
  Rev.} {\bfseries D83} (2011) 035020},
\href{http://arxiv.org/abs/1010.5108}{{\ttfamily arXiv:1010.5108 [hep-ph]}}.
%%CITATION = ARXIV:1010.5108;%%.

\bibitem{Jaeckel:2010xx}
J.~Jaeckel and S.~Roy, ``{Spectroscopy as a test of Coulomb's law: A Probe of
  the hidden sector},''
  \href{http://dx.doi.org/10.1103/PhysRevD.82.125020}{{\em Phys. Rev.}
  {\bfseries D82} (2010) 125020},
\href{http://arxiv.org/abs/1008.3536}{{\ttfamily arXiv:1008.3536 [hep-ph]}}.
%%CITATION = ARXIV:1008.3536;%%.

\bibitem{Schwob:1999zz}
C.~Schwob, L.~Jozefowski, B.~de~Beauvoir, L.~Hilico, F.~Nez, L.~Julien,
  F.~Biraben, O.~Acef, J.~J. Zondy, and A.~Clairon, ``{Optical Frequency
  Measurement of the S-2- D-12 Transitions in Hydrogen and Deuterium: Rydberg
  Constant and Lamb Shift Determinations},''
\href{http://dx.doi.org/10.1103/PhysRevLett.82.4960}{{\em Phys. Rev. Lett.}
  {\bfseries 82} (1999) 4960--4963}.
%%CITATION = PRLTA,82,4960;%%.

\bibitem{Simon:1980hu}
G.~G. Simon, C.~Schmitt, F.~Borkowski, and V.~H. Walther, ``{Absolute electron
  Proton Cross-Sections at Low Momentum Transfer Measured with a High Pressure
  Gas Target System},''
\href{http://dx.doi.org/10.1016/0375-9474(80)90104-9}{{\em Nucl. Phys.}
  {\bfseries A333} (1980) 381--391}.
%%CITATION = NUPHA,A333,381;%%.

\bibitem{Burrage:2015lya}
C.~Burrage and E.~J. Copeland, ``{Using Atom Interferometry to Detect Dark
  Energy},'' \href{http://dx.doi.org/10.1080/00107514.2015.1060058}{{\em
  Contemp. Phys.} {\bfseries 57} (2016) 164},
\href{http://arxiv.org/abs/1507.07493}{{\ttfamily arXiv:1507.07493
  [astro-ph.CO]}}.
%%CITATION = ARXIV:1507.07493;%%.

\bibitem{Elder:2016yxm}
B.~Elder, J.~Khoury, P.~Haslinger, M.~Jaffe, H.~Müller, and P.~Hamilton,
  ``{Chameleon Dark Energy and Atom Interferometry},''
\href{http://arxiv.org/abs/1603.06587}{{\ttfamily arXiv:1603.06587
  [astro-ph.CO]}}.
%%CITATION = ARXIV:1603.06587;%%.

\bibitem{Hamilton:2015zga}
P.~Hamilton, M.~Jaffe, P.~Haslinger, Q.~Simmons, H.~Müller, and J.~Khoury,
  ``{Atom-interferometry constraints on dark energy},''
  \href{http://dx.doi.org/10.1126/science.aaa8883}{{\em Science} {\bfseries
  349} (2015) 849--851},
\href{http://arxiv.org/abs/1502.03888}{{\ttfamily arXiv:1502.03888
  [physics.atom-ph]}}.
%%CITATION = ARXIV:1502.03888;%%.

\bibitem{Jaffe:2016fsh}
M.~Jaffe, P.~Haslinger, V.~Xu, P.~Hamilton, A.~Upadhye, B.~Elder, J.~Khoury,
  and H.~Müller, ``{Testing sub-gravitational forces on atoms from a
  miniature, in-vacuum source mass},''
\href{http://arxiv.org/abs/1612.05171}{{\ttfamily arXiv:1612.05171
  [physics.atom-ph]}}.
%%CITATION = ARXIV:1612.05171;%%.

\bibitem{Brax:2011hb}
P.~Brax and G.~Pignol, ``{Strongly Coupled Chameleons and the Neutronic Quantum
  Bouncer},'' \href{http://dx.doi.org/10.1103/PhysRevLett.107.111301}{{\em
  Phys. Rev. Lett.} {\bfseries 107} (2011) 111301},
\href{http://arxiv.org/abs/1105.3420}{{\ttfamily arXiv:1105.3420 [hep-ph]}}.
%%CITATION = ARXIV:1105.3420;%%.

\bibitem{Ivanov:2012cb}
A.~N. Ivanov, R.~Hollwieser, T.~Jenke, M.~Wellenzohen, and H.~Abele,
  ``{Influence of the chameleon field potential on transition frequencies of
  gravitationally bound quantum states of ultracold neutrons},''
  \href{http://dx.doi.org/10.1103/PhysRevD.87.105013}{{\em Phys. Rev.}
  {\bfseries D87} no.~10, (2013) 105013},
\href{http://arxiv.org/abs/1207.0419}{{\ttfamily arXiv:1207.0419 [hep-th]}}.
%%CITATION = ARXIV:1207.0419;%%.

\bibitem{Jenke:2014yel}
T.~Jenke {\em et~al.}, ``{Gravity Resonance Spectroscopy Constrains Dark Energy
  and Dark Matter Scenarios},''
  \href{http://dx.doi.org/10.1103/PhysRevLett.112.151105}{{\em Phys. Rev.
  Lett.} {\bfseries 112} (2014) 151105},
\href{http://arxiv.org/abs/1404.4099}{{\ttfamily arXiv:1404.4099 [gr-qc]}}.
%%CITATION = ARXIV:1404.4099;%%.

\bibitem{Cronenberg:2015bol}
G.~Cronenberg, H.~Filter, M.~Thalhammer, T.~Jenke, H.~Abele, and P.~Geltenbort,
  ``{A Gravity of Earth Measurement with a qBOUNCE Experiment},'' {\em PoS}
  {\bfseries EPS-HEP2015} (2015) 408,
\href{http://arxiv.org/abs/1512.09134}{{\ttfamily arXiv:1512.09134 [hep-ex]}}.
%%CITATION = ARXIV:1512.09134;%%.

\bibitem{Pokotilovski:2013pra}
{\relax Yu}.~N. Pokotilovski, ``{Strongly coupled chameleon fields: Possible
  test with a neutron Lloydʼs mirror interferometer},''
  \href{http://dx.doi.org/10.1016/j.physletb.2013.01.022}{{\em Phys. Lett.}
  {\bfseries B719} (2013) 341--345},
\href{http://arxiv.org/abs/1203.5017}{{\ttfamily arXiv:1203.5017 [nucl-ex]}}.
%%CITATION = ARXIV:1203.5017;%%.

\bibitem{Brax:2013cfa}
P.~Brax, G.~Pignol, and D.~Roulier, ``{Probing Strongly Coupled Chameleons with
  Slow Neutrons},'' \href{http://dx.doi.org/10.1103/PhysRevD.88.083004}{{\em
  Phys. Rev.} {\bfseries D88} (2013) 083004},
\href{http://arxiv.org/abs/1306.6536}{{\ttfamily arXiv:1306.6536 [quant-ph]}}.
%%CITATION = ARXIV:1306.6536;%%.

\bibitem{Brax:2014gja}
P.~Brax, ``{Testing Chameleon Fields with Ultra Cold Neutron Bound States and
  Neutron Interferometry},''
\href{http://dx.doi.org/10.1016/j.phpro.2013.12.017}{{\em Phys. Procedia}
  {\bfseries 51} (2014) 73--77}.
%%CITATION = INSPIRE-1324524;%%.

\bibitem{Lemmel:2015kwa}
H.~Lemmel, P.~Brax, A.~N. Ivanov, T.~Jenke, G.~Pignol, M.~Pitschmann,
  T.~Potocar, M.~Wellenzohn, M.~Zawisky, and H.~Abele, ``{Neutron
  Interferometry constrains dark energy chameleon fields},''
  \href{http://dx.doi.org/10.1016/j.physletb.2015.02.063}{{\em Phys. Lett.}
  {\bfseries B743} (2015) 310--314},
\href{http://arxiv.org/abs/1502.06023}{{\ttfamily arXiv:1502.06023 [hep-ph]}}.
%%CITATION = ARXIV:1502.06023;%%.

\bibitem{Li:2016tux}
K.~Li {\em et~al.}, ``{Neutron Limit on the Strongly-Coupled Chameleon
  Field},'' \href{http://dx.doi.org/10.1103/PhysRevD.93.062001}{{\em Phys.
  Rev.} {\bfseries D93} no.~6, (2016) 062001},
\href{http://arxiv.org/abs/1601.06897}{{\ttfamily arXiv:1601.06897
  [astro-ph.CO]}}.
%%CITATION = ARXIV:1601.06897;%%.

\bibitem{Brax:2013uh}
P.~Brax, A.-C. Davis, and J.~Sakstein, ``{Pulsar Constraints on Screened
  Modified Gravity},''
  \href{http://dx.doi.org/10.1088/0264-9381/31/22/225001}{{\em Class. Quant.
  Grav.} {\bfseries 31} (2014) 225001},
\href{http://arxiv.org/abs/1301.5587}{{\ttfamily arXiv:1301.5587 [gr-qc]}}.
%%CITATION = ARXIV:1301.5587;%%.

\bibitem{Zhang:2017srh}
X.~Zhang, T.~Liu, and W.~Zhao, ``{Gravitational radiation from compact binary
  systems in screened modified gravity},''
  \href{http://dx.doi.org/10.1103/PhysRevD.95.104027}{{\em Phys. Rev.}
  {\bfseries D95} no.~10, (2017) 104027},
\href{http://arxiv.org/abs/1702.08752}{{\ttfamily arXiv:1702.08752 [gr-qc]}}.
%%CITATION = ARXIV:1702.08752;%%.

\bibitem{1999IAUS..183...48S}
S.~{Sakai}, ``{The Tip of the Red Giant Branch as a Population II Distance
  Indicator},'' in {\em Cosmological Parameters and the Evolution of the
  Universe}, K.~{Sato}, ed., vol.~183 of {\em IAU Symposium}, p.~48.
\newblock 1999.

\bibitem{Freedman:2010xv}
W.~L. Freedman and B.~F. Madore, ``{The Hubble Constant},''
  \href{http://dx.doi.org/10.1146/annurev-astro-082708-101829}{{\em
  Ann.Rev.Astron.Astrophys.} {\bfseries 48} (2010) 673--710},
\href{http://arxiv.org/abs/1004.1856}{{\ttfamily arXiv:1004.1856
  [astro-ph.CO]}}.
%%CITATION = ARXIV:1004.1856;%%.

\bibitem{Beaton:2016nsw}
R.~L. Beaton {\em et~al.}, ``{The Carnegie-Chicago Hubble Program. I. An
  Independent Approach to the Extragalactic Distance Scale Using only
  Population II Distance Indicators},''
  \href{http://dx.doi.org/10.3847/0004-637X/832/2/210}{{\em Astrophys. J.}
  {\bfseries 832} no.~2, (2016) 210},
\href{http://arxiv.org/abs/1604.01788}{{\ttfamily arXiv:1604.01788
  [astro-ph.CO]}}.
%%CITATION = ARXIV:1604.01788;%%.

\bibitem{Jain:2012tn}
B.~Jain, V.~Vikram, and J.~Sakstein, ``{Astrophysical Tests of Modified
  Gravity: Constraints from Distance Indicators in the Nearby Universe},''
  \href{http://dx.doi.org/10.1088/0004-637X/779/1/39}{{\em Astrophys. J.}
  {\bfseries 779} (2013) 39},
\href{http://arxiv.org/abs/1204.6044}{{\ttfamily arXiv:1204.6044
  [astro-ph.CO]}}.
%%CITATION = ARXIV:1204.6044;%%.

\bibitem{Vikram:2014uza}
V.~Vikram, J.~Sakstein, C.~Davis, and A.~Neil, ``{Astrophysical Tests of
  Modified Gravity: Stellar and Gaseous Rotation Curves in Dwarf Galaxies},''
\href{http://arxiv.org/abs/1407.6044}{{\ttfamily arXiv:1407.6044
  [astro-ph.CO]}}.
%%CITATION = ARXIV:1407.6044;%%.

\bibitem{Wilcox:2016guw}
H.~Wilcox, R.~C. Nichol, G.-B. Zhao, D.~Bacon, K.~Koyama, and A.~K. Romer,
  ``{Simulation tests of galaxy cluster constraints on chameleon gravity},''
  \href{http://dx.doi.org/10.1093/mnras/stw1617}{{\em Mon. Not. Roy. Astron.
  Soc.} {\bfseries 462} no.~1, (2016) 715--725},
\href{http://arxiv.org/abs/1603.05911}{{\ttfamily arXiv:1603.05911
  [astro-ph.CO]}}.
%%CITATION = ARXIV:1603.05911;%%.

\bibitem{Terukina:2013eqa}
A.~Terukina, L.~Lombriser, K.~Yamamoto, D.~Bacon, K.~Koyama, and R.~C. Nichol,
  ``{Testing chameleon gravity with the Coma cluster},''
  \href{http://dx.doi.org/10.1088/1475-7516/2014/04/013}{{\em JCAP} {\bfseries
  1404} (2014) 013},
\href{http://arxiv.org/abs/1312.5083}{{\ttfamily arXiv:1312.5083
  [astro-ph.CO]}}.
%%CITATION = ARXIV:1312.5083;%%.

\bibitem{Lombriser:2014dua}
L.~Lombriser, ``{Constraining chameleon models with cosmology},''
  \href{http://dx.doi.org/10.1002/andp.201400058}{{\em Annalen Phys.}
  {\bfseries 526} (2014) 259--282},
\href{http://arxiv.org/abs/1403.4268}{{\ttfamily arXiv:1403.4268
  [astro-ph.CO]}}.
%%CITATION = ARXIV:1403.4268;%%.

\bibitem{Smith:2009fn}
T.~L. Smith, ``{Testing gravity on kiloparsec scales with strong gravitational
  lenses},''
\href{http://arxiv.org/abs/0907.4829}{{\ttfamily arXiv:0907.4829
  [astro-ph.CO]}}.
%%CITATION = ARXIV:0907.4829;%%.

\bibitem{Schmidt:2008tn}
F.~Schmidt, M.~V. Lima, H.~Oyaizu, and W.~Hu, ``{Non-linear Evolution of f(R)
  Cosmologies III: Halo Statistics},''
  \href{http://dx.doi.org/10.1103/PhysRevD.79.083518}{{\em Phys. Rev.}
  {\bfseries D79} (2009) 083518},
\href{http://arxiv.org/abs/0812.0545}{{\ttfamily arXiv:0812.0545 [astro-ph]}}.
%%CITATION = ARXIV:0812.0545;%%.

\bibitem{Schmidt:2009sv}
F.~Schmidt, ``{Cosmological Simulations of Normal-Branch Braneworld Gravity},''
  \href{http://dx.doi.org/10.1103/PhysRevD.80.123003}{{\em Phys. Rev.}
  {\bfseries D80} (2009) 123003},
\href{http://arxiv.org/abs/0910.0235}{{\ttfamily arXiv:0910.0235
  [astro-ph.CO]}}.
%%CITATION = ARXIV:0910.0235;%%.

\bibitem{Lombriser:2011zw}
L.~Lombriser, F.~Schmidt, T.~Baldauf, R.~Mandelbaum, U.~Seljak, and R.~E.
  Smith, ``{Cluster Density Profiles as a Test of Modified Gravity},''
  \href{http://dx.doi.org/10.1103/PhysRevD.85.102001}{{\em Phys. Rev.}
  {\bfseries D85} (2012) 102001},
\href{http://arxiv.org/abs/1111.2020}{{\ttfamily arXiv:1111.2020
  [astro-ph.CO]}}.
%%CITATION = ARXIV:1111.2020;%%.

\bibitem{Schmidt:2009am}
F.~Schmidt, A.~Vikhlinin, and W.~Hu, ``{Cluster Constraints on f(R) Gravity},''
  \href{http://dx.doi.org/10.1103/PhysRevD.80.083505}{{\em Phys. Rev.}
  {\bfseries D80} (2009) 083505},
\href{http://arxiv.org/abs/0908.2457}{{\ttfamily arXiv:0908.2457
  [astro-ph.CO]}}.
%%CITATION = ARXIV:0908.2457;%%.

\bibitem{Cataneo:2014kaa}
M.~Cataneo, D.~Rapetti, F.~Schmidt, A.~B. Mantz, S.~W. Allen, D.~E. Applegate,
  P.~L. Kelly, A.~von~der Linden, and R.~G. Morris, ``{New constraints on
  $f(R)$ gravity from clusters of galaxies},''
  \href{http://dx.doi.org/10.1103/PhysRevD.92.044009}{{\em Phys. Rev.}
  {\bfseries D92} no.~4, (2015) 044009},
\href{http://arxiv.org/abs/1412.0133}{{\ttfamily arXiv:1412.0133
  [astro-ph.CO]}}.
%%CITATION = ARXIV:1412.0133;%%.

\bibitem{Zhang:2005vt}
P.~Zhang, ``{Testing $f(R)$ gravity against the large scale structure of the
  universe.},'' \href{http://dx.doi.org/10.1103/PhysRevD.73.123504}{{\em Phys.
  Rev.} {\bfseries D73} (2006) 123504},
\href{http://arxiv.org/abs/astro-ph/0511218}{{\ttfamily arXiv:astro-ph/0511218
  [astro-ph]}}.
%%CITATION = ASTRO-PH/0511218;%%.

\bibitem{Song:2007da}
Y.-S. Song, H.~Peiris, and W.~Hu, ``{Cosmological Constraints on f(R)
  Acceleration Models},''
  \href{http://dx.doi.org/10.1103/PhysRevD.76.063517}{{\em Phys. Rev.}
  {\bfseries D76} (2007) 063517},
\href{http://arxiv.org/abs/0706.2399}{{\ttfamily arXiv:0706.2399 [astro-ph]}}.
%%CITATION = ARXIV:0706.2399;%%.

\bibitem{Dossett:2014oia}
J.~Dossett, B.~Hu, and D.~Parkinson, ``{Constraining models of f(R) gravity
  with Planck and WiggleZ power spectrum data},''
  \href{http://dx.doi.org/10.1088/1475-7516/2014/03/046}{{\em JCAP} {\bfseries
  1403} (2014) 046},
\href{http://arxiv.org/abs/1401.3980}{{\ttfamily arXiv:1401.3980
  [astro-ph.CO]}}.
%%CITATION = ARXIV:1401.3980;%%.

\bibitem{Raveri:2014cka}
M.~Raveri, B.~Hu, N.~Frusciante, and A.~Silvestri, ``{Effective Field Theory of
  Cosmic Acceleration: constraining dark energy with CMB data},''
  \href{http://dx.doi.org/10.1103/PhysRevD.90.043513}{{\em Phys. Rev.}
  {\bfseries D90} no.~4, (2014) 043513},
\href{http://arxiv.org/abs/1405.1022}{{\ttfamily arXiv:1405.1022
  [astro-ph.CO]}}.
%%CITATION = ARXIV:1405.1022;%%.

\bibitem{Silvestri:2011ch}
A.~Silvestri, ``{Scalar radiation from Chameleon-shielded regions},''
  \href{http://dx.doi.org/10.1103/PhysRevLett.106.251101}{{\em Phys. Rev.
  Lett.} {\bfseries 106} (2011) 251101},
\href{http://arxiv.org/abs/1103.4013}{{\ttfamily arXiv:1103.4013
  [astro-ph.CO]}}.
%%CITATION = ARXIV:1103.4013;%%.

\bibitem{Upadhye:2013nfa}
A.~Upadhye and J.~H. Steffen, ``{Monopole radiation in modified gravity},''
\href{http://arxiv.org/abs/1306.6113}{{\ttfamily arXiv:1306.6113
  [astro-ph.CO]}}.
%%CITATION = ARXIV:1306.6113;%%.

\bibitem{Jennings:2012pt}
E.~Jennings, C.~M. Baugh, B.~Li, G.-B. Zhao, and K.~Koyama, ``{Redshift space
  distortions in f(R) gravity},''
  \href{http://dx.doi.org/10.1111/j.1365-2966.2012.21567.x}{{\em Mon. Not. Roy.
  Astron. Soc.} {\bfseries 425} (2012) 2128--2143},
\href{http://arxiv.org/abs/1205.2698}{{\ttfamily arXiv:1205.2698
  [astro-ph.CO]}}.
%%CITATION = ARXIV:1205.2698;%%.

\bibitem{Bose:2016qun}
B.~Bose and K.~Koyama, ``{A Perturbative Approach to the Redshift Space Power
  Spectrum: Beyond the Standard Model},''
  \href{http://dx.doi.org/10.1088/1475-7516/2016/08/032}{{\em JCAP} {\bfseries
  1608} no.~08, (2016) 032},
\href{http://arxiv.org/abs/1606.02520}{{\ttfamily arXiv:1606.02520
  [astro-ph.CO]}}.
%%CITATION = ARXIV:1606.02520;%%.

\bibitem{Bose:2017dtl}
B.~Bose and K.~Koyama, ``{A Perturbative Approach to the Redshift Space
  Correlation Function: Beyond the Standard Model},''
\href{http://arxiv.org/abs/1705.09181}{{\ttfamily arXiv:1705.09181
  [astro-ph.CO]}}.
%%CITATION = ARXIV:1705.09181;%%.

\bibitem{Yamamoto:2010ie}
K.~Yamamoto, G.~Nakamura, G.~Hutsi, T.~Narikawa, and T.~Sato, ``{Constraint on
  the cosmological f(R) model from the multipole power spectrum of the SDSS
  luminous red galaxy sample and prospects for a future redshift survey},''
  \href{http://dx.doi.org/10.1103/PhysRevD.81.103517}{{\em Phys. Rev.}
  {\bfseries D81} (2010) 103517},
\href{http://arxiv.org/abs/1004.3231}{{\ttfamily arXiv:1004.3231
  [astro-ph.CO]}}.
%%CITATION = ARXIV:1004.3231;%%.

\bibitem{Xu:2014wda}
L.~Xu, ``{Constraint on $f(R)$ Gravity through the Redshift Space
  Distortion},'' \href{http://dx.doi.org/10.1103/PhysRevD.91.063008}{{\em Phys.
  Rev.} {\bfseries D91} no.~6, (2015) 063008},
\href{http://arxiv.org/abs/1411.4353}{{\ttfamily arXiv:1411.4353
  [astro-ph.CO]}}.
%%CITATION = ARXIV:1411.4353;%%.

\bibitem{Zavattini:2005tm}
{\bfseries PVLAS} Collaboration, E.~Zavattini {\em et~al.}, ``{Experimental
  observation of optical rotation generated in vacuum by a magnetic field},''
  \href{http://dx.doi.org/10.1103/PhysRevLett.99.129901,
  10.1103/PhysRevLett.96.110406}{{\em Phys. Rev. Lett.} {\bfseries 96} (2006)
  110406}, \href{http://arxiv.org/abs/hep-ex/0507107}{{\ttfamily
  arXiv:hep-ex/0507107 [hep-ex]}}.
[Erratum: Phys. Rev. Lett.99,129901(2007)].
%%CITATION = HEP-EX/0507107;%%.

\bibitem{Raffelt:1987im}
G.~Raffelt and L.~Stodolsky, ``{Mixing of the Photon with Low Mass
  Particles},''
\href{http://dx.doi.org/10.1103/PhysRevD.37.1237}{{\em Phys. Rev.} {\bfseries
  D37} (1988) 1237}.
%%CITATION = PHRVA,D37,1237;%%.

\bibitem{Brax:2007ak}
P.~Brax, C.~van~de Bruck, and A.-C. Davis, ``{Compatibility of the
  chameleon-field model with fifth-force experiments, cosmology, and PVLAS and
  CAST results},'' \href{http://dx.doi.org/10.1103/PhysRevLett.99.121103}{{\em
  Phys. Rev. Lett.} {\bfseries 99} (2007) 121103},
\href{http://arxiv.org/abs/hep-ph/0703243}{{\ttfamily arXiv:hep-ph/0703243
  [HEP-PH]}}.
%%CITATION = HEP-PH/0703243;%%.

\bibitem{Brax:2007hi}
P.~Brax, C.~van~de Bruck, A.-C. Davis, D.~F. Mota, and D.~J. Shaw, ``{Testing
  Chameleon Theories with Light Propagating through a Magnetic Field},''
  \href{http://dx.doi.org/10.1103/PhysRevD.76.085010}{{\em Phys. Rev.}
  {\bfseries D76} (2007) 085010},
\href{http://arxiv.org/abs/0707.2801}{{\ttfamily arXiv:0707.2801 [hep-ph]}}.
%%CITATION = ARXIV:0707.2801;%%.

\bibitem{Gies:2007su}
H.~Gies, D.~F. Mota, and D.~J. Shaw, ``{Hidden in the Light: Magnetically
  Induced Afterglow from Trapped Chameleon Fields},''
  \href{http://dx.doi.org/10.1103/PhysRevD.77.025016}{{\em Phys. Rev.}
  {\bfseries D77} (2008) 025016},
\href{http://arxiv.org/abs/0710.1556}{{\ttfamily arXiv:0710.1556 [hep-ph]}}.
%%CITATION = ARXIV:0710.1556;%%.

\bibitem{Ahlers:2007st}
M.~Ahlers, A.~Lindner, A.~Ringwald, L.~Schrempp, and C.~Weniger, ``{Alpenglow -
  A Signature for Chameleons in Axion-Like Particle Search Experiments},''
  \href{http://dx.doi.org/10.1103/PhysRevD.77.015018}{{\em Phys. Rev.}
  {\bfseries D77} (2008) 015018},
\href{http://arxiv.org/abs/0710.1555}{{\ttfamily arXiv:0710.1555 [hep-ph]}}.
%%CITATION = ARXIV:0710.1555;%%.

\bibitem{Upadhye:2009iv}
A.~Upadhye, J.~H. Steffen, and A.~Weltman, ``{Constraining chameleon field
  theories using the GammeV afterglow experiments},''
  \href{http://dx.doi.org/10.1103/PhysRevD.81.015013}{{\em Phys. Rev.}
  {\bfseries D81} (2010) 015013},
\href{http://arxiv.org/abs/0911.3906}{{\ttfamily arXiv:0911.3906 [hep-ph]}}.
%%CITATION = ARXIV:0911.3906;%%.

\bibitem{Chou:2008gr}
{\bfseries GammeV} Collaboration, A.~S. Chou {\em et~al.}, ``{A Search for
  chameleon particles using a photon regeneration technique},''
  \href{http://dx.doi.org/10.1103/PhysRevLett.102.030402}{{\em Phys. Rev.
  Lett.} {\bfseries 102} (2009) 030402},
\href{http://arxiv.org/abs/0806.2438}{{\ttfamily arXiv:0806.2438 [hep-ex]}}.
%%CITATION = ARXIV:0806.2438;%%.

\bibitem{Steffen:2010ze}
{\bfseries GammeV} Collaboration, J.~H. Steffen, A.~Upadhye, A.~Baumbaugh,
  A.~S. Chou, P.~O. Mazur, R.~Tomlin, A.~Weltman, and W.~Wester, ``{Laboratory
  constraints on chameleon dark energy and power-law fields},''
  \href{http://dx.doi.org/10.1103/PhysRevLett.105.261803}{{\em Phys. Rev.
  Lett.} {\bfseries 105} (2010) 261803},
\href{http://arxiv.org/abs/1010.0988}{{\ttfamily arXiv:1010.0988
  [astro-ph.CO]}}.
%%CITATION = ARXIV:1010.0988;%%.

\bibitem{Upadhye:2012ar}
A.~Upadhye, J.~H. Steffen, and A.~S. Chou, ``{Designing dark energy afterglow
  experiments},'' \href{http://dx.doi.org/10.1103/PhysRevD.86.035006}{{\em
  Phys. Rev.} {\bfseries D86} (2012) 035006},
\href{http://arxiv.org/abs/1204.5476}{{\ttfamily arXiv:1204.5476 [hep-ph]}}.
%%CITATION = ARXIV:1204.5476;%%.

\bibitem{Brax:2013tsa}
P.~Brax and A.~Upadhye, ``{Chameleon Fragmentation},''
  \href{http://dx.doi.org/10.1088/1475-7516/2014/02/018}{{\em JCAP} {\bfseries
  1402} (2014) 018},
\href{http://arxiv.org/abs/1312.2747}{{\ttfamily arXiv:1312.2747 [hep-ph]}}.
%%CITATION = ARXIV:1312.2747;%%.

\bibitem{Asztalos:2009yp}
{\bfseries ADMX} Collaboration, S.~J. Asztalos {\em et~al.}, ``{A SQUID-based
  microwave cavity search for dark-matter axions},''
  \href{http://dx.doi.org/10.1103/PhysRevLett.104.041301}{{\em Phys. Rev.
  Lett.} {\bfseries 104} (2010) 041301},
\href{http://arxiv.org/abs/0910.5914}{{\ttfamily arXiv:0910.5914
  [astro-ph.CO]}}.
%%CITATION = ARXIV:0910.5914;%%.

\bibitem{Asztalos:2003px}
{\bfseries ADMX} Collaboration, S.~J. Asztalos {\em et~al.}, ``{An Improved RF
  cavity search for halo axions},''
  \href{http://dx.doi.org/10.1103/PhysRevD.69.011101}{{\em Phys. Rev.}
  {\bfseries D69} (2004) 011101},
\href{http://arxiv.org/abs/astro-ph/0310042}{{\ttfamily arXiv:astro-ph/0310042
  [astro-ph]}}.
%%CITATION = ASTRO-PH/0310042;%%.

\bibitem{Sikivie:1983ip}
P.~Sikivie, ``{Experimental Tests of the Invisible Axion},''
  \href{http://dx.doi.org/10.1103/PhysRevLett.51.1415,
  10.1103/PhysRevLett.52.695.2}{{\em Phys. Rev. Lett.} {\bfseries 51} (1983)
  1415--1417}.
[Erratum: Phys. Rev. Lett.52,695(1984)].
%%CITATION = PRLTA,51,1415;%%.

\bibitem{Rybka:2010ah}
{\bfseries ADMX} Collaboration, G.~Rybka {\em et~al.}, ``{A Search for Scalar
  Chameleons with ADMX},''
  \href{http://dx.doi.org/10.1103/PhysRevLett.105.051801}{{\em Phys. Rev.
  Lett.} {\bfseries 105} (2010) 051801},
\href{http://arxiv.org/abs/1004.5160}{{\ttfamily arXiv:1004.5160
  [astro-ph.CO]}}.
%%CITATION = ARXIV:1004.5160;%%.

\bibitem{Zioutas:2004hi}
{\bfseries CAST} Collaboration, K.~Zioutas {\em et~al.}, ``{First results from
  the CERN Axion Solar Telescope (CAST)},''
  \href{http://dx.doi.org/10.1103/PhysRevLett.94.121301}{{\em Phys. Rev. Lett.}
  {\bfseries 94} (2005) 121301},
\href{http://arxiv.org/abs/hep-ex/0411033}{{\ttfamily arXiv:hep-ex/0411033
  [hep-ex]}}.
%%CITATION = HEP-EX/0411033;%%.

\bibitem{Brax:2010xq}
P.~Brax and K.~Zioutas, ``{Solar Chameleons},''
  \href{http://dx.doi.org/10.1103/PhysRevD.82.043007}{{\em Phys. Rev.}
  {\bfseries D82} (2010) 043007},
\href{http://arxiv.org/abs/1004.1846}{{\ttfamily arXiv:1004.1846
  [astro-ph.SR]}}.
%%CITATION = ARXIV:1004.1846;%%.

\bibitem{Anastassopoulos:2015yda}
{\bfseries CAST} Collaboration, V.~Anastassopoulos {\em et~al.}, ``{Search for
  chameleons with CAST},''
  \href{http://dx.doi.org/10.1016/j.physletb.2015.07.049}{{\em Phys. Lett.}
  {\bfseries B749} (2015) 172--180},
\href{http://arxiv.org/abs/1503.04561}{{\ttfamily arXiv:1503.04561
  [astro-ph.SR]}}.
%%CITATION = ARXIV:1503.04561;%%.

\bibitem{Baum:2014rka}
S.~Baum, G.~Cantatore, D.~H.~H. Hoffmann, M.~Karuza, Y.~K. Semertzidis,
  A.~Upadhye, and K.~Zioutas, ``{Detecting solar chameleons through radiation
  pressure},'' \href{http://dx.doi.org/10.1016/j.physletb.2014.10.055}{{\em
  Phys. Lett.} {\bfseries B739} (2014) 167--173},
\href{http://arxiv.org/abs/1409.3852}{{\ttfamily arXiv:1409.3852
  [astro-ph.IM]}}.
%%CITATION = ARXIV:1409.3852;%%.

\bibitem{Karuza:2015sia}
M.~Karuza, G.~Cantatore, A.~Gardikiotis, D.~H.~H. Hoffmann, Y.~K. Semertzidis,
  and K.~Zioutas, ``{KWISP: an ultra-sensitive force sensor for the Dark Energy
  sector},'' \href{http://dx.doi.org/10.1016/j.dark.2016.02.004}{{\em Phys.
  Dark Univ.} {\bfseries 12} (2016) 100--104},
\href{http://arxiv.org/abs/1509.04499}{{\ttfamily arXiv:1509.04499
  [physics.ins-det]}}.
%%CITATION = ARXIV:1509.04499;%%.

\bibitem{Brax:2009aw}
P.~Brax, C.~Burrage, A.-C. Davis, D.~Seery, and A.~Weltman, ``{Collider
  constraints on interactions of dark energy with the Standard Model},''
  \href{http://dx.doi.org/10.1088/1126-6708/2009/09/128}{{\em JHEP} {\bfseries
  09} (2009) 128},
\href{http://arxiv.org/abs/0904.3002}{{\ttfamily arXiv:0904.3002 [hep-ph]}}.
%%CITATION = ARXIV:0904.3002;%%.

\bibitem{Burrage:2008ii}
C.~Burrage, A.-C. Davis, and D.~J. Shaw, ``{Detecting Chameleons: The
  Astronomical Polarization Produced by Chameleon-like Scalar Fields},''
  \href{http://dx.doi.org/10.1103/PhysRevD.79.044028}{{\em Phys. Rev.}
  {\bfseries D79} (2009) 044028},
\href{http://arxiv.org/abs/0809.1763}{{\ttfamily arXiv:0809.1763 [astro-ph]}}.
%%CITATION = ARXIV:0809.1763;%%.

\bibitem{Steffen:2006px}
A.~T. Steffen, I.~Strateva, W.~N. Brandt, D.~M. Alexander, A.~M. Koekemoer,
  B.~D. Lehmer, D.~P. Schneider, and C.~Vignali, ``{The x-ray-to-optical
  properties of optically-selected active galaxies over wide luminosity and
  redshift ranges},'' \href{http://dx.doi.org/10.1086/503627}{{\em Astron. J.}
  {\bfseries 131} (2006) 2826--2842},
\href{http://arxiv.org/abs/astro-ph/0602407}{{\ttfamily arXiv:astro-ph/0602407
  [astro-ph]}}.
%%CITATION = ASTRO-PH/0602407;%%.

\bibitem{Young:2009jc}
M.~Young, M.~Elvis, and G.~Risaliti, ``{The DR5 SDSS/XMM-Newton Quasar
  Survey},'' \href{http://dx.doi.org/10.1088/0067-0049/183/1/17}{{\em
  Astrophys. J. Suppl.} {\bfseries 183} (2009) 17},
\href{http://arxiv.org/abs/0905.0496}{{\ttfamily arXiv:0905.0496
  [astro-ph.HE]}}.
%%CITATION = ARXIV:0905.0496;%%.

\bibitem{Burrage:2009mj}
C.~Burrage, A.-C. Davis, and D.~J. Shaw, ``{Active Galactic Nuclei Shed Light
  on Axion-like-Particles},''
  \href{http://dx.doi.org/10.1103/PhysRevLett.102.201101}{{\em Phys. Rev.
  Lett.} {\bfseries 102} (2009) 201101},
\href{http://arxiv.org/abs/0902.2320}{{\ttfamily arXiv:0902.2320
  [astro-ph.CO]}}.
%%CITATION = ARXIV:0902.2320;%%.

\bibitem{Pettinari:2010ay}
G.~W. Pettinari and R.~Crittenden, ``{On the Evidence for Axion-like Particles
  from Active Galactic Nuclei},''
  \href{http://dx.doi.org/10.1103/PhysRevD.82.083502}{{\em Phys. Rev.}
  {\bfseries D82} (2010) 083502},
\href{http://arxiv.org/abs/1007.0024}{{\ttfamily arXiv:1007.0024
  [astro-ph.CO]}}.
%%CITATION = ARXIV:1007.0024;%%.

\bibitem{Avgoustidis:2010ju}
A.~Avgoustidis, C.~Burrage, J.~Redondo, L.~Verde, and R.~Jimenez,
  ``{Constraints on cosmic opacity and beyond the standard model physics from
  cosmological distance measurements},''
  \href{http://dx.doi.org/10.1088/1475-7516/2010/10/024}{{\em JCAP} {\bfseries
  1010} (2010) 024},
\href{http://arxiv.org/abs/1004.2053}{{\ttfamily arXiv:1004.2053
  [astro-ph.CO]}}.
%%CITATION = ARXIV:1004.2053;%%.

\bibitem{Schelpe:2010he}
C.~A.~O. Schelpe, ``{Chameleon-Photon Mixing in a Primordial Magnetic Field},''
  \href{http://dx.doi.org/10.1103/PhysRevD.82.044033}{{\em Phys. Rev.}
  {\bfseries D82} (2010) 044033},
\href{http://arxiv.org/abs/1003.0232}{{\ttfamily arXiv:1003.0232
  [astro-ph.CO]}}.
%%CITATION = ARXIV:1003.0232;%%.

\bibitem{Davis:2010nj}
A.-C. Davis, C.~A.~O. Schelpe, and D.~J. Shaw, ``{The Chameleonic Contribution
  to the SZ Radial Profile of the Coma Cluster},''
  \href{http://dx.doi.org/10.1103/PhysRevD.83.044006}{{\em Phys. Rev.}
  {\bfseries D83} (2011) 044006},
\href{http://arxiv.org/abs/1008.1880}{{\ttfamily arXiv:1008.1880
  [astro-ph.CO]}}.
%%CITATION = ARXIV:1008.1880;%%.

\bibitem{Jain:2011ji}
B.~Jain and J.~VanderPlas, ``{Tests of Modified Gravity with Dwarf Galaxies},''
  \href{http://dx.doi.org/10.1088/1475-7516/2011/10/032}{{\em JCAP} {\bfseries
  1110} (2011) 032},
\href{http://arxiv.org/abs/1106.0065}{{\ttfamily arXiv:1106.0065
  [astro-ph.CO]}}.
%%CITATION = ARXIV:1106.0065;%%.

\bibitem{Vikram:2013uba}
V.~Vikram, A.~Cabré, B.~Jain, and J.~T. VanderPlas, ``{Astrophysical Tests of
  Modified Gravity: the Morphology and Kinematics of Dwarf Galaxies},''
  \href{http://dx.doi.org/10.1088/1475-7516/2013/08/020}{{\em JCAP} {\bfseries
  1308} (2013) 020},
\href{http://arxiv.org/abs/1303.0295}{{\ttfamily arXiv:1303.0295
  [astro-ph.CO]}}.
%%CITATION = ARXIV:1303.0295;%%.

\bibitem{Hellwing:2014nma}
W.~A. Hellwing, A.~Barreira, C.~S. Frenk, B.~Li, and S.~Cole, ``{A clear and
  measurable signature of modified gravity in the galaxy velocity field},''
  \href{http://dx.doi.org/10.1103/PhysRevLett.112.221102}{{\em Phys. Rev.
  Lett.} {\bfseries 112} (2014) 221102},
\href{http://arxiv.org/abs/1401.0706}{{\ttfamily arXiv:1401.0706
  [astro-ph.CO]}}.
%%CITATION = ARXIV:1401.0706;%%.

\end{thebibliography}\endgroup
\bibliographystyle{utphys}
\addcontentsline{toc}{section}{References}

\end{document}